\documentclass{aa}  

\usepackage{graphicx}
\usepackage{txfonts}

 \newcommand{\izw}{I\,Zw\,18}
\newcommand{\1}{{~\sc i}}
\newcommand{\2}{{~\sc ii}}
\newcommand{\3}{{~\sc iii}}

\newcommand{\5}{{~\sc v}}
\newcommand{\vect}[1]{\protect\overrightarrow{#1}}
\newcommand{\mic}{{\,$\mu$m}}

\usepackage{natbib}
\bibpunct{(}{)}{;}{a}{}{,} 

\defcitealias{Lebouteiller2017a}{L17}
\defcitealias{Pequignot2008a}{P08}

\begin{document}

\title{Topological models to infer multiphase interstellar medium properties}


   \author{V. Lebouteiller
          \inst{1}
          \and
          L. Ramambason\inst{1}
          }

          \institute{AIM, CEA, CNRS, Université Paris-Saclay, Université Paris Diderot, Sorbonne Paris Cité, F-91191 Gif-sur-Yvette, France
            \email{vianney.lebouteiller@cea.fr}
             }

   \date{Received 26/04/2022; accepted 01/07/2022}

 
  \abstract
  {Spectroscopic observations of high-redshift galaxies slowly reveal the same complexity of the interstellar medium (ISM) as expected from resolved observations in nearby galaxies. While providing, in principle, a wealth of diagnostics concerning galaxy evolution, star formation, or the nature and influence of compact objects, such high-z spectra are often spatially and spectrally unresolved, and inferring reliable diagnostics represents a major obstacle. Bright, nearby, unresolved galaxies observed in the optical and infrared domains provide many constraints to design methods to infer ISM properties, but they have so far been limited to deterministic methods and/or with simple topological assumptions (e.g., single 1D model). }
   {It is urgent to build upon previous ISM multiphase and multicomponent methods by using a probabilistic approach that makes it possible to derive probability density functions for relevant parameters while also enabling a large number of free parameters with potential priors. The goal is to provide a flexible statistical framework that is agnostic to the model grid and that considers either a few discrete components defined by their parameter values and/or statistical distributions of parameters. In this paper, we present a first application with the objective to infer probability distributions of several physical parameters (e.g., the mass of H$^0$, H$_2$, escape fraction of ionizing photons, and metallicity) for the star-forming regions of the metal-poor dwarf galaxy \izw\ in order to confirm the low molecular gas content and high escape fraction of ionizing photons from H\2\ regions. }
   {We present a Bayesian approach to model a suite of spectral lines using a sequential Monte Carlo method provided by the Python package PyMC which combines several concepts such as tempered likelihoods, importance sampling, and independent Metropolis-Hastings chains. The algorithm, provided by the associated code MULTIGRIS, accepts a few components which can be represented as sectors around one or several stellar clusters, or continuous (e.g., power-law, normal) distributions for any given parameter. We applied this approach to a grid of models calculated with the photoionization and photodissociation code Cloudy in order to produce topological models of \izw. }
   {The statistical framework we present makes it possible to consider a large number of spectroscopic tracers, with the extinction and systematic uncertainties as potential additional random variables. We applied this technique to the galaxy \izw\ in order to reproduce and go beyond previous topological models specifically tailored to this object. While our grid is designed for global properties of low-metallicity star-forming galaxies, we were able to calculate accurate values for the metallicity, number of ionizing photons, masses of ionized and neutral hydrogen, as well as the dust mass and the dust-to-gas mass ratio in \izw. We find a relatively modest amount of H$_2$ ($\sim10^5$\,M$_\odot$) which is predominantly CO-dark and traced by C$^+$ rather than C$^0$. Nevertheless, more than $90$\%\ of the [C\2] emission is associated with the neutral atomic gas. Our models confirm the necessity to include an X-ray source with an inferred luminosity in good agreement with direct X-ray observations. Finally, we investigate the escape fraction of ionizing photons for different energy ranges. While the escape fraction for the main H\2\ region lies around $50-65$\%, we show that most of the soft X-ray photons are able to escape and may play a role in the ionization and heating of the circumgalactic or intergalactic medium.  }
   {Multicomponent ISM models associate a complex enough distribution of matter and phases with a simple enough topological description to be constrained with probabilistic frameworks. Despite ignoring effects such as reflected light, the diffuse radiation field, or ionization by several non-cospatial sources, they remain well adapted to individual H\2\ regions and to star-forming galaxies dominated by one or a few H\2\ regions, and the improvement due to the combination of several components largely compensates for other secondary effects. }


   \maketitle

%

\section{Introduction}

Spectroscopic observations of very high redshift ($z\gtrsim7$) galaxies have become routinely available (e.g., \citealt{Harikane2018a,Wagg2020a,Lee2021a}), with an increasing number of tracers arising from different phases of the interstellar medium (ISM), enabling numerous diagnostics (e.g., star formation rate, active galactic nuclei, and molecular gas content) that had been so far limited to galaxy surveys around or after the cosmic noon ($z\lesssim2-3$). The integrated emission-line spectrum of galaxies, in particular, makes it possible to infer chemical abundances, physical conditions, and relative contributions of energetic sources at work in galaxies, which in turn provide useful constraints on cosmological evolution of galaxies (e.g., \citealt{Spinoglio2017a}) or to examine specific processes such as the escape of ionizing photons (e.g., \citealt{Harikane2019a}) and the nature and influence of compact objects in low-metallicity environments (e.g., \citealt{Lebouteiller2017a}).

The integrated spectrum reflects a combination of (1) many interstellar components with different properties (e.g., ionization parameter, abundances, column density, and incident radiation field) that need to be disentangled, and (2) tracers with different emission conditions (e.g., critical density and excitation temperature) that need to be consistently accounted for. It is often assumed either that integrated measurements can be linked to some kind of loosely defined ``average'' parameter or that some regions with specific properties (e.g., H\2\ regions) dominate the galaxy emission. The modeling of integrated spectra is thus often limited to a single cloud hypothesis (i.e., assuming that the physical conditions of a single cloud correspond to the average conditions of the galaxy). Specific diagnostics involving only a few tracers are sometimes considered to isolate a specific parameter (e.g., density measurements with optical [S\2] lines, ionization parameter, or metallicity; e.g., \citealt{Kewley2019a}) or to isolate a specific region or phase (e.g., low-density ionized gas with far-IR [N\2] lines; e.g., \citealt{Lee2019b}). Nevertheless, the derived conditions may not always represent a meaningful average galaxy value due to multiple nonlinear effects that produce integrated emission lines.

Possible ways forward include global approaches treating all observables coherently, such as what is done for the full spectral energy distribution (SED) with, for example, MAGPHYS \citep{daCunha2008a} or BEAST \citep{Gordon2016a}. While many tools are available to infer physical conditions from the SED (see review in \citealt{Walcher2011a}), the treatment of emission lines is, however, not systematic and emission lines are sometimes included only to decontaminate observed photometry bands for full SED modeling \citep{Burgarella2005a}. Using a large number of emission lines as constraints is a difficult task due to the complex, multiphase nature of the ISM. For instance, efforts are underway to consider photodissociation regions or molecular gas in CIGALE \citep{Boquien2019a}. Concerning emission lines models specifically, efforts to treat many lines at once have been mostly focused on the ionized gas, including with modern statistical frameworks (e.g., BOND, \citealt{ValeAsari2016a}, NebulaBayes, \citealt{Thomas2019a}, and WARPFIELD, \citealt{Kang2022a}). Still, some excitation mechanisms are often not considered (e.g., shocks and X-rays) and a single-sector approach is often preferred due to the quickly overwhelming number of free parameters and the difficulty in defining the distribution of parameters.

It is thus urgent to continue and build upon such efforts especially since multiwavelength and multiphase ISM studies in external galaxies have been facilitated by the advent of infrared space instruments enabling access to neutral gas tracers such as the Infrared Spectrograph (IRS; \citealt{Houck2004a}) onboard \textit{Spitzer} \citep{Werner2004a}, the Photodetector Array Camera and Spectrometer (PACS; \citealt{Poglitsch2010a}) onboard \textit{Herschel} \citep{Pilbratt2010a} and airborne ones such as FIFI-LS onboard SOFIA \citep{Fischer2018a}, as well as millimeter observatories that provide multiphase tracers at high-z (e.g., ALMA and IRAM/NOEMA). 

Several projects have been undergone over recent years to infer physical parameters from IR and optical spectra of nearby galaxies. Studies focusing on low-metallicity dwarf galaxies from the Dwarf Galaxy Survey \citep{Madden2013a} pioneered the multiphase models of the ISM in external galaxies. We modeled simultaneously the ionized and neutral (atomic and molecular) gas with the photoionization and photodissociation code Cloudy \citep{Ferland2013a} while considering increasingly complex arrangements of regions with different characteristics. Since a realistic description of the ISM structure for a given galaxy is impossible to infer, simplifications are necessary. For instance, one can assume that the ionizing sources are cospatial. Assuming galaxies are dominated by one or a few massive H\2\ regions, our approach has then been to consider a stellar cluster surrounded by a small number of sectors defined by their covering factor (fraction of the total solid angle seen by the cluster), with the incident radiation field propagating into each -- independent -- sector. The various iterations are described in detail in Section\,\ref{sec:previous}. This approach had been devised by \cite{Pequignot2008a}, who showed that difficulties in explaining $T_e$([O\3]) in the dwarf galaxy \izw\ can be lifted using radiation-bounded condensations embedded in a low-density matter-bounded diffuse medium. Following the approach of \cite{Pequignot2008a}, the principle is to constrain an ISM topology, which represents the relative contribution to the total integrated emission from discrete sectors having different physical conditions. In other words, same results will be obtained for a given topology representing different geometries (i.e., the way in which these sectors are distributed, intertwined, or split; Sect.\,\ref{sec:ismtopo}).

In most cases, the number of available tracers and their different origins (e.g., ionized/neutral and diffuse/dense phases) require at least two sectors with different properties even when the integrated spectrum appears to be dominated by a single H\2\ region. In our studies, the reflected and diffuse light were not accounted for but the associated uncertainties are largely compensated for by improvements enabled by a multiphase/sector approach. While the ISM topology description above is adapted to H\2\ region-dominated galaxies, several efforts are underway to extend the sectors into a diffuse component and to combine several H\2\ regions. Our approach has been so far to combine a rather small number of discrete sectors, but another approach considers that the integrated emission-line spectrum effectively corresponds to an vast ensemble of clouds whose properties (i.e., distribution of density and distance to source) are described and linked through a specific parameterized law (e.g., power law; \citealt{Richardson2014a,Richardson2016a}). The observed emission is then the result of strong selection effects due to the fact that some lines emit preferentially under some physical conditions. 

As such models become increasingly complex, it is interesting to realize how they complement 3D simulations. While simulations do consider self-consistent and realistic ISM structures, the treatment of the radiative transfer is often too calculation-intensive. On the other hand, state-of-the-art radiative transfer codes (e.g., \citealt{Ercolano2005a,Bisbas2012a,Bisbas2015a,Jin2022a}) are difficult to scale to 3D \citep{Morisset2011a,Morisset2013a} or even to adapt to multiple components since the exact distribution of matter and ionizing/energetic sources has to be predefined. Simulations and models could eventually meet halfway provided the latter are able to consider complex geometries. As the complexity increases (e.g., by increasing the number of sectors, stellar populations, or excitation mechanisms), however, it becomes important to implement a statistically robust method to identify high-probability regions in the parameter space. Previous efforts to model galaxies with multiple sectors and phases have mostly relied on $\chi^2$ methods, with several limitations described in Section\,\ref{sec:previous}. We have thus designed a new statistical framework to evaluate solutions in a multidimensional grid and to quantify confidence intervals in a multiphase/sector approach. The approach is close to NebulaBayes \citep{Thomas2019a} but here adapted to complex configurations with several sectors with varying escape fraction of ionizing photons, that is with high dimensionality and several continuous variables that can be implemented in a customized environment. The tool we present is agnostic to the grid of models used as input (i.e., it is possible to use grids from different photoionization codes or any grid for which sets of input parameters and output values can be tabulated). We present here ISM applications, with a focus on the low-metallicity dwarf galaxy \izw. The latter has been extensively studied and open questions include in particular the amount and distribution of the elusive molecular gas. 

We summarize the various steps and results using topological models of galaxies in Section\,\ref{sec:previous}. We then describe the new probabilistic method in Section\,\ref{sec:method} and present specific applications to ISM studies in Section\,\ref{sec:ismapp}. Benchmark results are presented in Section\,\ref{sec:benchmark}. Finally, we illustrate the use of the new method with the dwarf galaxy \izw\ in Section\,\ref{sec:i18app}.

\section{Modeling approach and previous results}\label{sec:previous}

\begin{figure}
   \centering
\includegraphics[width=0.5\textwidth]{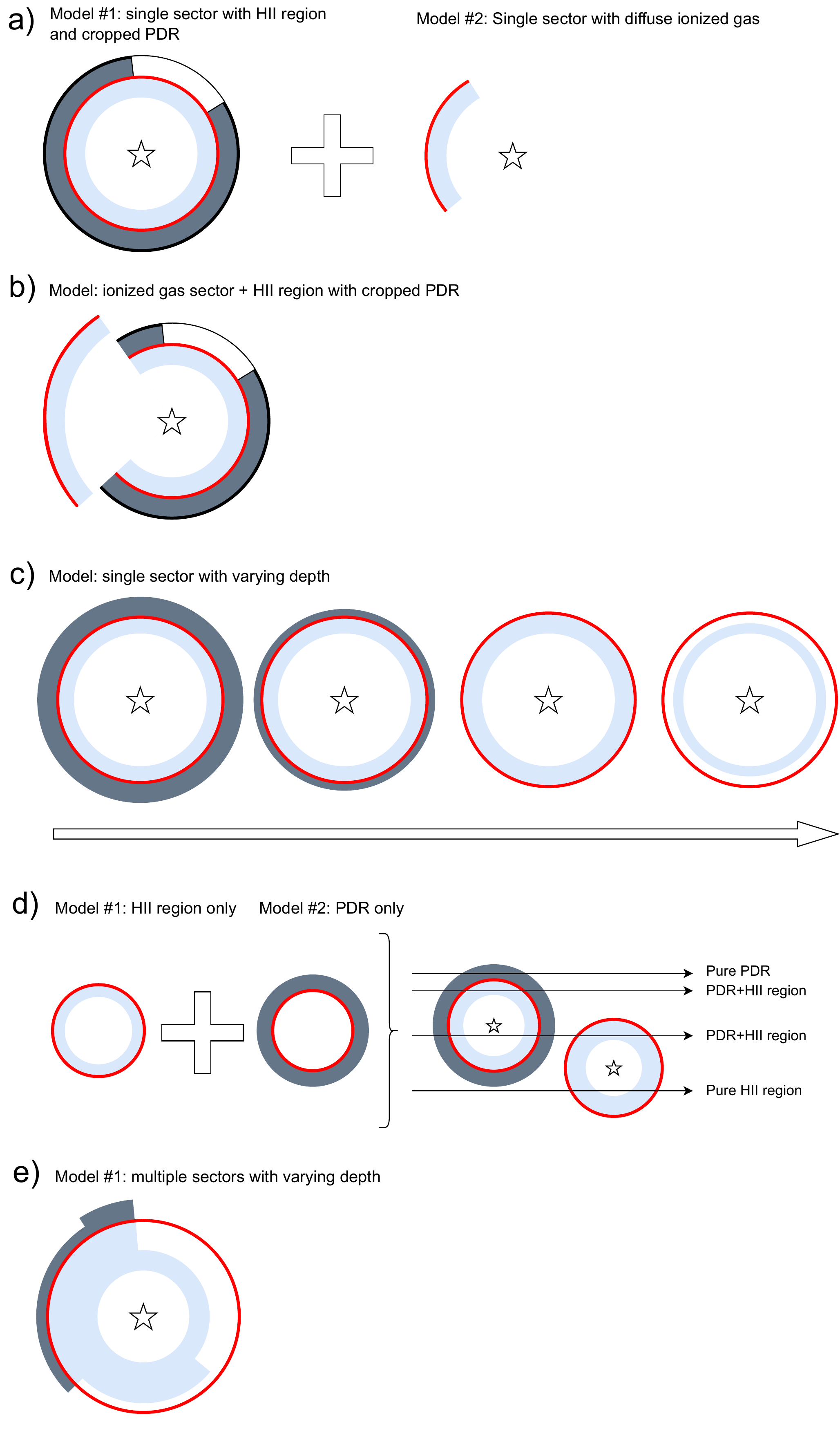}
\caption{Illustration of the topological configurations considered in previous studies. For each approach, we show the relative contributions of sectors each corresponding to a specific Cloudy model. These contributions are equivalent to the fraction of the total solid angle in 3D. The light blue shell corresponds to the ionized gas, the dark gray shell to the neutral gas, and the red arc to the photoionization front. In \cite{Cormier2012a} (a), the model used an H\2\ region connected with a PDR, with the PDR covering factor scaled a posteriori (cropped shell in white), and an additional model is considered a posteriori to account for the diffuse ionized gas. In \cite{Cormier2019a} (b), two sectors are considered with the PDR associated only with the high-ionization parameter sector. In \cite{Polles2019a} (c), models use a single sector with a varying depth. In \cite{LambertHuyghe2022a} (d), observations are compared to a mix of pure H\2\ region and PDR components drawn from the same Cloudy radiation-bounded model. In \cite{Lebouteiller2017a} (e), up to four sectors were considered with different depths. }
         \label{fig:topos}
       \end{figure}

In order to motivate the general approach presented in this study, we illustrate below recent progresses in the modeling approach with multiple phases and sectors (summarized in Fig.\,\ref{fig:topos}). 
The first step was to study spatially-unresolved emission lines in the dwarf galaxy Haro\,11 \citep{Cormier2012b}. We were able to construct a model with Cloudy that matches the flux of $17$ IR emission lines. The full model implements the propagation of the radiation field into a single sector from a dense ionized gas into the photodissociation region (PDR). Since the model includes only one sector, the modeled ``PDR covering factor'' is one by definition. However, the model is constrained using only unambiguous H\2\ region lines and the predictions for the PDR lines [C\2] and [O\1] are compared to the observations a posteriori to calculate an effective PDR covering factor. The model suggests that the PDR covering factor is about $10\%$ in this galaxy. A diffuse ionized/neutral phase was included a posteriori to fully explain the emission of [N\2], [Ne\2], and [C\2]. Structure-wise, the volume filling factors of the diffuse component, H\2\ region, and the PDR are $\gtrsim90\%$, $\sim0.2\%$, and $\lesssim0.01\%$ respectively, hinting at a ``porous'' ISM (small filling factor of the dense phase) which could be linked to the relatively low-metallicity environment.
   
In \cite{Cormier2019a}, the technique was improved and applied to most of the DGS targets which were fully covered by the IRS and PACS spectrograph apertures ($38$ spatially-unresolved galaxies; \citealt{Madden2013a}). The IR emission lines are modeled with two radiation-bounded sectors distinguished by their ionization parameter and densities. The PDR is assumed to be associated with the high-ionization parameter sector (i.e., the PDR is clipped from the low-ionization parameter sector). Possible PDR covering factor values, corresponding to a scaling factor of all PDR lines, are then tabulated along with other physical parameters (e.g., density and ionization parameter). Contrary to the iterative process to characterize the components needed in Haro\,11, all sector parameter values are determined simultaneously. The main result shows that the PDR covering factor increases with metallicity, strengthening the result that the ISM porosity decreases at low metallicity. However, it was not possible to convert this parameter into a fraction of escaping ionizing photons.
   
In \cite{Polles2019a}, we studied IR spectral cubes of the nearby ($\approx700$\,kpc) galaxy IC\,10. The H\2\ region lines are modeled with a single sector, as in Haro\,11, but the sector is allowed to be matter-bounded. As such, the physical depth of the sector is another free parameter in addition to cluster age, density etc... This new parameter allowed us to determine that regions are more and more radiation-bounded as the spatial aperture over which the emission lines are integrated increases. In other words, H\2\ regions are systematically matter-bounded and the full optically body scale is mostly radiation-bounded. While this should translate into a similar conclusion for the escape of ionizing photons, the calculation of the latter was unfortunately not possible.

We have also examined IR emission lines in star-forming regions in the Magellanic Clouds \citep{LambertHuyghe2022a}. The observations are spatially resolved and individual pixels are modeled with Cloudy with a single sector that can be matter-bounded. The particularity of the approach is that each Cloudy model is split between its H\2\ region and PDR counterparts and each pixel is modeled with a mix of these. As an illustration, a pixel can be a naked H\2\ region (matter-bounded) or a pure PDR with no associated ionized gas emission (reflecting the fact that the ionizing sources may be located in another pixel). Any combination in between is possible, but, for the sake of this specific project, the H\2\ region and PDR counterparts for any given pixel originate from the same Cloudy model. 

Finally, the study of \izw\ \citep{Lebouteiller2017a} was inspired by the work of \cite{Pequignot2008a}, with a conversion of the Nebu radiative transfer code to Cloudy, and with the inclusion of new IR emission lines, X-ray constraints, and dust treatment. Several sectors are used, some of them matter-bounded, and both optical and IR emission lines were used. While the model is relatively more complex compared to other efforts, the topology has not been inferred statistically but rather from a partly-manual and specific convergence process. Structure-wise, the model shows that the covering factor of the sector reaching into the molecular gas must be lower than $\sim0.1\%$, corresponding to subparsec size clumps, thereby confirming the low porosity of the ISM in an extremely metal-poor environment. 

The studies described above allowed us to make considerable progress for ISM studies. As far as the technique is concerned, calibration uncertainties and other correlated uncertainties have been successfully accounted for with a covariance matrix. However, several limitations have been identified, concerning in particular the $\chi^2$ minimization technique to evaluate the solution and associated error bars. Firstly, upper limits were often ignored or else treated in a binary way (i.e., either the model is above or below the upper limit). Secondly, the number of degrees of freedom is difficult to estimate, even for linear models \citep{Andrae2010b}, and some observations may trace the same physical parameter while a given observation may constrain several parameters (this difficulty increases when priors need to be considered). Degeneracy and correlation between parameters are generally difficult to account for. Due to the large number of free parameters and to the difficulty in assessing similar or degenerate solutions, it was difficult to consider combinations of more than two components (i.e., either dealing with more sectors and/or more stellar populations). Thirdly, the type of constraints (ratios and/or absolute values) and choice of tracers (ignoring or adding some tracers) often produce erratic probability density functions with solutions that are not compatible with each other within errors. Fourthly, the $\chi^2$ distribution is difficult to translate robustly into a confidence/credible interval. Finally, priors could not be used to inform the model of known quantities with uncertainties.

In the present study, we present a new probabilistic tool called MULTIGRIS\footnote{\url{https://gitlab.com/multigris}} to solve most of the issues above. The underlying objective is to derive automatically a configuration as complex as the one illustrated in Figure\,\ref{fig:topos}e without the need to tabulate some parameters (in particular the mixing weight of components (i.e., their relative contributions) which can be treated as continuous variables instead. For a more general discussion of deterministic and Bayesian approaches in astrophysics we refer to \cite{Galliano2022a}.

\section{Methodology}\label{sec:method}

\subsection{General principle}

The goal is to evaluate the posterior distribution of parameters from a multidimensional grid of individual precalculated models, with a series of observations (measured values with uncertainties) as constraints. Primary parameters that define a unique set of models are used for the inference, with each parameter value drawn from a Monte-Carlo Markov Chain (MCMC). Mixing weights may be used as additional inference parameters to combine model prediction sets. Predicted observations are then computed as a linear combination of prediction sets and compared to measured values accounting for their error bars (and potentially upper/lower limits)..
MULTIGRIS is completely agnostic to the grid of precomputed models used as long as predicted observations can be expressed as a linear combination of prediction sets. In practice, the code can therefore perform on any kind of grid for which input (primary) parameters and output values (predictions) are tabulated.

Although one would prefer to independently compute predicted values on-the-fly for each set of drawn parameter values, the computing time may be too long for a time-efficient MCMC sampling. In our approach, the sampler explores the grid of precomputed models either with nearest neighbor or linear interpolation and with several additional continuous variables (e.g., mixing weights for the relative contribution of components or any nuisance variables). The sampling process, which is required to draw continuous random variables such as the mixing weight or any nuisance variable, is much slower than algorithms such as NebulaBayes \citep{Thomas2019a}, for which likelihoods are evaluated at a predefined number of grid points, with or without interpolation. Such ``brute-force'' algorithms remain ideal when the dimensionality is relatively low and/or when all combination of parameters can be precalculated and tabulated. The main advantage of the sampling is to consider continuous or discrete variables without the need to tabulate them and the ability to implement new on-the-fly distributions using the same input grid. 

\subsection{Inference}\label{sec:inference}

The Bayesian inference is performed with PyMC\footnote{\url{https://www.pymc.io/}} which implements and enables gradient-based MCMC algorithms in an intuitive and relatively abstract way \citep{Salvatier2016a}. The predicted value for each observation $i$ is defined as:
\begin{equation}
  M_i \sim s \sum\limits_{c=1}^{N_{\rm comps}} w_c f(\vect{\theta_c}),
\end{equation}
where $f$ is a function that retrieves the predicted observation for any set of primary parameters from a grid using a given interpolation method, $w_c$ are the mixing weights of each component, $\vect{\theta_c}$ are the primary parameters uniquely identifying a precomputed model in the grid, and the scaling factor $s$ is described as a Normal distribution with mean $\mu$ and standard deviation $\sigma$:
\begin{equation}\label{eq:scaling}
  s \sim \mathcal{N}(\mu, \sigma^2),
\end{equation}
with
\begin{equation}
\mu = \frac{1}{N_{\rm obs}} \sum\limits_{i=1}^{N_{\rm obs}} O_i - \widetilde{M_i},
\end{equation}
where $N_{\rm obs}$ is the total number of observations, $O_i$ is the measured value of observation $i$ and $\widetilde{M_i}$ is the median of the predicted values for the corresponding observation, and 
\begin{equation}
\sigma = \sqrt{ \frac{1}{N_{\rm obs}-1} \sum\limits_{i=1}^{N_{\rm obs}} [O_i - \widetilde{M}_i - \mu]^2 }.
\end{equation}
By default, the model is scaled globally to match the set of observed values, making the scaling factor one of the inferred parameters (see Table\,\ref{tab:sumparam} for the list of parameters used for inference). The use of the median of predicted values ensures that quantities do not reach large numbers while optimizing the computing time for sampling random variables. 

\noindent The mixing weights are described as a Dirichlet distribution:
\begin{equation}
w_c \sim \mathcal{D}(\sum\limits_{c=1}^{N_{\rm comps}} w_c = 1).
\end{equation}

Primary parameters are described with Normal distributions truncated to the minimum/maximum values in the grid:
\begin{equation}
\theta_j \sim  \min(\theta) < \mathcal{N}(\mu_j, \sigma_j^2) < \max(\theta),
\end{equation}
with $\mu_j$ the average parameter value in the grid and $\sigma_j$ spans the entire parameter value range by default for a weakly informative prior when no user-provided prior is set. The model likelihood is calculated as the product of asymmetric Student-T distributions $\mathcal{S}$ centered on the observation for each draw of model values $M_i$:
\begin{equation}\label{eq:l}
\mathcal{L} = \prod\limits_{i=1}^{N_{\rm obs}} \mathcal{S}(\nu, \mu=O_i, \sigma^2=U_i^2),
\end{equation}
with $\nu$ the normality parameter (with a default value corresponding to a Normal distribution) and $U_i$ the asymmetric uncertainty on the observed value. The normality parameter $\nu$ may be changed to account for outliers, but if there are no outliers, decreasing the normality parameter effectively increases the number of high-probability regions. The model is then allowed to scan wider ranges and find that other solutions are acceptable.

\begin{table}\caption{Summary of inference model parameters.}\label{tab:sumparam}  \begin{tabular}{lc}
    \hline
    \hline
    Inference model parameters & \\
    \hline
    Primary parameters for each component $c$ & $\theta_c$ \\
    Mixing weight for each component $c$ & $w_c$ \\
    Scaling factor & $s$ \\
    \hline
    \end{tabular}
  \end{table}

  The posterior probability distribution for a given model $\mathcal{M}$ is then defined as
\begin{equation}\label{eqn:bayes}
p(\vect{\theta}|\vect{O}, \mathcal{M}) = \frac{ p(\vect{O}|\vect{\theta}, \mathcal{M}) p(\vect{\theta}|\mathcal{M}) }{ p(\vect{O}|\mathcal{M}) },
\end{equation}
with $p(\vect{O}|\vect{\theta}, \mathcal{M})$ the likelihood, $p(\vect{\theta}, \mathcal{M})$ the prior probability, and $p(\vect{O}, \mathcal{M})$ the marginal likelihood which integrates all parameter combinations:
\begin{equation}
p(\vect{O}|\mathcal{M}) = \int_\theta p(\vect{O}|\vect{\theta}, \mathcal{M}) p(\vect{\theta}|\mathcal{M}) d\theta.
  \end{equation}

  For lower and upper limits we use a half-Student-T likelihood with the same normality parameter $\nu$ as for Equation\,\ref{eq:l}, with a default value corresponding to a Normal distribution.
  Systematic uncertainties on observations are treated as nuisance variables encompassing a set of one or several observations, distributed by default as a Normal distribution.

Priors on, or between, parameters are meant to be implemented in an intuitive way. Built-in priors currently allow constraining parameter values between components (e.g., value in component 1 lower than in component 2), constraining absolute parameter values as Normal distributions, upper or lower limits, and using literal equations binding parameter values to each other.

\subsection{Input data grid}\label{sec:grid}

A model table is required with each row defining a unique precomputed model predictions. The model table is then transformed to produce a grid spanning all possible combination of parameters. If some parameter combinations are not available, the invalid data is accounted for during the inference process with zero likelihood. A mask is used to track the regions where the modeled values are not finite, either due to grid completion (but also due to model upper limits).

A preprocessing script may be used to manipulate the grid before the inference step. This is useful, for instance to create new observations based on existing ones in order to be later used as constraints.

It is possible to upsample the grid, for instance with a linear interpolation. The inference (Sect.\,\ref{sec:inference}) enables on-the-fly interpolation for each MCMC draw but nonlinear interpolation that are calculation-intensive would need to be performed beforehand in order to generate a refined grid. 
The default grid interpolation method uses nearest neighbors, as it allows a relatively quick solution to be found. A multidimensional linear interpolation is also implemented that can be used for either all parameters or a subset. For performance issues the code samples fractional grid indices by default with nearest neighbor or linear interpolation but it is possible to force sampling real values instead.  

The conversion of the initial model table into a multidimensional grid may produce many nonvalid values. For this reason, the linear interpolation often deals with an n-dimensional cell with nonfinite vertices. The current performance-driven workaround is to perform a discrete interpolation (upsample) in order to force the sampler to probe vertices instead of exploring around it. This is done by introducing a categorical random variable. 

It is also expected that the grid is sampled well enough in each dimension so that the variation of the likelihood function across each grid node is accurately described with a linear interpolation (see an illustration and the discussion in \citealt{Galliano2021a}). Potential solutions to better sample the likelihood function include computing new models possibly thanks to an adaptative grid, using a quadratic interpolation to force smooth variations of the likelihood function, or to perform a linear regression using Machine Learning techniques (Morisset et al.\ in prep). 

As for all grid search algorithms, there can be some edge effects. The inference model needs to probe around the high probability regions, and if the latter happen to lie on the edge of the grid or on the edge of an invalid data region (which is more likely to occur as the number of dimensions increases), the posterior mean will be slightly offset from the edge even if the maximum likelihood lies in principle at the edge. Edge effects can be mitigated in part by using the median or the mode of the chain. 

Finally, the inference is performed on the primary parameters themselves by default, but it is possible to consider power-law, broken power-law, normal, or double-normal distributions instead. In such cases, the slopes and pivot point (power-law distributions) or means and standard deviations (normal distributions) as well as lower- and upper-boundaries are described as random variables.

\subsection{Sampling methods}\label{sec:stepmethod}

Several MCMC step methods are implemented in PyMC, some of which we have benchmarked for our purpose. The Sequential Monte Carlo (SMC) sampler is a method that progresses by a series of successive annealed sequences from the prior to the posterior. The PyMC implementation is based on \cite{Ching2007a} and \cite{Minson2013a}. The No U-turn sampler (NUTS) works on continuous variables with Hamiltonian mechanics \citep{Hoffman2011a}. Other methods include Metropolis-Hastings (e.g., \citealt{Robert2015a}) and Adaptive Differential Evolution Metropolis, the latter which uses the past chain values to inform and generate jumps \citep{terBraak2008a}. Other methods may be investigated in the future.

Random walkers such as NUTS need long enough chains to authorize excursions either to or from the high-probability region. In comparison, SMC identifies the high-likelihood regions by running in parallel a large number of chains starting from random values sampled from the prior distribution. Among the many advantages of SMC, it can sample from distributions with multiple peaks and does not have a burn-in phase. Random walkers are more likely to be stuck around one peak. While several methods have been included in MULTIGRIS, the envisioned use for ISM studies implies that the probability distribution is expected to be multipeaked and we thus describe the SMC method further below. 

\begin{figure}
   \centering
\includegraphics[width=0.5\textwidth]{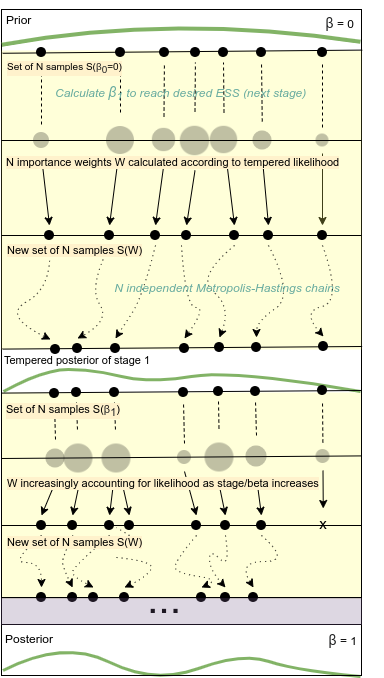}
\caption{Illustration of Sequential Importance Resampling Particle Filter. From top to bottom, the distribution is sampled from the prior, reweighted and resampled according to the tempered posterior density. Surviving samples seed new Markov chains and walk through a number of Metropolis steps.}
         \label{fig:smc}
       \end{figure}

SMC uses a combination of various techniques, with in particular tempered likelihoods and importance sampling. The tempered posterior is proportional to the product of the tempered likelihood and the prior. For a given model $\mathcal{M}$ this translates as
\begin{equation}
p(\vect{\theta}|\vect{O}, \mathcal{M})_\beta \propto p(\vect{O}|\vect{\theta}, \mathcal{M})^\beta p(\vect{\theta}).
\end{equation}
SMC first samples the prior ($\beta=0$), calculates a new $\beta$ value to match a predefined effective sample size (by default half the number of draws), computes importance weights using the tempered likelihood $p(\vect{O}|\vect{\theta}, \mathcal{M})^\beta$, computes a new set of samples by resampling according to importance weights, computes the mean/covariance of the proposed distribution, runs relatively short (on the order of a few to a few dozens) independent Metropolis-Hastings chains from the proposed distribution to explore the tempered posterior. The importance weights increasingly account for the likelihood as the stages increase. The stages continue until the true posterior is sampled ($\beta\geq1$). The overall process is illustrated in Figure\,\ref{fig:smc}.

       Since SMC samples a large number of chains exploring the entire parameter space instead of a single chain, it is important to have enough samples to probe the prior parameter space. Furthermore, since each individual chain walks with independent Metropolis-Hastings, it is in principle possible to have only few samples per parameter but better results are obtained when using a large number of chains per parameter combination to either obtain an average probability density function (PDF) if the chains have significantly different distributions and/or to evaluate whether all chains have converged to the same distribution. While the number of samples may seem appropriate for the considered prior distribution, they also need to be numerous enough to probe the posterior distribution, which may show multiple peaks. If a small number of samples is considered, this may lead to significant stochasticity between independent model runs. SMC runs several jobs in parallel, and in the following we always show the combined chain result. The inference result contains the posterior distribution of the parameters, which are then processed for further analysis.

\subsection{Component identification}

When the configuration requires more than one component, an important postsampling step is to identify them. By default, drawn values of primary parameters (used for the inference) are not assigned to any specific component (i.e., components are not explicitly identified or tied to any given parameter or parameter set). This is meant to assert whether inferred parameters have indeed significantly different values between each component. If parameter values are not significantly different between components, the components are thus allowed to switch back and forth while the solution should remain stable (specifically, for the NUTS/Metropolis samplers this means that draws may switch with time, while for the SMC sampler this means that each draw at any given stage may probe any component). 

If the main objective is to derive PDFs of secondary parameters (those not used for inference and calculated in the postprocessing step; Sect.\,\ref{sec:pp}), there is no need to identify explicitly each component. Nevertheless, if the parameter value for each component is difficult to disentangle, some samples may be wasted to switch from one component to the other and the probability distributions may not be as smooth as possible.

\begin{figure}
   \centering
\includegraphics[width=0.24\textwidth,clip,trim=420 870 0 0]{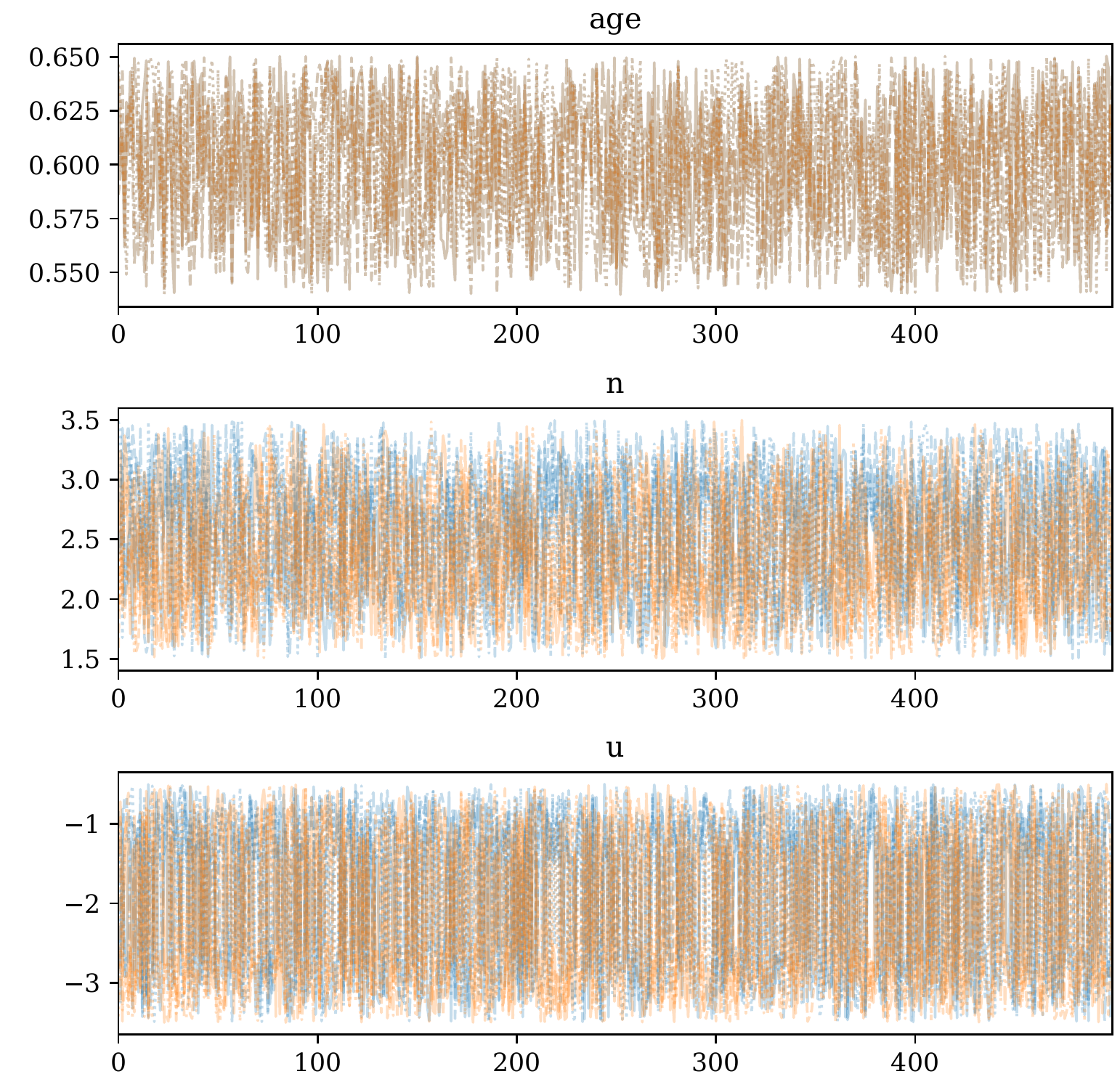}
\includegraphics[width=0.24\textwidth,clip,trim=420 870 0 0]{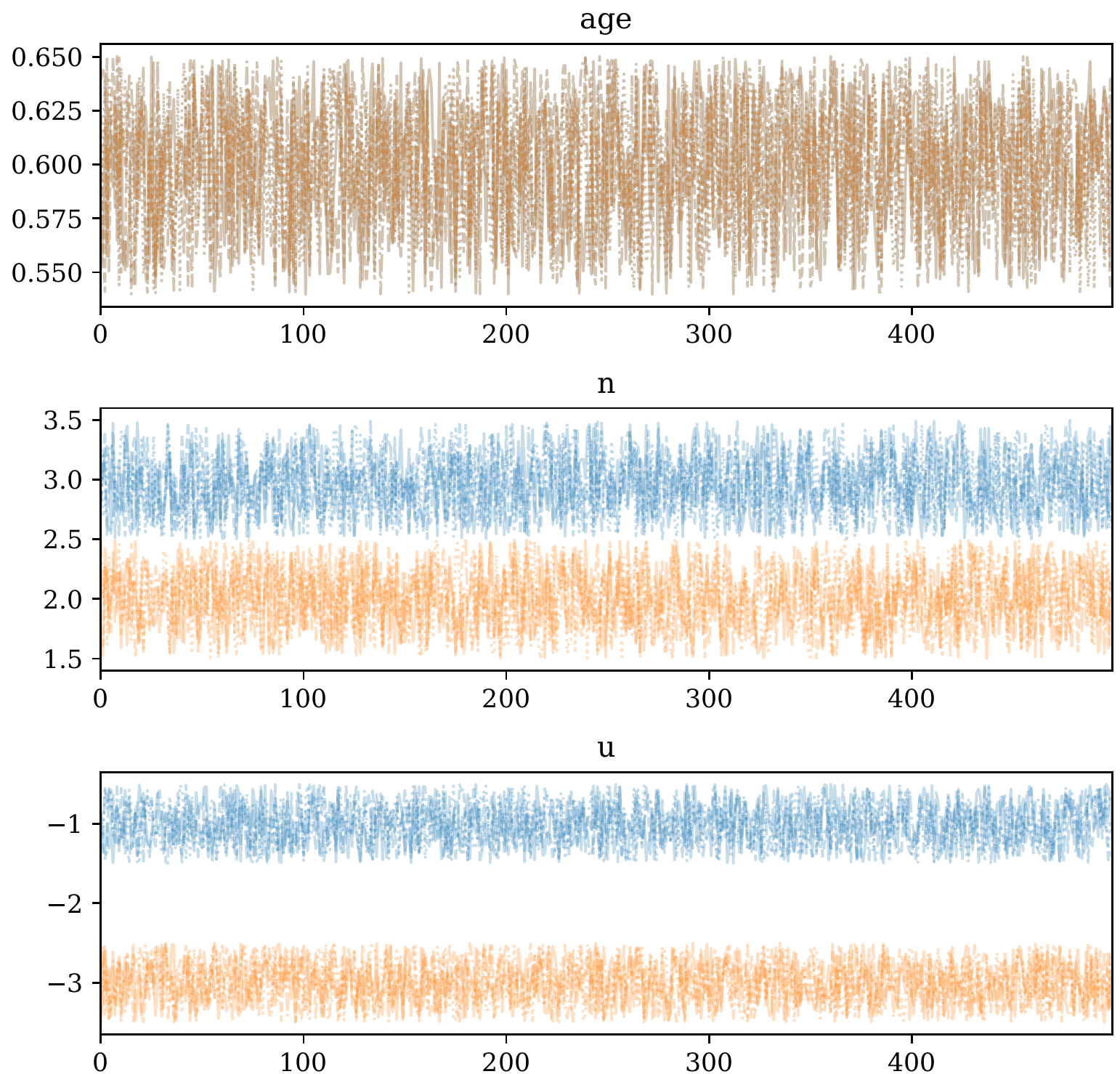}
\caption{Posterior distribution of parameters for a model with two components. One parameter (age) has been purposefully forced to be the same of the two components. The raw posterior distribution is shown on the left, for which the two components may switch during inference, and the final result is shown on the right, in which the individual component parameter values have been redistributed a posteriori. For visual purposes only a subset of the parameters are shown }
         \label{fig:compsort}
       \end{figure}

In some cases it may be interesting to examine the posterior distribution of primary parameters for each component individually. MULTIGRIS performs a processing step to identify and characterize individual components from the chain, by redistributing components through a minimization of the standard deviation for all parameters for individual chains and components (Fig.\,\ref{fig:compsort}). This is an iterative process, redistributing components for each draw along the entire chain several times. This redistribution bears some uncertainties and the individual component properties may still have some level of degeneracy. The global solution (combination of the components) is, however, not affected by the redistribution of components since all parameters are switched together.

Depending on the solution, it may be difficult to redistribute components a posteriori, which may indicate a need for more constraints. For this reason, it may be useful to force the identification of components a priori. In MULTIGRIS, components can be sorted using a prior constraint on any given parameter.
Sorting with a prior constraint will obviously affect the inference, and the result will be different depending on which parameter is used for sorting.

\subsection{Diagnostics}

\begin{figure}
  \centering
\includegraphics[width=0.5\textwidth,clip]{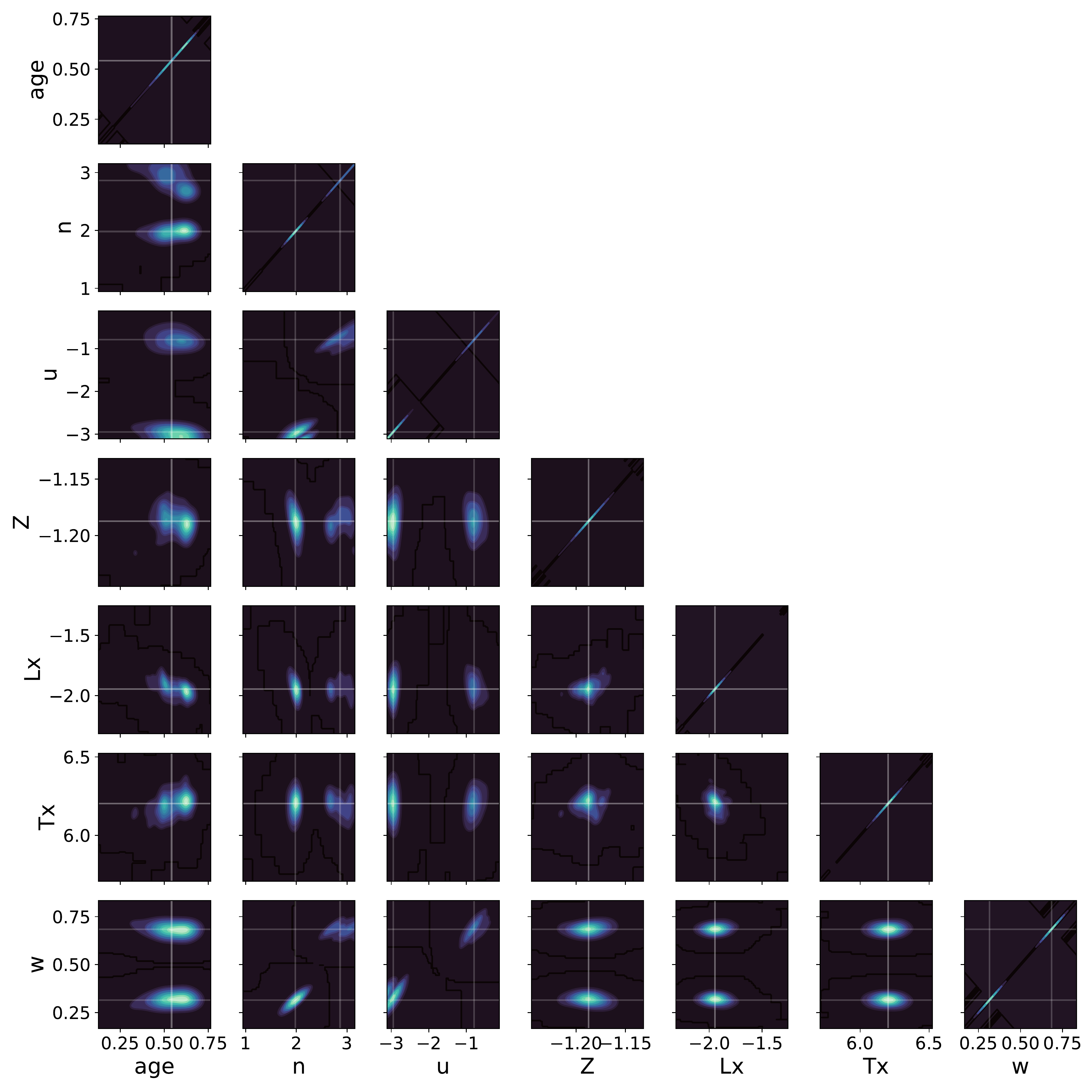}
\includegraphics[width=0.48\textwidth,clip]{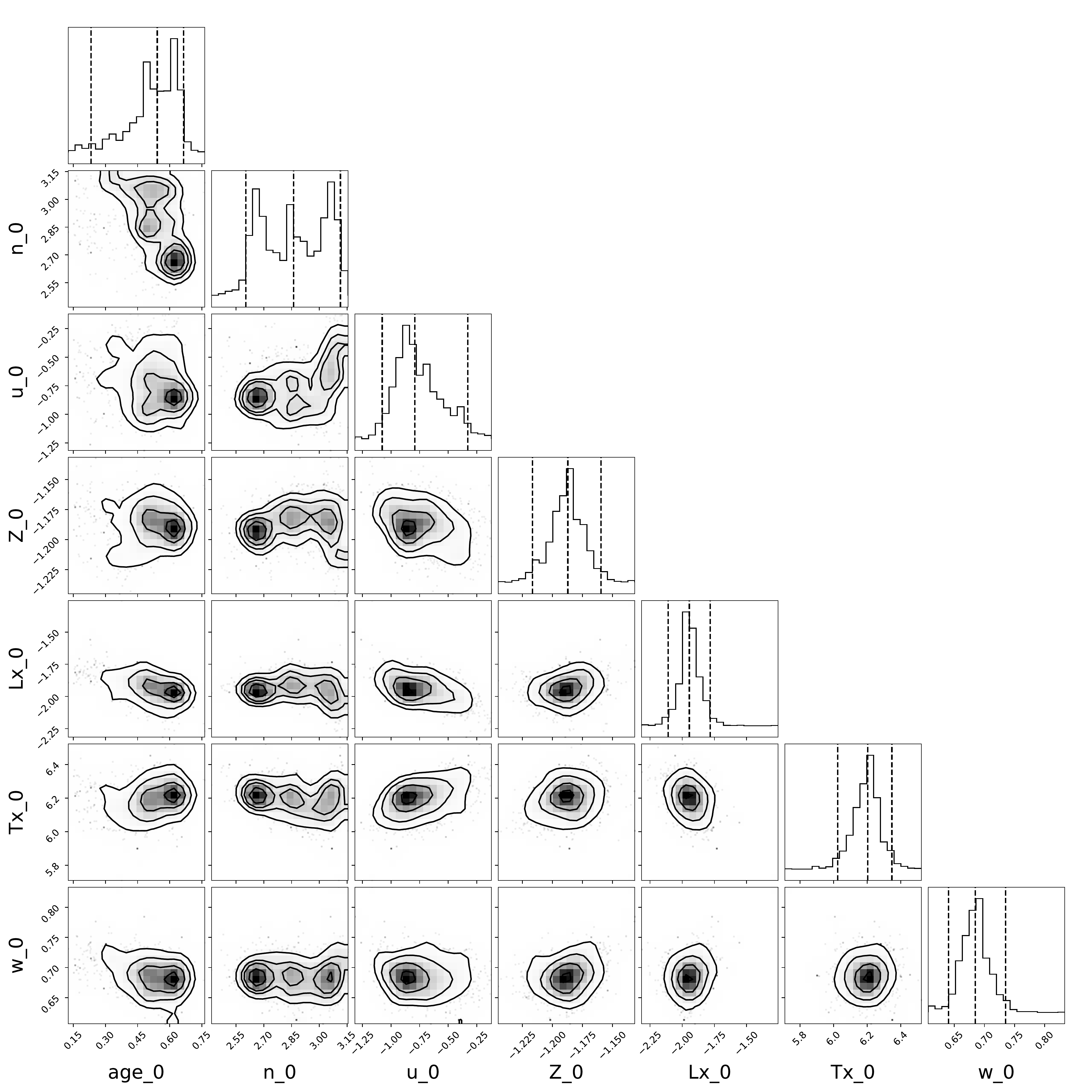}
\caption{Example of results for a 2-component model. The kernel density estimate (nonparametric method to estimate a PDF) is shown on top and a corner plot for one of the two components on the bottom. We note that some parameters have been forced to have identical values for the two components. }
         \label{fig:example_2d}
       \end{figure}

\begin{figure*}
   \centering
\includegraphics[width=0.49\textwidth,clip]{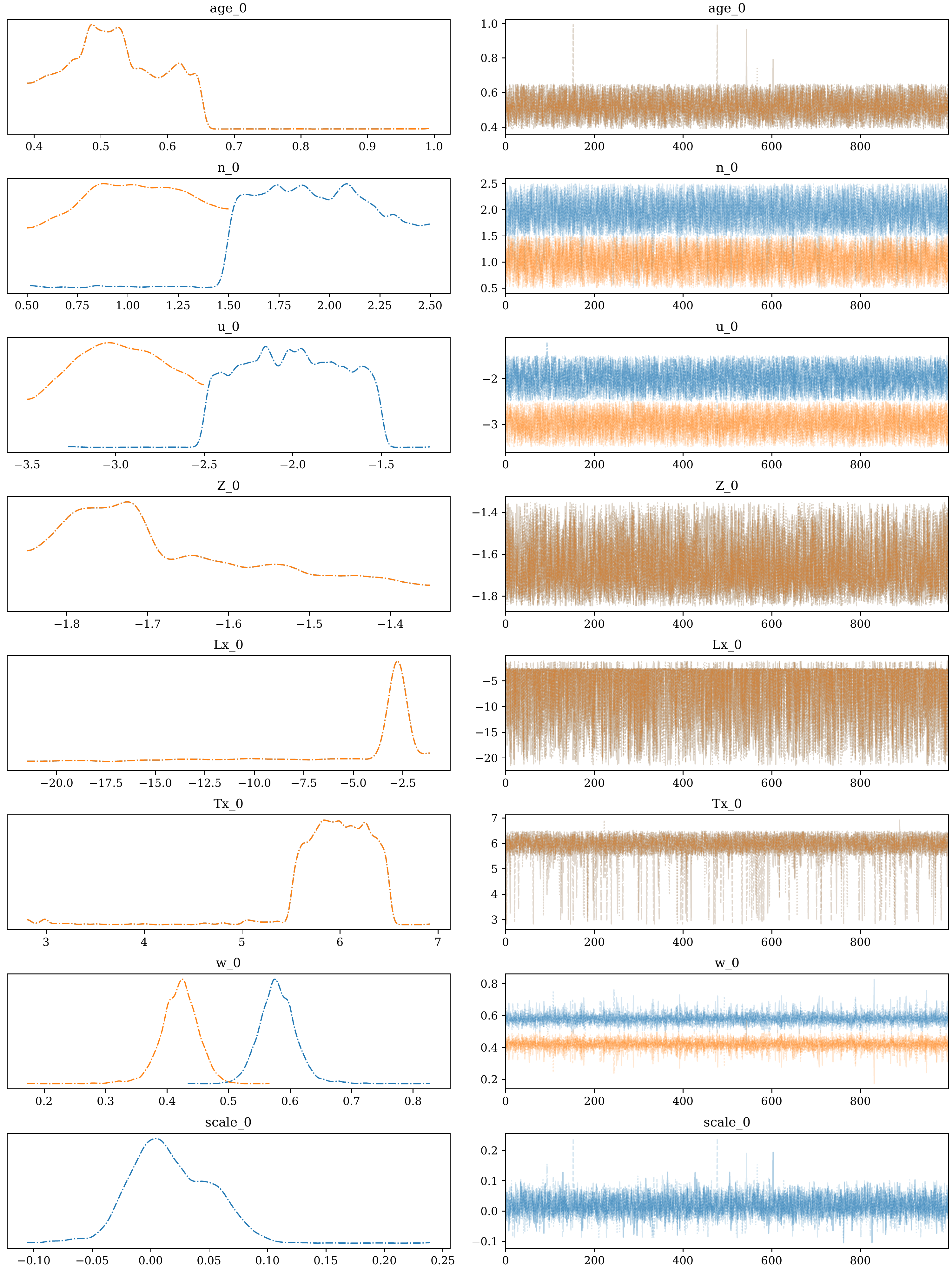}
\includegraphics[width=0.49\textwidth,clip,trim=0 0 0 720]{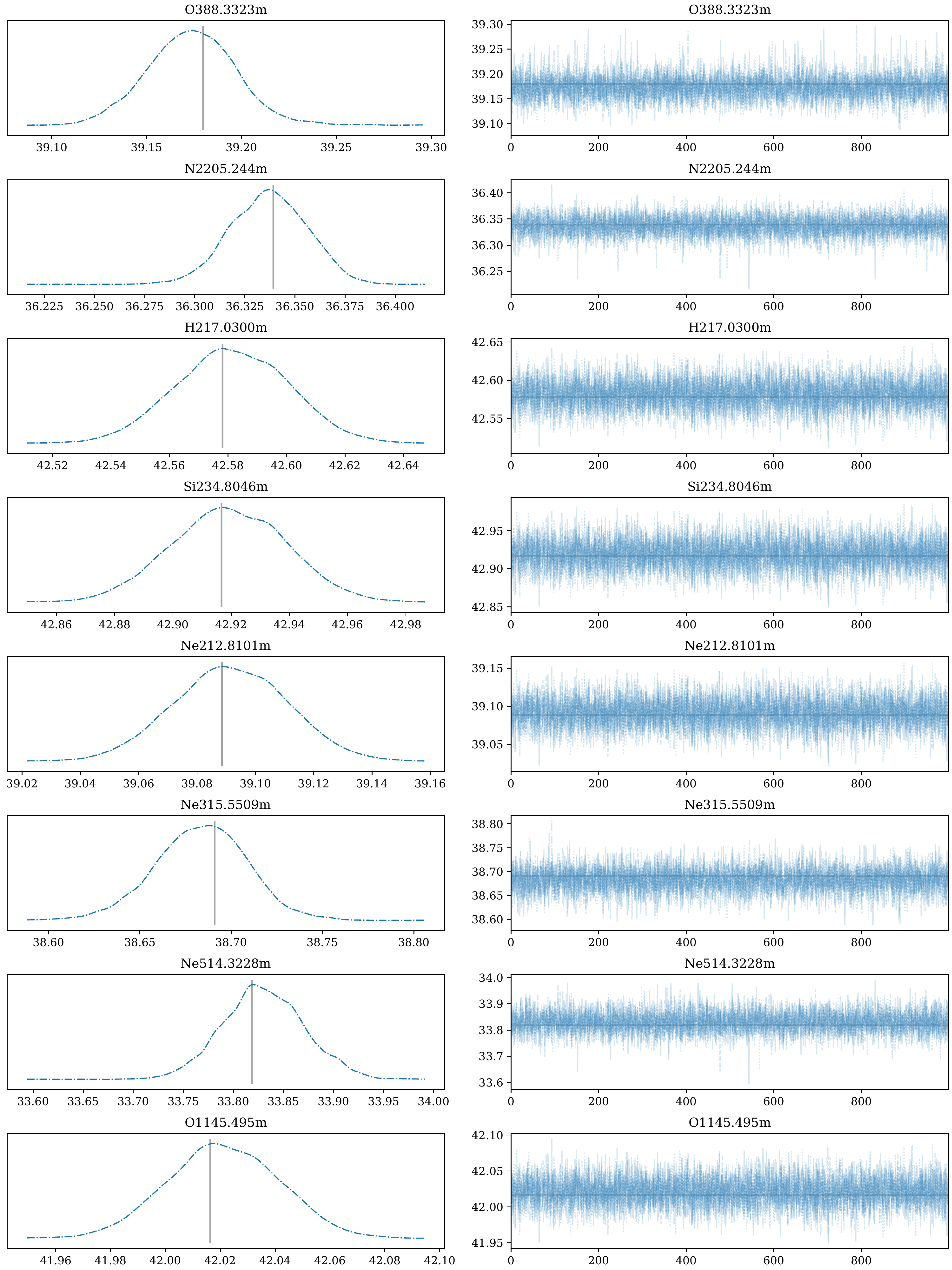}
\caption{Example of the posterior distributions for the primary parameters (two leftmost columns) and for the observations (two rightmost columns). For each quantity, the left plot shows the PDF (with the vertical line indicating the reference measured value for observations), and the right plot shows the sequence of samples. For these plots individual jobs are combined. We note that some parameters have been forced to have identical values for the two components. }
         \label{fig:example_trace}
       \end{figure*}

PyMC automatically returns useful statistics for each parameter (e.g., effective sample size, autocorrelation, Gelman-Rubin convergence test using multiple chains, and marginal likelihood for SMC). Several diagnostics are available in particular to check the model convergence. To calculate the final posterior distribution, we draw a random set of values from the chains (and, depending on the step method, removing the tuning steps).
We also compute posterior predictive $p$-values for each parameter and for all parameters combined following the method in \cite{Galliano2021a}. 
Figures\,\ref{fig:example_2d} and \ref{fig:example_trace} provide some illustrations of some available diagnostic plots (see also Sect.\,\ref{sec:i18app}).

\subsection{Model selection and comparison}\label{sec:model_comparison}

Model selection methods in PyMC include WAIC (widely-applicable Information Criterion; \citealt{Watanabe2010a}) and LOO (Leave-one-out cross-validation; \citealt{Vehtari2017a}). LOO cross-validation, in particular, repeatedly extracts a training set to fit the model and evaluates the fit using the remainder data. Approximations based on importance sampling make it possible to perform the iterations without having to refit the data.

The SMC sampler also provides the marginal likelihood $p(\vect{O}|\mathcal{M})$ (Eqn\,\ref{eqn:bayes}), that is the probability of the observed data given the model (where parameters are integrated out). 
While the $\chi^2$ asymptotically reaches a minimum value when the number of components increases, reflecting the fact that the fit does not improve significantly anymore (the reduced $\chi^2$ is difficult to estimate; Sect.\,\ref{sec:previous}), the marginal likelihood reaches a maximum and decreases as the number of potential solutions increases through the larger number of components (see details in Sect.\,\ref{sec:ncomps}). Even though increasing the number of components increases the number of potential solutions, such models become less and less likely as the overall number of parameter sets increases (spreading the prior), implying that the probability of the data given the model decreases. 

To compare models, we are eventually interested in computing for each model:
\begin{equation}
p(\mathcal{M}|\vect{O}) \propto p(\vect{O}|\mathcal{M}) p(\mathcal{M}).
\end{equation}
One can then compare models directly with different priors or models with a different number of components by using the marginal likelihood $p(\vect{O}|\mathcal{M})$ as long as models have the same a priori probability $p(\mathcal{M})$ under all circumstances. 

Model comparison may be used to select one preferred model or to combine models through averaging:
\begin{equation}
p(\vect{\theta}|\vect{O}) = \sum_{i=0}^{K} p(\vect{\theta}|\vect{O}, \mathcal{M}_i) p(\mathcal{M}_i|\vect{O}),
\end{equation}
where $K$ is the number of models. An alternative to weighing through the marginal likelihood is to use information criteria methods such as WAIC or LOO with the Akaike weight for a model $i$ as
\begin{equation}
w_i = \frac{e^{-\frac{1}{2} dIC_i}}{\sum_{j=0}^{K} e^{-\frac{1}{2} dIC_j}},
\end{equation}
with $dIC = -2\times{\rm elpd}_{\rm LOO,WAIC}$ where ${\rm elpd}$ is the expected log pointwise predictive density for either LOO or WAIC.

\subsection{Postprocessing}\label{sec:pp}

Primary parameters, that uniquely define precomputed model predictions in the grid, are the ones used for inference. From the posterior distribution of primary parameters, we can obtain the corresponding distribution for any other parameter in the grid. The postprocessing step uses the PDF of primary parameters in order to deduce secondary parameters that completely depend on the primary parameters, or new observations that were not used as constraints. The semantic distinction between new observations to predict and secondary parameters is only due to the assumption that secondary parameters can never be observed.

\section{Application to ISM studies}\label{sec:ismapp}

While MULTIGRIS is agnostic to the grid of precomputed models, it has been initially designed for the purpose of combining ISM components corresponding to sectors around stellar clusters or to several clusters surrounded with one or several sectors. In that vein, the mixing weight described in Section\,\ref{sec:inference} is interpreted as the integrated covering factor of lines of sight (or equivalently as the fraction of the total solid angle) corresponding to the same 1D model. More complex configurations are described in Section\,\ref{sec:ismtopo}. If a single cluster is considered with several ISM components, we refer to the components as sectors in the following.

Other configurations can also be considered, for instance considering a large number of clouds whose parameters are linked with a distribution (e.g., power-law and normal; \citealt{Richardson2014a}). In the latter cases the distribution parameters (e.g., slope and boundaries) need to be inferred (instead of primary parameters for a discrete number of clouds/components). It is of course possible to simply tabulate the statistical cloud distribution parameters as a preprocessing step but the ability to infer these parameters on-the-fly through the sampling process of continuous random variables makes it easier to consider many dimensions (e.g., varying the lower- and upper-boundaries). 

Although MULTIGRIS can consider power-law or normal distributions, we illustrate here the applications for a multisector approach building on previous works using the same approach. Following our previous efforts, we consider that star-forming dwarf galaxies are dominated by one or a few giant H\2\ regions. The ISM topology is then defined as a group of sectors around a given cluster including thermal and possibly nonthermal sources. 

MULTIGRIS includes by default part of the BOND database \citep{ValeAsari2016a} and a specific model database calculated with the photoionization/photodissociation code Cloudy (SFGX; \citealt{Ramambason2022a}). The SFGX database is especially adapted to low-metallicity ISM studies and includes predictions for UV to IR emission lines and X-ray sources. This model database consists in a table with each row corresponding to a specific model outcome computed with Cloudy \citep{Ferland2017a}. Primary parameters defined in MULTIGRIS are the luminosity and age of the stellar cluster, the luminosity and temperature of a potential X-ray source, the ionization parameter, the gas density, the metallicity, the dust-to-gas mass ratio, and the physical depth of the cloud. Secondary parameters include in particular masses of dust and various gas phases as well as the escape fraction of ionizing photons.  We briefly describe the database below with the details available in \cite{Ramambason2022a}.

\subsection{Cluster properties}\label{sec:clusters}

Stellar clusters are defined by their age and luminosity, as well as the temperature/luminosity of a potential X-ray component. The latter is described by a multicolor blackbody parameterized by the temperature at the inner radius of the accretion disk and by the total luminosity.

We consider Cloudy runs with a $10^7$ or $10^9$\,L$_\odot$ cluster.
A single scaling factor is used to match the observed fluxes (the scaling factor may use all observations like in \cite{Cormier2019a} or a single observation, such as the total IR emission). The scaling factor together with the model luminosity provide the effective luminosity. This means that a $10^{10}$\,L$_\odot$ galaxy is effectively considered to be a sum of a thousand $10^7$\,L$_\odot$ or ten $10^9$\,L$_\odot$ H\2\ regions rather than a $10^{10}$\,L$_\odot$ H\2\ region. These two reference luminosities are considered to avoid variations due to shape and degeneracies between luminosity and scaling factor (filled sphere vs.\ thin shell; see \citealt{Ramambason2022a}).

There is no degeneracy between the stellar luminosity and the scaling factor used to reproduce the lines, since the relative line fluxes do change with luminosity while the scaling factor simply scales up or down all lines from the inferred luminosity. Line ratios are not affected by the scaling factor, contrary to topological factors like the depth of each sector (Sect.\,\ref{sec:ismprops}).

\subsection{ISM properties}\label{sec:ismprops}

The ISM properties for each sector are defined by the density, ionization parameter, and cloud depth. The density parameter is set by the density at the illuminated front. The density profile is then constant until the ionization front and then scales with the total H column density.

An ISM parameter of interest to study the neutral gas, low-ionization H\2\ region species, or the escape of ionizing photons is the physical depth of the shells.
We need a general definition for the depth cuts that applies to all models, in particular because the scaling of parameters such as visual extinction $A_V$ as a function of depth varies depending on metallicity. We introduce in the SFGX model database a ``cut'' parameter whose variation depends on observables, described as 
\begin{itemize}    
\item cut=0: inner radius (illuminated edge).
\item cut=1: H$^+$--H$^0$ transition (i.e., ionization front; electron fraction $0.5$).
\item cut=2: H$^0$--H$_2$ transition (photodissociation front; molecular gas fraction $0.5$).
\item cut=3: C$^0$--CO transition (photodissociation front; molecular gas fraction $0.5$).
\item cut=4: full depth corresponding to stopping criterion $A_V=10$.
\end{itemize}
Fractional cut values are considered by sampling the extinction $A_V$ profile. With this definition, radiation-bounded models have cut$\gtrsim1$. The last cut value is somewhat less trivial as the corresponding stopping criterion is the maximum model $A_V=10$ which vary dramatically between models with different metallicities.

\subsection{ISM topology}\label{sec:ismtopo}
    
Possible configurations enabled by MULTIGRIS include any combination of components, either as a discrete number of sectors around several clusters with the inferred mixing weight defining the relative contribution of each sector or as a statistical distribution of clouds parameterized by, for instance, a power-law or a normal distribution. In the following, we focus on a discrete number of sectors. 

The contexts enabled by MULTIGRIS make it possible to use predefined sets of priors, for instance, one cluster (defined by the age of the stellar population and by the X-ray source properties) and $n$ sectors (possibly matter-bounded), a set of $n$ clusters, each with one sector (in practice, we mix models with different stellar population ages and/or different X-ray source properties), or a combination of several clusters and several sectors around each cluster. When several sectors are considered, their covering factor is not tabulated a priori, contrary to \cite{Cormier2019a}. It is instead inferred on-the-fly through the mixing weight continuous random variable.

\begin{figure}
   \centering
\includegraphics[width=0.24\textwidth]{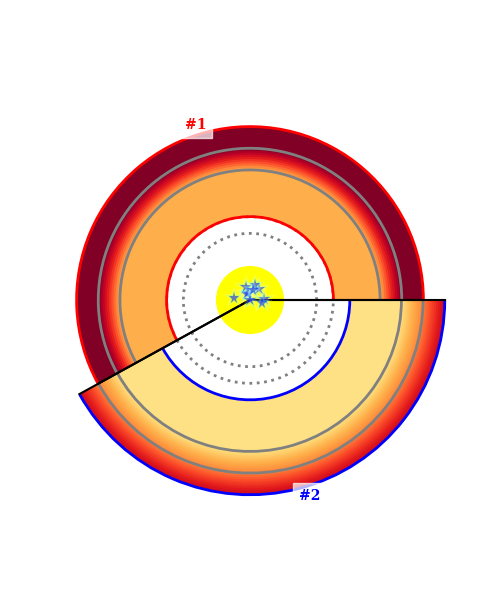}
\includegraphics[width=0.24\textwidth]{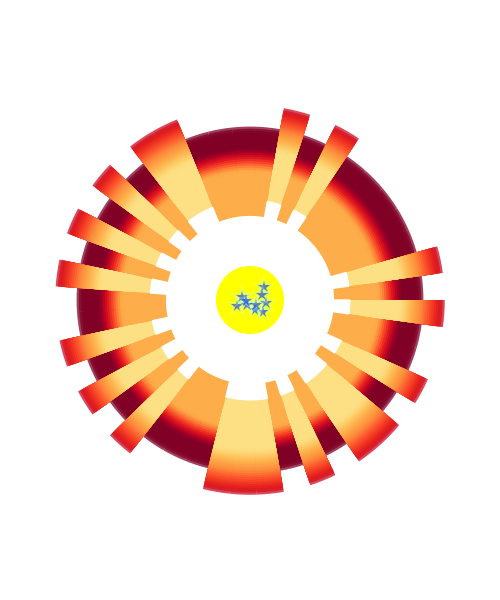}\\
\includegraphics[width=0.4\textwidth,clip]{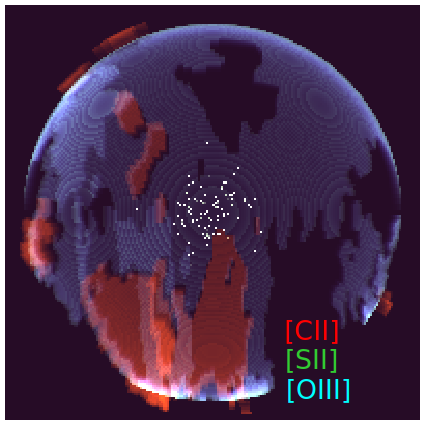}
\caption{Configuration of inferred sectors for an example model. \textit{Top}: the left plot shows the integrated covering factors while the right plot shows the same topology but with a different, random, geometry. The color scales from yellow to red as the gas volume density increases. Different inner radii indicate different ionization parameter ($U$) values, with dotted arcs showing $U=0,1,...$. The gray arcs show the photoionization and photodissociation fronts. The depth is not drawn to scale. \textit{Bottom}: sectors are distributed randomly in 3D according to their relative covering factors and then projected in 2D. Screening/extinction effects are ignored. Stars are added for visual purposes only. }
         \label{fig:example_sectors}
       \end{figure}

Figure\,\ref{fig:example_sectors} shows an illustration of inferred mixing weights. The default representation is simply a pie chart indicating the contributions of each sector, that is the fraction of solid angle around the cluster that corresponds to a given sector. Another representation randomly distributes the sectors around the cluster according to their covering factor and reflects the fact that the ISM topology alone is constrained and that all the random distributions of sectors (different geometries) are thus equivalent. Finally, it is possible to illustrate a 3D distribution by creating a spherical random map to distribute all sectors based on their covering factor and by populating a line emissivity cube with each line of sight from the cube center toward a given angle corresponding to a given sector. The projected result is shown in Figure\,\ref{fig:example_sectors}.

       A caveat exists when dealing with two sectors including one with negligible flux contribution for all lines (e.g., very small depth). There is then a degeneracy between the covering factor of the main sector and the model scaling factor. This is important only for parameters that heavily depend on the topology (see also Sect.\,\ref{sec:ncompsinfluence}). This problem is exacerbated if using a single observation that arises mostly from one sector. For these reasons, it is advised to check whether a second sector is in fact necessary through model selection methods (Sect.\,\ref{sec:model_comparison}).

       As discussed in \cite{Ramambason2022a}, the overall topology and the depth of each sector may have a significant impact on the metallicity and SFR determinations, and an even stronger impact on the escape fraction of ionizing photons (see also Sect.\,\ref{sec:ncompsinfluence}). The escape fraction is constrained by line ratios from tracers originating at various depths and also depends on the total cluster luminosity determination (Sect.\,\ref{sec:clusters}).

\subsection{Observations}

Any spectral line or photometric band can be used as observational constraint as long as Cloudy is able to compute it. The SFGX database includes several hundred tracers mostly in the optical and IR domains. While extinction-corrected fluxes can be used as constraints, MULTIGRIS may include extinction or attenuation law parameters as random variables, therefore consistently matching observations and models. This is especially useful when comparing optical and IR datasets because optical datasets only consider extinction within the layer from which optical lines arise while IR datasets may probe more embedded regions. 

Nuisance variables can be added to reflect systematic uncertainties on observations. For instance, a systematic uncertainty can be considered for elemental abundances. On first order, line fluxes scale with the elemental abundance for small variations that do not modify significantly the physical conditions (e.g., cooling rates; see \citealt{Cormier2019a}) and a systematic factor can thus be used to scale all species of a given element. Other nuisance variables can consider, such as calibration errors between several groups of tracers.

Several observation sets can be defined with some tracer potentially appearing in several sets (e.g., observed with different instruments and requiring a different scaling factor). As new observations become available, it is therefore possible to update the posterior distributions.

\section{Benchmark}\label{sec:benchmark}

MULTIGRIS is designed to run on standard multicore personal computers, with the ability to use large grids. The SFGX grid currently contains more than a billion values and can be read in less than a second.

With the SMC sampler, relatively short individual chains are sampled with independent Metropolis-Hastings (Sect.\,\ref{sec:stepmethod}). Depending on the parameter variation across the grid, the minimum number of chains per parameter may range from two to a few. From our benchmarks studies, good results are obtained with the SFGX context with about two chains per parameter. For a configuration with a single cluster with two or three matter-bounded sectors, this translates into a few $10^{3-4}$ samples. The inference time scales with the number of parameters to explore, and, to a somewhat lesser extent, with the number of observations due to the likelihood calculation.

\subsection{Multimodal solutions}

A common issue with ISM grids is the existence of multiple potential solutions across the grid. Random MCMC walkers are not particularly adapted because the sampler may not be able to probe multiple peaks if the latter are too distant from each other, unless the chain is sufficiently long. On the other hand, the SMC step method initially samples from the prior and is able to probe multiple peaks through the convergence process to the posterior distribution (Sect.\,\ref{sec:stepmethod}).  

\begin{figure}
   \centering
\includegraphics[clip,width=0.24\textwidth,trim=0 0 450 0]{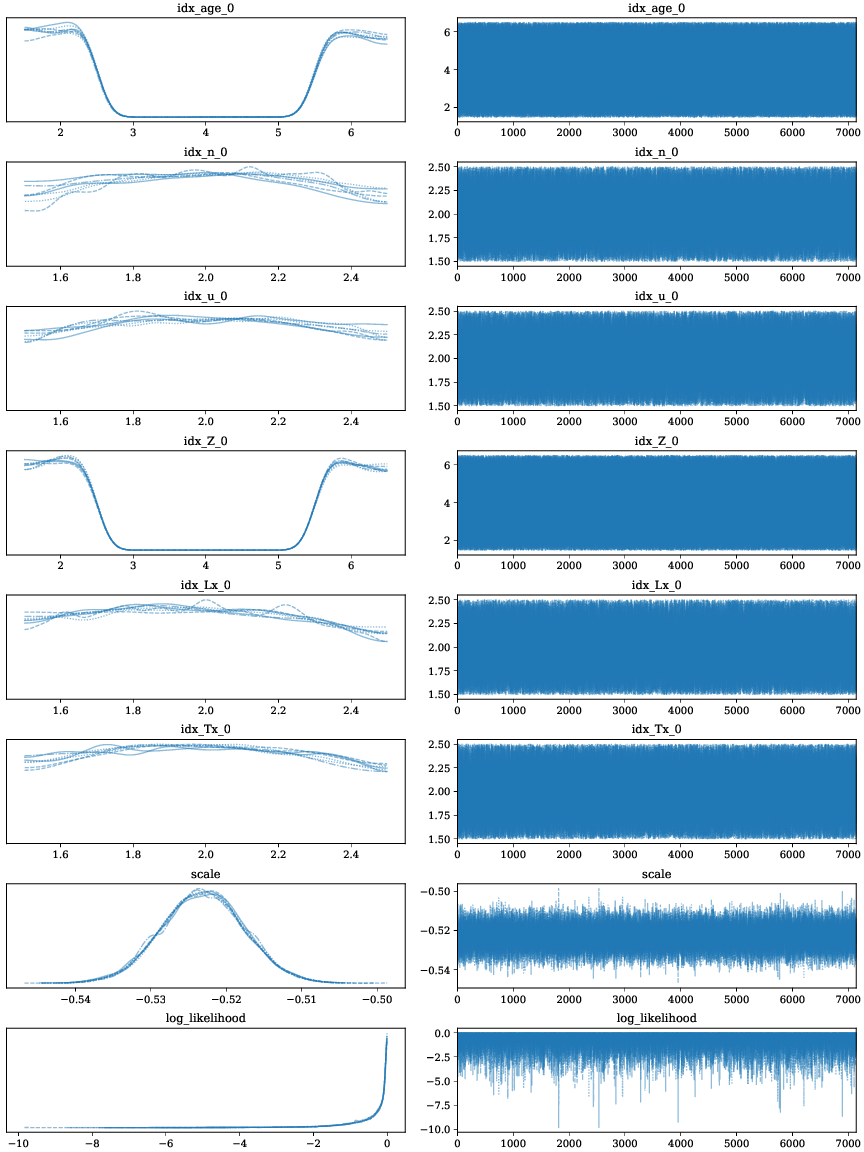}
\includegraphics[clip,width=0.24\textwidth,trim=0 0 450 0]{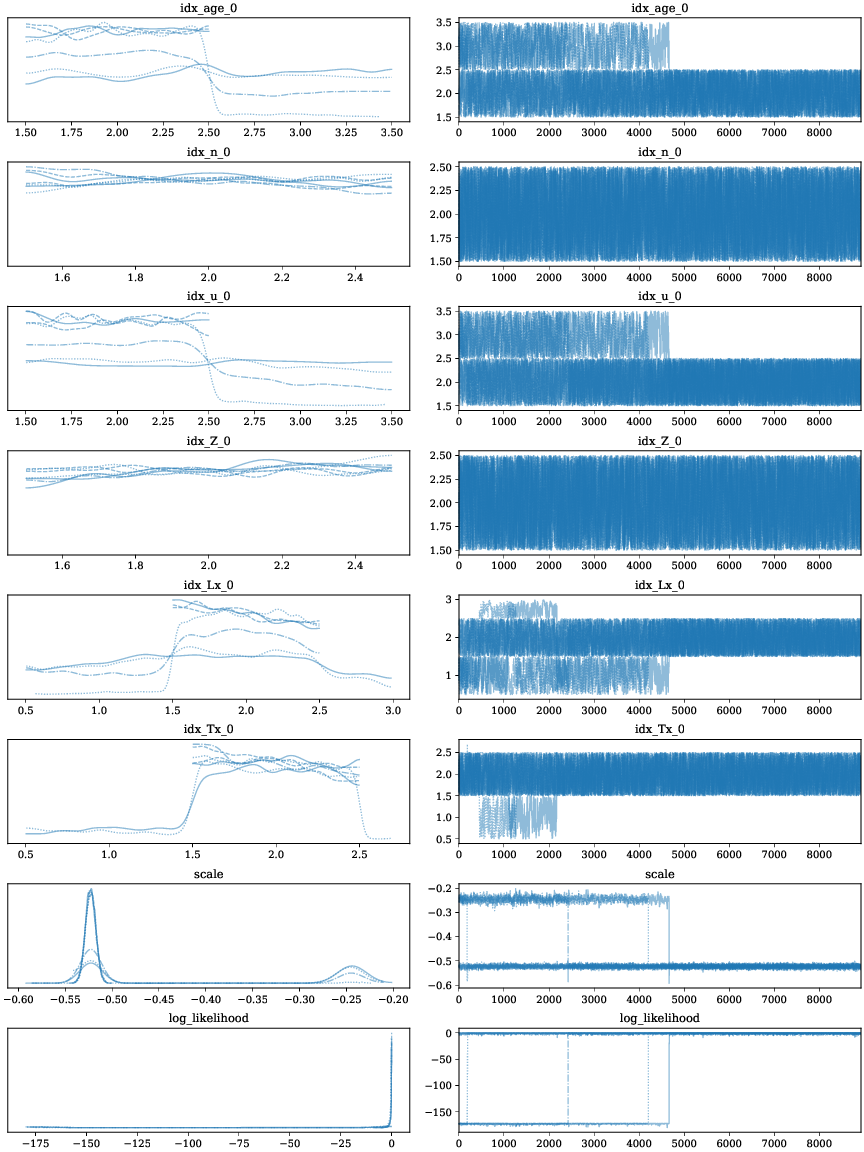}
\caption{Benchmark model run with a degeneracy artificially introduced for the age (${\rm idx\_age\_0}$) and metallicity (${\rm idx\_Z\_0}$). The SMC step method is shown on the left (each individual line corresponding to a given job) and NUTS on the right (each line corresponding to a given chain). The x-axes in the two columns have different ranges. }
         \label{fig:bimodal}
       \end{figure}

We show in Figure\,\ref{fig:bimodal} a model run with the SMC and NUTS samplers with similar execution time. The grid was modified to introduce a degeneracy for the age and metallicity with two equivalent solutions. SMC is able to explore the two peaks in the posterior distribution while individual chains in NUTS may be stuck to one or the other peak. 

\subsection{Number of sectors}\label{sec:ncomps}

Another common issue with multisector models is to estimate the most likely, or at least minimum, number of sectors required. 
Model comparison metrics help in making a choice to estimate how many components are necessary (Sect.\,\ref{sec:model_comparison}). One can either choose a frequentist tests by computing likelihood ratios (testing how well the models explains the data using the most likely parameter set) or hypothesis tests by computing the marginal likelihood (testing how much belief one should have in each model given the data and the priors). In \cite{Cormier2019a} the decision was based on the $\chi^2$ value but this only indicated the minimum number of sectors to obtain a satisfactory agreement between observations and models, that is from a pure frequentist point of view.

In principle, the marginal likelihood reaches a maximum for the most likely number of sectors given the data and the priors. In other words, a higher number of sectors is less likely to reproduce the data because the latter is overfitted. In that vein, the Bayes factor, which computes the marginal likelihood ratio is a useful metrics, but the actual probability we are interested in to compare models is $p(\mathcal{M}|\vect{O})$, which supposes the knowledge of the overall a priori probability of the model, $p(\mathcal{M})$ (Sect.\,\ref{sec:model_comparison}). 

Another model comparison method is to run a model with $n$ sectors and infer whether up to $n-1$ sectors can be ignored by setting their covering factor to zero during inference. The $n-1$ sectors that can be ignored are described with a categorical random variable. This allows for an on-the-fly estimate of the most likely number of components, with the advantage of inferring a probability for each number of components. On important caveat though is that this method is adapted only in cases when potential additional sectors do not change significantly the properties of the other sectors in order to prevent erratic jumps during inference. 

\begin{figure*}
   \centering
\includegraphics[width=0.33\textwidth,clip]{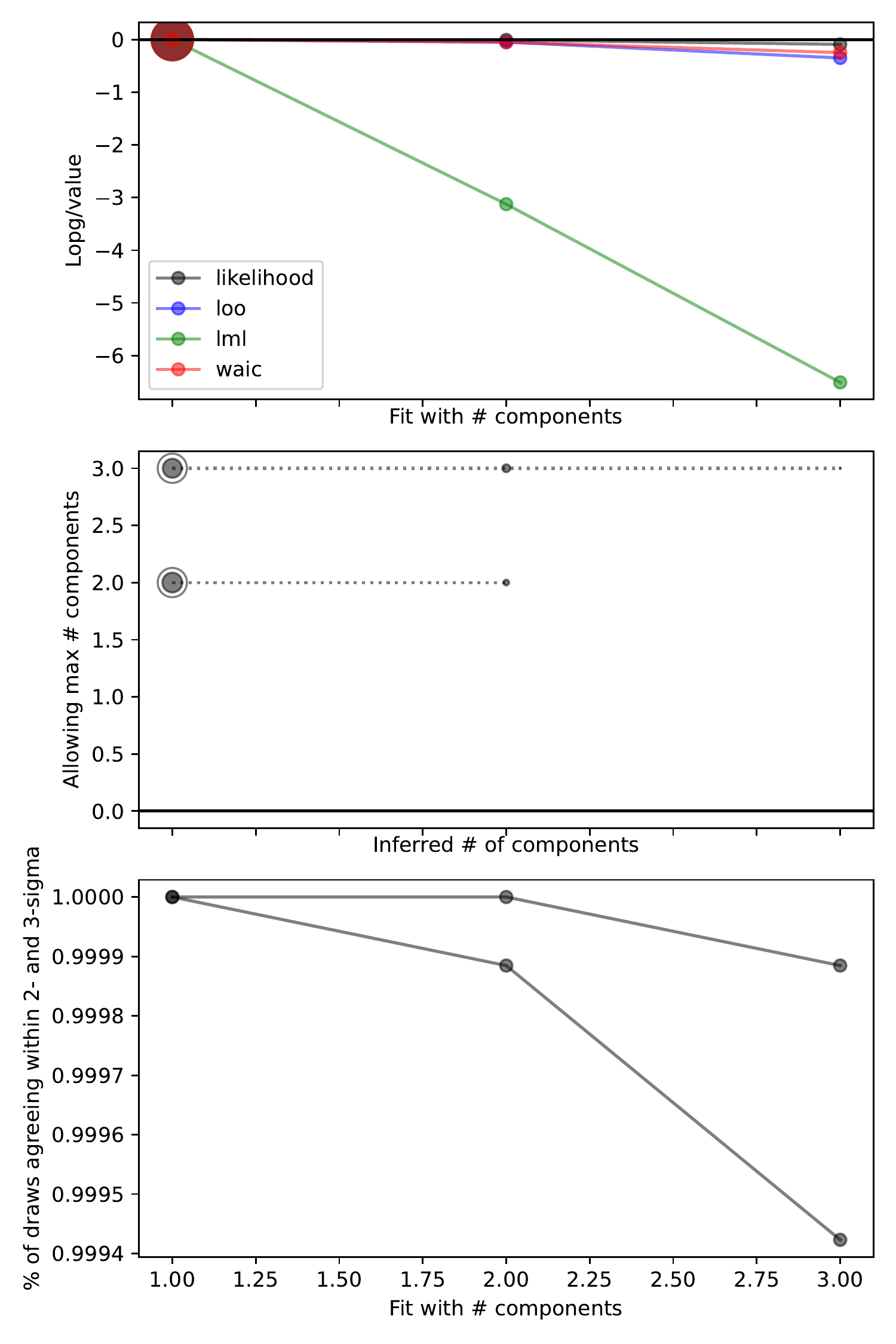}
\includegraphics[width=0.33\textwidth,clip]{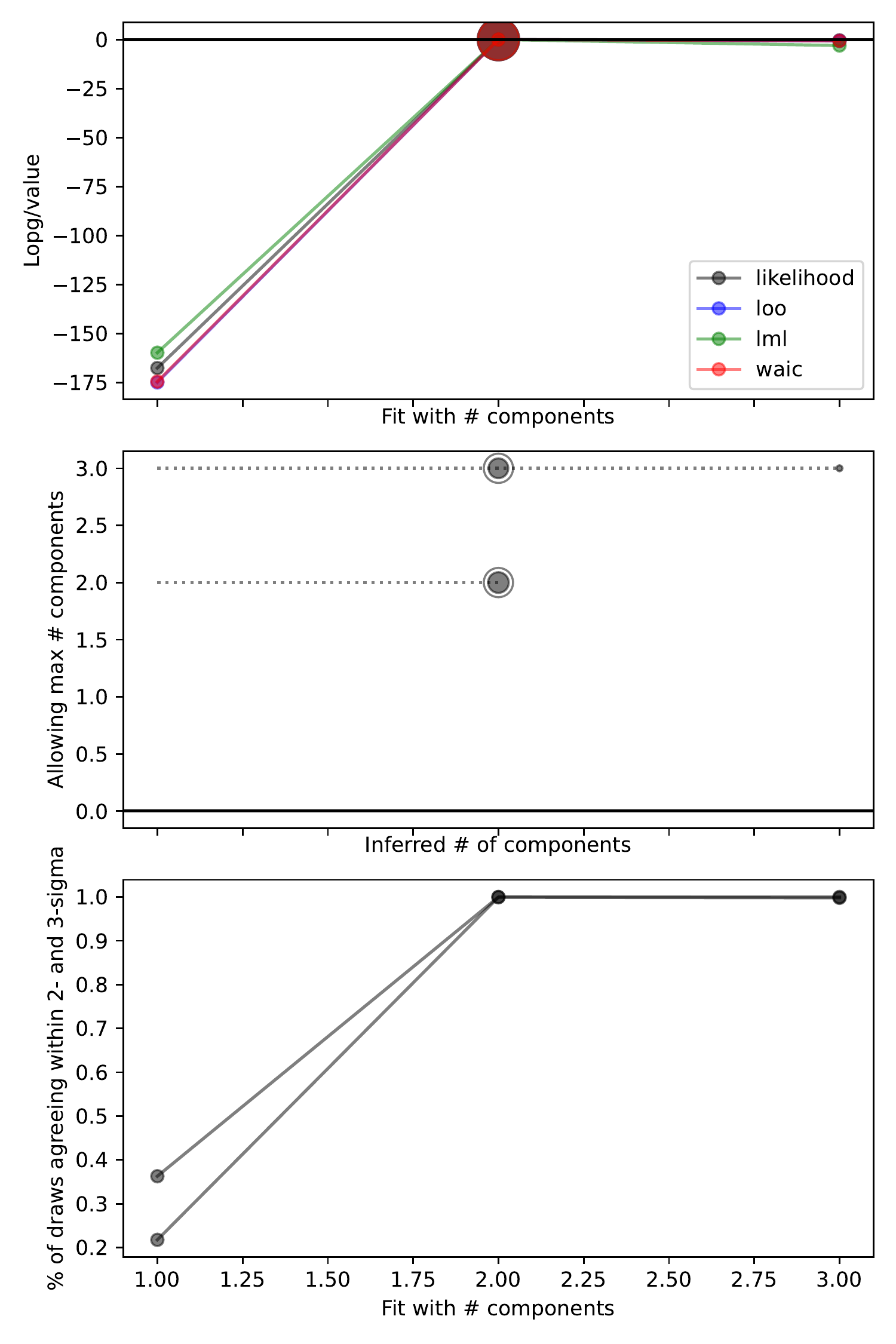}
\includegraphics[width=0.33\textwidth,clip]{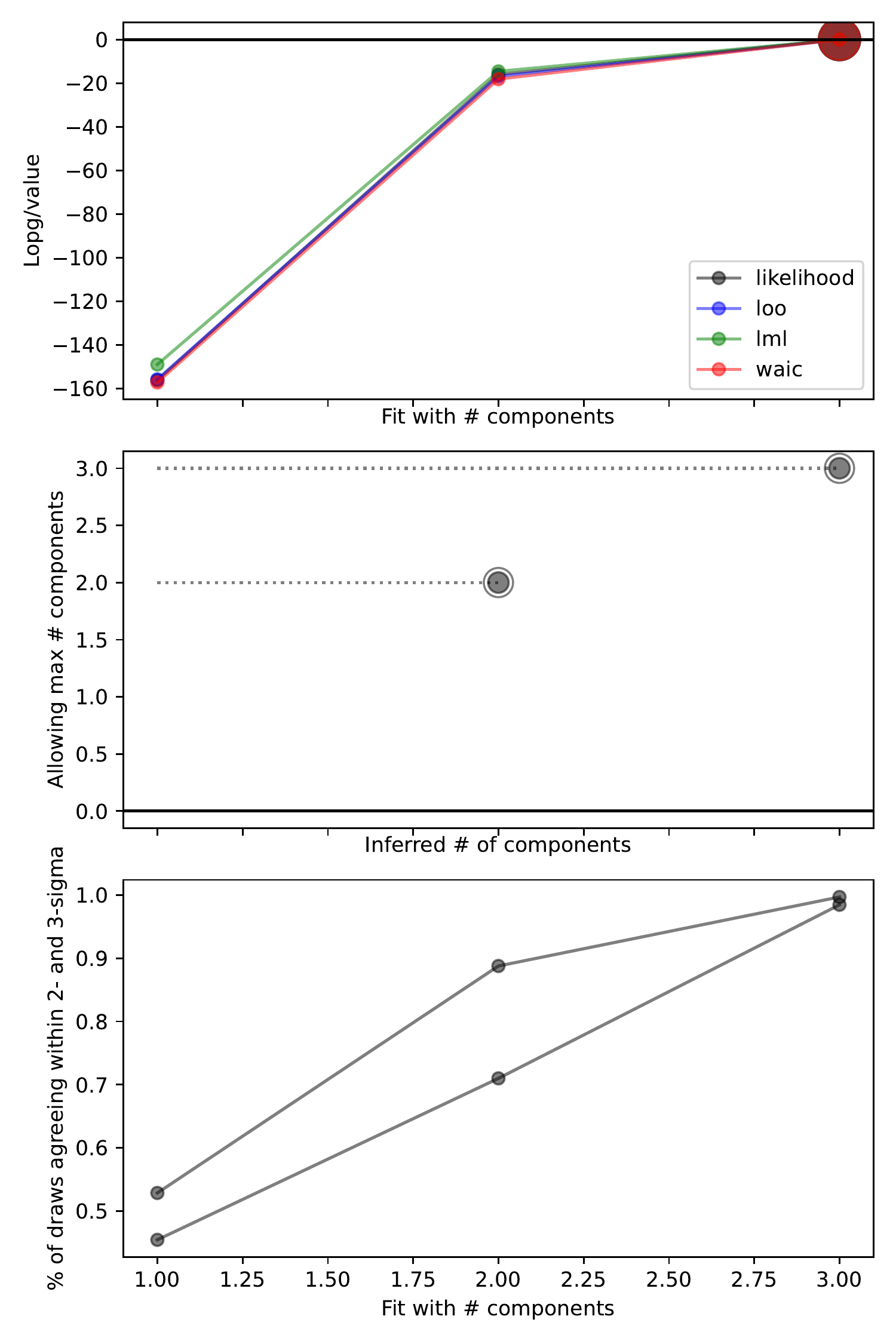}
\caption{Model comparison to estimate the number of sectors required. We show models with an input of one, two, and three sectors from left to right. In the \textit{top} panel we show the log values of LOO, WAIC, and the SMC marginal likelihood (``lml'') for MCMC runs with a number of sectors from one to three. In the \textit{middle} panel we show the likelihood weights (not to be confused with the mixing weight) of each sector when the number of sectors is allowed to vary. Each horizontal line corresponds to a model with the number of sectors allowed to vary up to two or three. For each horizontal line, the filled circles show the inferred weight and the circled one highlights the number of sectors with the largest weight. In the \textit{bottom} panel we show the fraction of draws that agree with the observed value within $2$- and $3$-$\sigma$. }
         \label{fig:ncompsfig}
       \end{figure*}

In Figure\,\ref{fig:ncompsfig} we show a benchmark test for estimating the number of sectors. We start from a model with a given number of sectors (from one to three) and run the MCMC model fixing the number of sectors from one to three. We also show the results when letting the number of sectors free (but limited to two or three). One can see that all metrics (LOO, WAIC, SMC marginal likelihood) reach a maximum for the expected number of sectors. When the number of sectors is itself allowed to vary, the expected number is inferred (obviously as long as the maximum number of sectors is enough). We note that such methods are reliable only if (1) all models have passed the burn-in phase (for random walkers), (2) all chains have converged toward an identical solution, and (3) there are significantly more observations that parameters. Furthermore, we made the implicit assumption that there is no intrinsic reason to prefer any specific configuration for all possible combinations of parameters and data (i.e., $p(\mathcal{M})$ is the same for all models).

While the benchmark test performs as expected, the application for ISM components is much less straightforward. Models are supposed to represent any given number of sectors around a single cluster or possibly even several clusters (Sect.\,\ref{sec:ismtopo}). However, depending on the complexity of the dataset (e.g., observations coming from different phases) and the complexity of the studied object (e.g., single H\2\ region or a full galaxy) the model may not be fully adapted a priori, resulting in posterior probabilities that need to be regarded with caution. Moreover, models may not have the same a priori probability $p(\mathcal{M})$. For a multisector approach, $p(\mathcal{M})$ can be somewhat subjective and cannot be computed rigorously since we know the model does not correspond to the reality of a galaxy or a cluster. At the very least, $p(\mathcal{M})$ is expected to be low for a single sector with a given density/pressure profile. 

The systematic uncertainties related to the model description as a few sectors may motivate the use of more statistically appropriate models with a large number of clouds whose physical conditions can be parameterized. In that vein, it is important to consider and compare various types of distributions (Sect.\,\ref{sec:i18app}). 

\subsection{Influence of topology on parameters}\label{sec:ncompsinfluence}

While it is possible to compare models with different number of sectors based on the available observations (Sect.\,\ref{sec:ncomps}), it must be noted that some parameters are particularly sensitive to the number of sectors used. First and foremost, it is generally assumed here that all sectors contribute significantly to the emission of at least one tracer. This is to avoid having to consider an empty or unconstrained sector (model with holes), which would effectively change the covering factors of the other sectors. If we were to assume empty sectors, it is possible to calculate the effective total covering factor of nonempty sectors provided the input stellar cluster luminosity (i.e., the bolometric luminosity) is constrained independently (Sect.\,\ref{sec:ismtopo}). In the following we always assume only nonempty sectors.

\begin{figure*}
   \centering
\includegraphics[clip,width=0.16\textwidth]{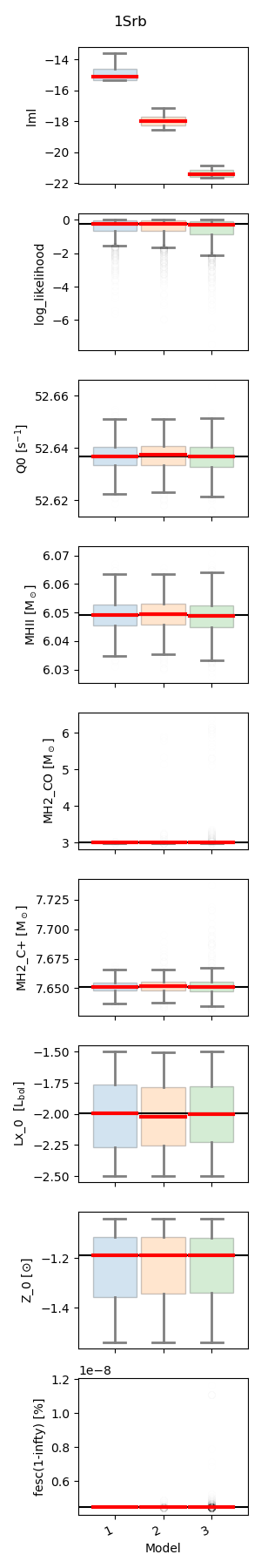}
\includegraphics[clip,width=0.16\textwidth]{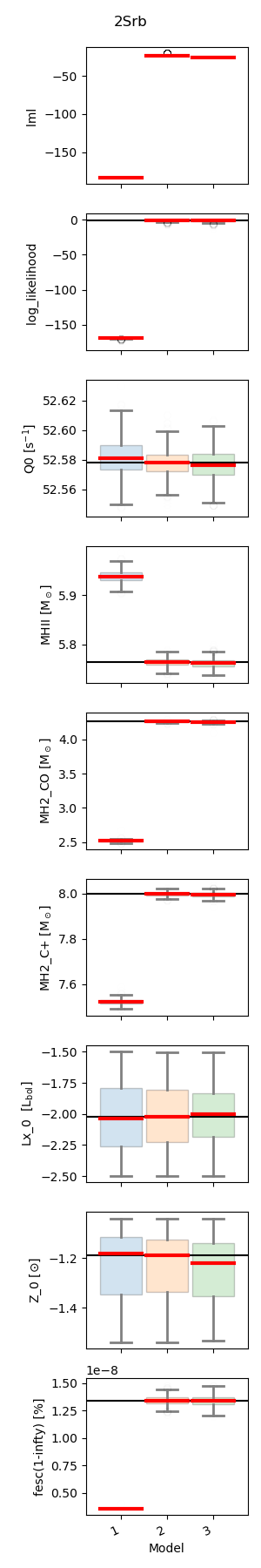}
\includegraphics[clip,width=0.16\textwidth]{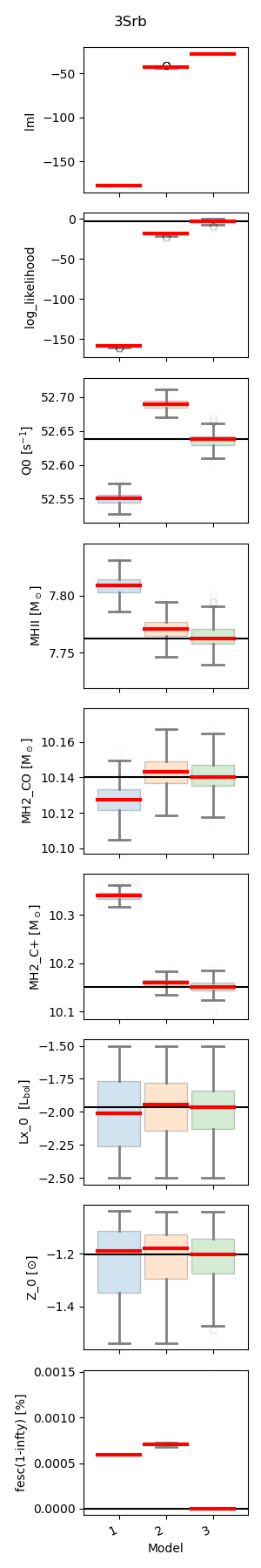}
\includegraphics[clip,width=0.16\textwidth]{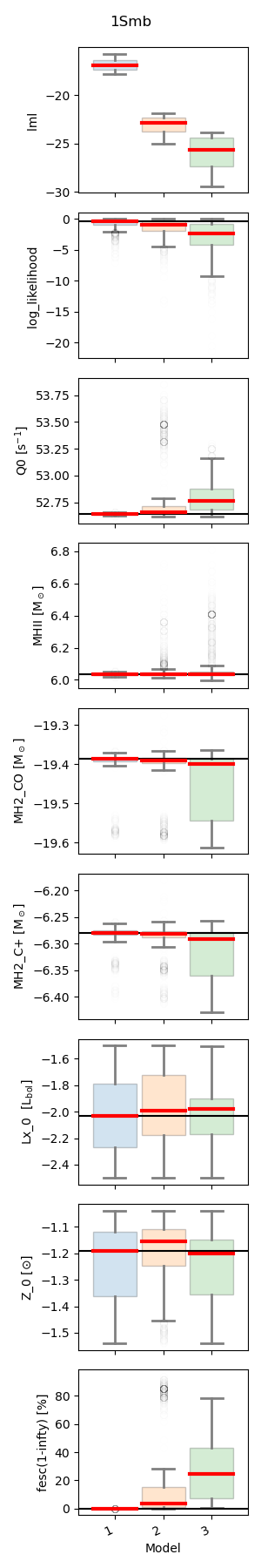}
\includegraphics[clip,width=0.16\textwidth]{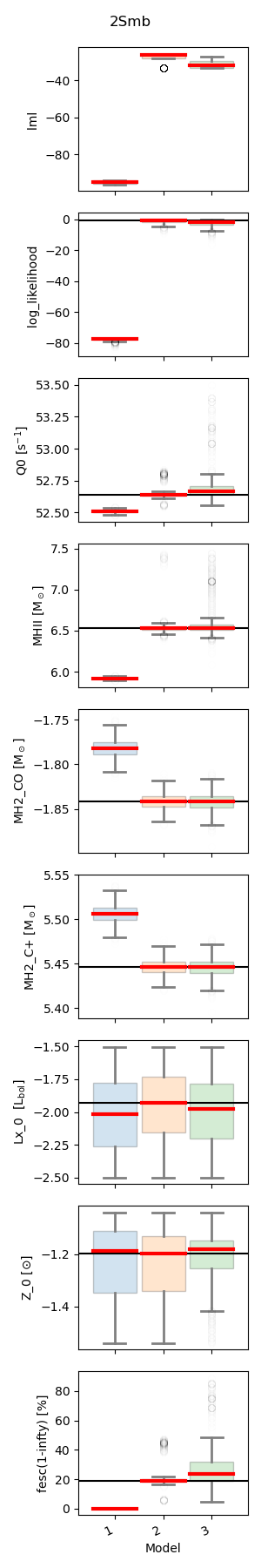}
\includegraphics[clip,width=0.16\textwidth]{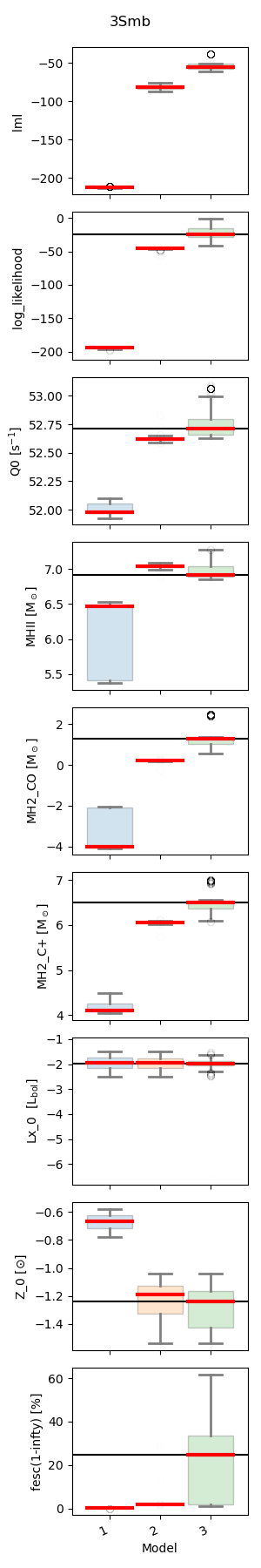}
\caption{Output parameters for the benchmark model comparison. We show models with an input of one, two, and three sectors and for radiation-bounded (``rb'') or matter-bounded (``mb'') template models. Each plot shows a standard boxplot configuration with the median value in red, with the box spanning from the first to third quartile, and the horizontal gray bars extending to $1.5$ times the inter-quartile range. Faint open circles show the data that extend beyond the horizontal gray bars.  In each panel we show the results forcing the number of components from one to three. The two top panels show the marginal likelihood (``lml'') and the likelihood while the other panels below show several parameters (from top to bottom the number of ionizing photons, the mass of H$^+$, the mass of H$_2$ traced by CO, the mass of H$_2$ traced by C$^+$, the X-ray luminosity, the metallicity, and the escape fraction of ionizing photons). The horizontal lines show the expected value, except for the marginal likelihood where it indicates the highest value. }
         \label{fig:paramsncomps}
       \end{figure*}

Figure\,\ref{fig:paramsncomps} shows the variation of inferred parameters as a function of the number of sectors used for the same series of typical benchmark models as in Section\,\ref{sec:ncomps}. When the true solution corresponds to a single sector, using a model with two or three sectors does not change significantly the parameter values except for the escape fraction of ionizing photons. However, such models become more and more unlikely as the number of sectors increases. Similarly, using three sectors for a true solution with two sectors leads to a good agreement with expected results, this time even for the escape fraction. More generally, using a larger than necessary number of sectors usually does not impact significantly the results except for the escape fraction in some cases. On the other hand, using a lower than necessary number of sectors may result in significant differences for all parameters, but keeping in mind that the model is likely unsatisfactory (Fig.\,\ref{fig:ncompsfig}). n summary and unsurprisingly, the parameters which depend the most on the topology are the ones which depend on the depth of each sector, that is the escape fraction of ionizing photons but also the mass of H$_2$.

\subsection{Selection of observations}\label{sec:tracers}

\begin{figure}
   \centering
\includegraphics[width=0.45\textwidth]{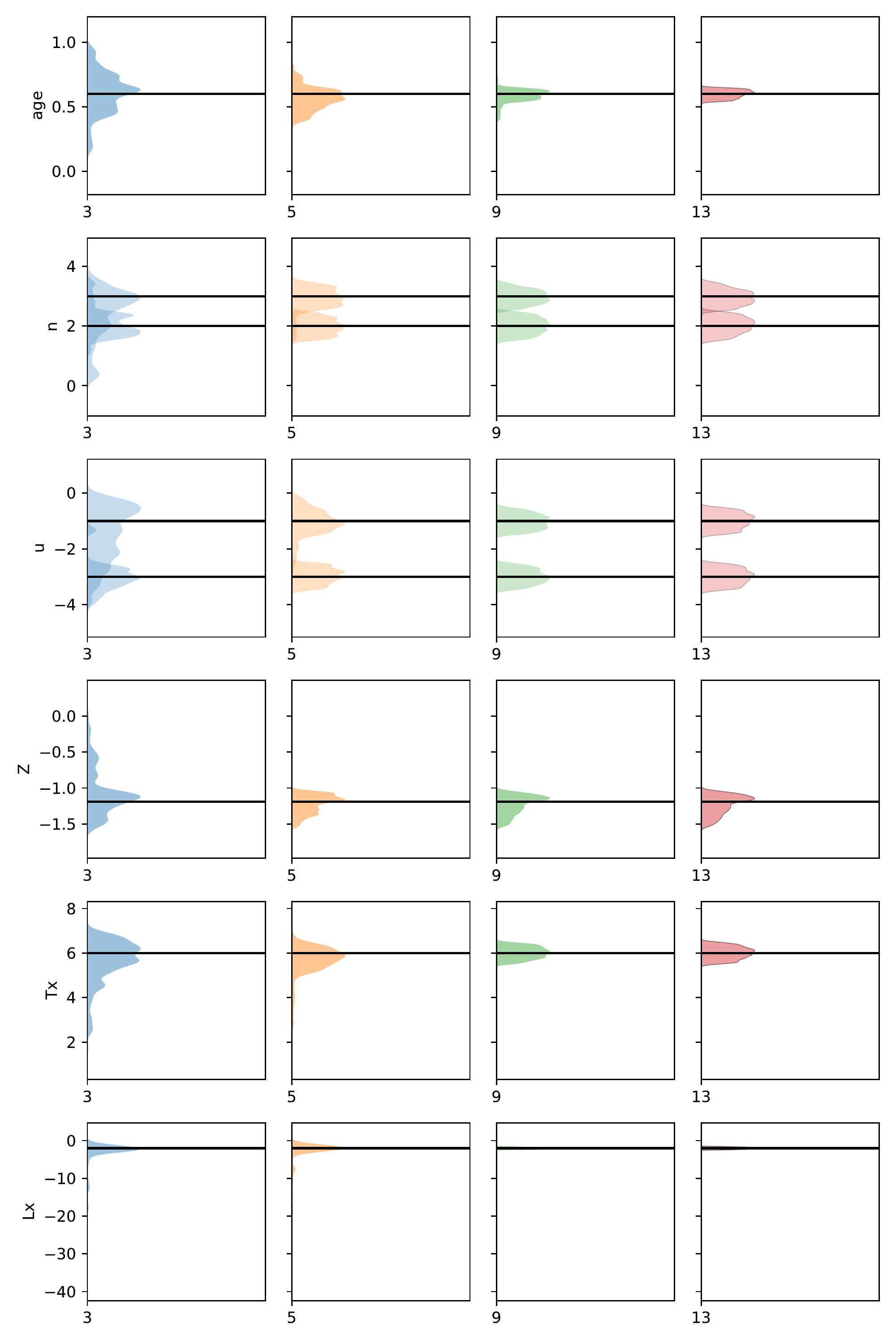}
\caption{Evolution of the solution with the number of lines considered for a benchmark run. Each row corresponds to a parameter, and each column represents a different number of lines from three to $13$ ([C\2]+[O\3] $88$\mic+[O\1] $63$\mic, +[N\2] $205$\mic+[O\1] $145$\mic, +[Ne\2] $12.8$\mic+[Ne\3] $15.6$\mic+[N\3] $57$\mic+[Ar\2] $7$\mic, +[Ar\3] $9$\mic+H$_2$ $17$\mic+[Si\2] $34$\mic+[Ne\5] $14$\mic. The horizontal black lines indicate the expected solution. The model used here is radiation-bounded. }
         \label{fig:nlines_rb}
       \end{figure}

\begin{figure}
   \centering
\includegraphics[width=0.45\textwidth]{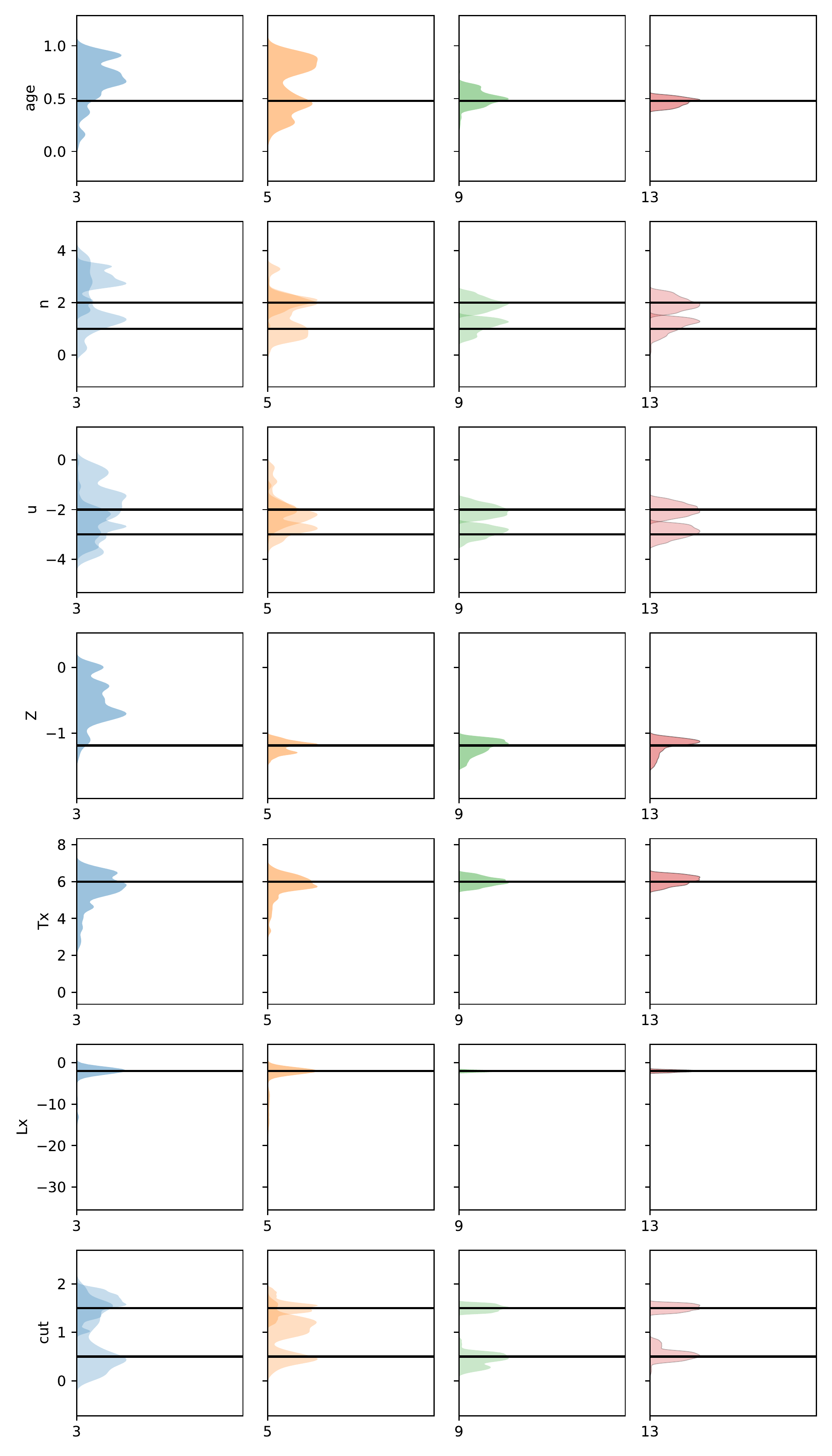}
\caption{Same as Figure\,\ref{fig:nlines_rb}, but for a matter-bounded model. }
         \label{fig:nlines_mb}
       \end{figure}

In Figures\,\ref{fig:nlines_rb} and \ref{fig:nlines_mb} we show a benchmark test analyzing the evolution of the solution with an increasing number of lines. For this illustration, the series of lines we consider is sorted based on their detection or potential detection in high-$z$ sources. The first set of three lines includes [C\2], [O\3] $88$\mic, and [O\1] $63$\mic. We then add [N\2] $205$\mic\ and [O\1] $145$\mic, then [Ne\2] $12.8$\mic, [Ne\3] $15.6$\mic, [N\3] $57$\mic\ and [Ar\2] $7$\mic, then [Ar\3] $9$\mic, H$_2$ $17$\mic, [Si\2] $34$\mic, and [Ne\5] $14$\mic. As the figures show, the solution naturally improves with the number of lines considered. In the radiation-bounded case (eight parameters; Fig.\,\ref{fig:nlines_rb}), a satisfactory solution is found even with three lines. For the matter-bounded case (ten parameters), for which there is an added degeneracy due to the inclusion of the depth parameter, a satisfactory solution is found with nine lines. 

       While it may seem that considering as many lines as possible would be the best option, there are several important caveats to take into account when dealing with actual observations. Since one is looking for an approximate/simplified topological model, it is may be somewhat illusory to search a model that will reproduce many observations at once. There is a risk of diluting the actual solution if the configuration is not adapted and/or if the systematic uncertainties (e.g., instrumental, atomic data, and abundances) are not properly accounted for. For instance, a model may be controlled by the likelihoods of a subset of many observations corresponding to lines arising in similar physical conditions -- though in slight disagreement -- at the expense of other few observations constraining potentially interesting/complementary parameters. Furthermore, outliers or observations with abnormally small error bars may produce a pathological posterior with local probability peaks where the sampler gets stuck (either preventing excursions for random walkers or affecting the computation of important samples for SMC). Unfortunately, this prevents the sampler from exploring other regions in the parameter space and, as a result, the sampler may be stuck in a global (all observations combined) low-probability region. Inversely, this also means that a global high-probability region may be dominated by an outlier or and observation with an abnormally small uncertainty. The $\hat{r}$ diagnostic (comparing means and variances of chains; \citealt{Vehtari2019a}) may be useful to examine whether one observation dominates the likelihood calculation.

Since these issues intrinsically depend on which observations are considered and with what uncertainties, the Bayesian approach by itself does not perform necessarily better than the $\chi^2$ approach and great care must be taken to adapt the model configuration to the set of observations and to account for all possible systematic uncertainties. Nevertheless, we can use the PDF to analyze the dependencies between parameters and predicted observations. Once the sampler converges, the values drawn according to the primary parameter PDFs make it possible to investigate correlations between parameters, between parameters and predicted observations, and between predicted observations themselves. With this in mind, it may be useful to include all available observations at first and, (1)
identify observations that are not well reproduced in order to consider potential missing systematic uncertainties and/or incorrect model assumptions (e.g., number of parameters and specific configuration), (2) check if some observations effectively constrain at least one parameter, and inversely, if every parameter is constrained by at least one observation, and (3) check if some observations constrain the same parameter(s), in which case redundant observations may be removed to prevent that the likelihood is controlled only by few parameters without changing significantly the solution.
  
\begin{figure}
   \centering
\includegraphics[width=0.52\textwidth]{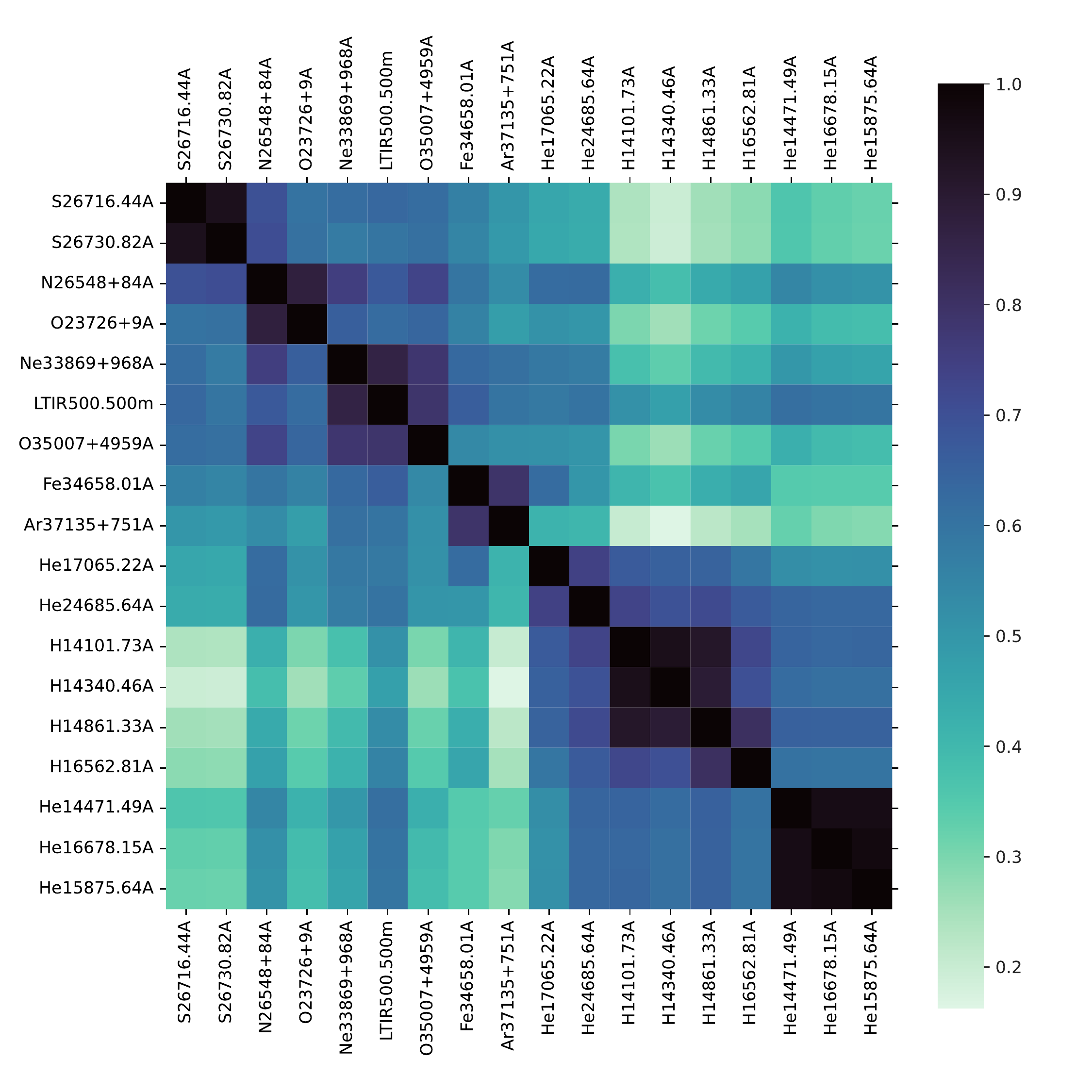}
\caption{Map of the correlation coefficients between predicted observations for a benchmark run. Here one can see that, according to the input grid, most H lines trace the same parameters around the solution (the same applies to He or [S\2]).
}
         \label{fig:corr_compa}
       \end{figure}

  Figure\,\ref{fig:corr_compa} shows the map of the correlation coefficients between predicted observations for a given run. For each set of two observations, we compare the correlation coefficients with each parameter for all MCMC draws (i.e., around the solution). The final correlation coefficient between the two predicted observations is zero if the two predicted observations constrain entirely different parameters and one if the two predicted observations constrain precisely the same parameters. The correlations do not account for the observed values/uncertainties and are calculated using the input grid values. Nevertheless, they may help in choosing an optimal set of observations. 

\section{Modeling of the low-metallicity star-forming dwarf galaxy \izw}\label{sec:i18app}

The low-metallicity star-forming dwarf galaxy \izw\ has been extensively studied with topological models in \citet[hereafter P08]{Pequignot2008a} and \citet[hereafter L17]{Lebouteiller2017a}. Among the most pressing questions concerning this object are the amount and distribution of the molecular gas and the influence of the known bright X-ray source on the ISM properties. 

We use the SFGX grid (Sect.\,\ref{sec:ismapp}) with a linear interpolation for the metallicity and luminosity parameters. The interpolation is particularly useful for the luminosity which is not well sampled with only two values (Sect.\,\ref{sec:clusters}). Output parameter values are provided as the median and highest density intervals with $94$\% confidence level. The full list of observations used is shown in Table\,\ref{tab:lines}. We note that the observed values for IR lines have been completely revised in \citetalias{Lebouteiller2017a} compared to \citetalias{Pequignot2008a}, but these lines had not been used for the convergence process in the latter. Furthermore, the fluxes used here are tailored to the study of the NW H\2\ region in \izw\ and differ slightly from the uniform dataset used for the DGS sample in \cite{Ramambason2022a}.

\subsection{Reproducing the modeling strategy of \cite{Pequignot2008a} with optical lines: Model $\mathcal{M}_{\rm I,opt}$}\label{sec:IGP08}

We first build model $\mathcal{M}_{\rm I,opt}$ to illustrate how MULTIGRIS may reproduce the results in \citetalias{Pequignot2008a} with the use of a general photoionization grid. \citetalias{Pequignot2008a} used an iterative and partly-manual convergence process specifically tailored to \izw. The observed spectrum used in \citetalias{Pequignot2008a} is close to \cite{Izotov1999a} but includes a correction from E(B-V)$=0.08$ to $0.04$, which in turn provides a different value and a constraint on the H$\alpha$/H$\beta$ ratio as compared to \cite{Izotov1999a}. 

While \citetalias{Pequignot2008a} used $12.97$\,Mpc for the distance, we use here $18.2$\,Mpc \citep{Aloisi2007b} to be consistent with \citetalias{Lebouteiller2017a}. Predicted physical quantities in \citetalias{Pequignot2008a} are modified accordingly here. Contrary to \citetalias{Pequignot2008a} we use error bars as they are naturally required to produce random draws in the MCMC process. In order to prevent the sampler from being stuck in local high-probability regions (Sect.\,\ref{sec:tracers}), we introduce a default minimum $10\%$ error on all lines. 

Instead of using the [S\2] line ratio as a specific observation to constrain the density around the ionization front as in \citetalias{Pequignot2008a}, we use the absolute fluxes of all lines including both [S\2] lines. The lines used for inference depend on the densities at various depths, which all translate into a constraint for the density at the illuminated front, which is the only parameter defining the density (Sect.\,\ref{sec:ismprops}). While we could use the known X-ray luminosity \citep{Kaaret2013a} and dust-to-gas mass ratio \citep{Galliano2021a} in \izw\ as observed values together with the spectral line fluxes, we chose to predict them instead and compare them with observations a posteriori. 

The helium abundance in the SFGX database is He/H$=0.098$ and does not scale with metallicity. We tentatively scale all helium line intensities using a nuisance variable with the assumption that (1) on first order all line fluxes scale with the abundance as long as the scaling factor is small enough and (2) the chemical and physical conditions of the model are not significantly impacted by the abundance variation. In all results below, we find He/H between $0.07-0.08$ for all runs, compared to $0.08$ used in \citetalias{Pequignot2008a}. Helium is the only element for which we force the abundance. The other elemental abundances scale with O/H depending on several prescriptions described in \cite{Ramambason2022a}.

Since our goal is to compare to \citetalias{Pequignot2008a}, we first use a similar three-sector configuration with an identical inner radius for all sectors, with no prior. We allow the cloud depth to reach slightly beyond the ionization front in order to fully consider the ionized layer. The primary parameters that we use for inference as well as predicted (secondary) parameters are listed in Table\,\ref{tab:params}.

\begin{table*}
  \caption{ISM model parameters.}
\begin{tabular}{l|ll}
  \hline\hline
  Parameter & Description & Values/range \\
  \hline
 \multicolumn{3}{l}{Primary parameters with distinct values for each component (sector)} \\
  \hline
  $w$ & Component (sector) mixing weight & $[0--1]$\\
  $\log U_0$ & Ionization parameter\tablefootmark{a} & $[-4,-3,-2,-1,0]$  \\
  $\log n_0$ [cm$^{-3}$] & Gas density\tablefootmark{a} & $[0,1,2,3,4]$  \\
  cut  & Depth into the cloud\tablefootmark{b} & $[0--4]$ step $0.25$ \\
  \hline
 \multicolumn{3}{l}{Primary parameters with a single value for all components (sectors)} \\
  \hline
  $\log Z$ & Metallicity & $[-2.19, -1.889, -1.19, -0.889, -0.667, -0.491, -0.19, 0.111]$ \\
  $\log Z_{\rm dust}$ & Dust-to-gas mass ratio\tablefootmark{c} & $[-0.68, 0, +0.70]$ \\
  $\log L_{\rm X}$ [L$_\odot$] & X-ray source luminosity & $[-\infty,-3,-2,-1]$ \\
  $\log T_{\rm X}$ [K] & X-ray source temperature & $[0,5,6,7]$ \\
  $\log L*$ [L$_\odot$] & Stellar cluster luminosity & $[7, 9]$ \\
  Age [Myr] & Burst age & $[1,2,3,4,5,6,8,10]$  \\
  \hline
 \multicolumn{3}{l}{Secondary (predicted) parameters in this study} \\
  \hline
  $\log Q({\rm H})$ [s$^{-1}$] & Number of ionizing photons & \\
  $f_{\rm esc}(1-\infty)$ [\%] & \multicolumn{2}{l}{Escape fraction of photons with energy range $[1-\infty]$\,Ryd} \\
  $f_{\rm esc}(1-1.8)$ [\%] & \multicolumn{2}{l}{Escape fraction of photons with energy range $[1-1.8]$\,Ryd} \\
  $f_{\rm esc}(1.8-4)$ [\%] & \multicolumn{2}{l}{Escape fraction of photons with energy range $[1.8-4]$\,Ryd} \\
  $f_{\rm esc}(4-20)$ [\%] & \multicolumn{2}{l}{Escape fraction of photons with energy range $[4-20]$\,Ryd} \\
  $f_{\rm esc}(20-\infty)$ [\%] & \multicolumn{2}{l}{Escape fraction of photons with energy range $[20-\infty]$\,Ryd} \\
  $M_{\rm H+}$ [$\log$ M$_\odot$] & Mass of H$^+$ \\
  $M_{\rm H0}$ [$\log$ M$_\odot$] & Mass of H$^0$ \\
  $M_{\rm H2}$ [$\log$ M$_\odot$] & Mass of H$_2$ \\
  $M_{\rm H2,C+}$ [$\log$ M$_\odot$] & Mass of H$_2$ traced by C$^+$ \\
  $M_{\rm H2,C0}$ [$\log$ M$_\odot$] & Mass of H$_2$ traced by C$^0$ \\
  $M_{\rm H2,CO}$ [$\log$ M$_\odot$] & Mass of H$_2$ traced by CO \\
  $M_{\rm dust}$ [$\log$ M$_\odot$] & Dust mass \\
  $f_{\rm CII,H+}$ & Fraction of [C\2] associated with H$^+$ \\
  $f_{\rm CII,H0}$ & Fraction of [C\2] associated with H$^0$ \\
  $f_{\rm CII,H2}$ & Fraction of [C\2] associated with H$_2$ \\
  \hline
\end{tabular}\\
\label{tab:params}
    \tablefoottext{a}{Calculated at the illuminated face of the cloud. }
    \tablefoottext{b}{See definition in Section\,\ref{sec:ismprops}. }
    \tablefoottext{c}{The value indicated corresponds to the 95\% envelop (in dex) around the median relationship in \cite{Galliano2021a}.}
  \end{table*}

\begin{table*}
  \caption{Observations and predictions for the ionized and neutral gas models of \izw.}
    \resizebox{\textwidth}{!}{%
\begin{tabular}{l|c|cccc}
\hline\hline
    Observations & Observed value  & $\mathcal{M_{\rm I,opt}}$  & $\mathcal{M_{\rm I,opt+IR}}$ & $\mathcal{M_{\rm NA}}$ &  $\mathcal{M_{\rm NM}}$  \\
  \hline
    S26716.44A           & ${22.50}\pm{7.00}$ & ${16.06}^{+4.62}_{-3.71}$  & ${15.29}^{+4.49}_{-3.08}$  & ${17.48}^{+5.50}_{-3.38}$  & ${18.27}^{+3.86}_{-4.89}$ \\ 
    S26730.82A           & ${16.90}\pm{7.00}$ & ${12.61}^{+2.85}_{-2.70}$  & ${11.88}^{+2.39}_{-3.07}$  & ${13.66}^{+4.18}_{-3.04}$  & ${14.29}^{+3.05}_{-3.16}$ \\ 
    H14101.73A           & ${266.00}\pm{26.60}$ & ${250.68}^{+41.85}_{-25.57}$  & ${249.06}^{+22.45}_{-27.66}$  & ${250.26}^{+22.62}_{-27.61}$  & ${241.69}^{+22.96}_{-26.11}$ \\ 
    H14340.46A           & ${461.00}\pm{46.10}$ & ${487.67}^{+43.61}_{-60.95}$  & ${486.39}^{+57.38}_{-40.49}$  & ${487.33}^{+51.74}_{-48.84}$  & ${477.37}^{+51.21}_{-45.53}$ \\ 
    H14861.33A           & ${1000.00}\pm{100.00}$ & ${1019.23}^{+87.81}_{-125.72}$  & ${1015.64}^{+119.14}_{-88.84}$  & ${1017.26}^{+113.61}_{-102.38}$  & ${1002.37}^{+104.68}_{-101.70}$ \\ 
    H16562.81A           & ${2860.00}\pm{286.00}$ & ${2901.65}^{+278.02}_{-366.10}$  & ${2958.23}^{+286.61}_{-342.18}$  & ${3089.73}^{+344.61}_{-357.65}$  & ${2993.75}^{+351.15}_{-292.05}$ \\ 
    He14471.49A          & ${21.40}\pm{2.14}$ & ${24.60}^{+3.06}_{-3.64}$  & ${23.94}^{+3.26}_{-3.30}$  & ${23.25}^{+3.78}_{-2.43}$  & ${24.08}^{+3.27}_{-3.10}$ \\ 
    He15875.64A          & ${67.70}\pm{6.77}$ & ${68.73}^{+7.09}_{-10.57}$  & ${66.95}^{+7.86}_{-8.92}$  & ${65.13}^{+8.88}_{-7.84}$  & ${66.42}^{+7.29}_{-9.12}$ \\ 
    He16678.15A          & ${25.30}\pm{2.53}$ & ${18.60}^{+2.44}_{-2.73}$  & ${18.01}^{+2.69}_{-2.19}$  & ${17.43}^{+2.29}_{-2.53}$  & ${17.74}^{+2.16}_{-2.28}$ \\ 
    He17065.22A          & ${24.40}\pm{2.44}$ & ${22.76}^{+4.29}_{-4.10}$  & ${22.10}^{+4.63}_{-3.99}$  & ${22.82}^{+5.47}_{-3.79}$  & ${22.70}^{+4.82}_{-3.73}$ \\ 
    He24685.64A          & ${36.80}\pm{3.68}$ & ${35.59}^{+8.56}_{-10.48}$  & ${38.49}^{+7.11}_{-10.18}$  & ${34.85}^{+9.22}_{-10.53}$  & ${35.54}^{+8.53}_{-9.30}$ \\ 
    N26548+84A           & ${9.20}\pm{2.00}$ & ${9.81}^{+2.95}_{-2.83}$  & ${9.18}^{+2.88}_{-2.62}$  & ${9.06}^{+4.16}_{-2.53}$  & ${9.93}^{+2.84}_{-2.66}$ \\ 
    O23726+9A            & ${238.00}\pm{23.80}$ & ${217.80}^{+36.13}_{-47.13}$  & ${213.20}^{+44.43}_{-39.32}$  & ${202.23}^{+51.98}_{-48.84}$  & ${207.29}^{+43.47}_{-43.02}$ \\ 
    O35007+4959A         & ${2683.00}\pm{268.30}$ & ${2541.64}^{+419.60}_{-372.86}$  & ${2382.20}^{+314.02}_{-274.36}$  & ${2407.67}^{+308.47}_{-277.28}$  & ${2446.83}^{+293.20}_{-348.72}$ \\ 
    Ne33869+968A         & ${191.00}\pm{19.10}$ & ${204.78}^{+37.24}_{-29.07}$  & ${192.99}^{+21.14}_{-25.22}$  & ${195.72}^{+20.34}_{-26.47}$  & ${194.82}^{+26.52}_{-22.43}$ \\ 
    Ar37135+751A         & ${23.50}\pm{2.35}$ & ${22.90}^{+3.82}_{-4.10}$  & ${23.54}^{+2.87}_{-2.86}$  & ${22.97}^{+2.32}_{-4.18}$  & ${22.21}^{+2.89}_{-3.07}$ \\ 
    Fe34658.01A          & ${4.50}\pm{2.00}$ & ${7.08}^{+1.78}_{-1.86}$  & ${7.12}^{+1.54}_{-1.90}$  & ${6.70}^{+2.00}_{-1.44}$  & ${6.30}^{+1.46}_{-1.79}$ \\ 
    O388.3323m           & ${200.00}\pm{30.00}$ & (${187.20}^{+47.45}_{-60.46}$)  & ${174.74}^{+48.43}_{-30.17}$  & ${175.54}^{+41.36}_{-31.22}$  & ${166.47}^{+31.80}_{-41.51}$ \\ 
    H112.3684m           & ${8.50}\pm{3.00}$ & (${6.38}^{+0.82}_{-0.99}$)  & ${6.30}^{+0.60}_{-0.76}$  & ${6.25}^{+0.65}_{-0.73}$  & ${6.00}^{+0.68}_{-0.67}$ \\ 
    O425.8832m           & ${38.00}\pm{8.00}$ & (${35.36}^{+18.04}_{-21.47}$)  & ${27.63}^{+6.75}_{-9.73}$  & ${27.58}^{+8.68}_{-9.89}$  & ${27.51}^{+8.95}_{-8.54}$ \\ 
    Ne212.8101m          & ${6.00}\pm{3.00}$ & (${2.36}^{+0.83}_{-0.85}$)  & ${2.31}^{+0.69}_{-0.56}$  & ${4.62}^{+3.30}_{-1.93}$  & ${5.03}^{+3.36}_{-2.35}$ \\ 
    Ne315.5509m          & ${47.00}\pm{10.00}$ & (${51.59}^{+8.90}_{-7.10}$)  & ${49.24}^{+4.96}_{-6.39}$  & ${49.82}^{+5.60}_{-5.72}$  & ${49.27}^{+6.10}_{-5.29}$ \\ 
    Ne336.0036m          & $<7.00$ & (${4.67}^{+0.87}_{-0.70}$)  & ${4.43}^{+0.45}_{-0.58}$  & ${4.48}^{+0.49}_{-0.53}$  & ${4.43}^{+0.55}_{-0.48}$ \\ 
    Ne514.3228m          & $<2.30$ & (${3.02}^{+8.75}_{-3.02}$)  & ${0.93}^{+1.14}_{-0.92}$  & ${1.50}^{+1.42}_{-1.17}$  & ${1.58}^{+1.17}_{-1.13}$ \\ 
    Ne524.2065m          & $<4.00$ & (${3.61}^{+10.13}_{-3.61}$)  & ${1.10}^{+1.33}_{-1.10}$  & ${1.83}^{+1.59}_{-1.41}$  & ${1.87}^{+1.26}_{-1.39}$ \\ 
    S318.7078m           & ${28.00}\pm{7.00}$ & (${37.01}^{+6.77}_{-6.25}$)  & ${37.98}^{+4.39}_{-4.75}$  & ${37.85}^{+3.69}_{-5.15}$  & ${37.12}^{+4.41}_{-4.44}$ \\ 
    S333.4704m           & ${40.00}\pm{7.00}$ & (${50.64}^{+13.24}_{-9.58}$)  & ${50.15}^{+8.59}_{-6.49}$  & ${51.27}^{+8.50}_{-7.77}$  & ${49.90}^{+8.26}_{-6.55}$ \\ 
    S410.5076m           & ${49.00}\pm{10.00}$ & (${65.65}^{+29.61}_{-21.61}$)  & ${53.05}^{+10.92}_{-9.87}$  & ${56.33}^{+9.74}_{-11.53}$  & ${56.86}^{+11.20}_{-10.97}$ \\ 
    Ar26.98337m          & $<5.00$ & (${0.22}^{+0.13}_{-0.11}$)  & ${0.18}^{+0.06}_{-0.06}$  & ${0.47}^{+0.45}_{-0.24}$  & ${0.57}^{+0.42}_{-0.33}$ \\ 
    Ar321.8253m          & $<8.00$ & (${0.60}^{+0.11}_{-0.13}$)  & ${0.60}^{+0.07}_{-0.08}$  & ${0.58}^{+0.09}_{-0.09}$  & ${0.56}^{+0.08}_{-0.08}$ \\ 
    Ar38.98898m          & $<10.00$ & (${8.25}^{+1.62}_{-1.27}$)  & ${8.45}^{+0.76}_{-1.37}$  & ${8.19}^{+1.15}_{-1.24}$  & ${7.92}^{+1.07}_{-1.09}$ \\ 
    Fe322.9190m          & $<6.00$ & (${3.75}^{+0.89}_{-1.23}$)  & ${3.53}^{+1.05}_{-1.00}$  & ${3.35}^{+1.30}_{-0.78}$  & ${3.12}^{+0.96}_{-0.87}$ \\ 
    LTIR500.500m         & ${28084.00}\pm{2808.40}$ & (${5425.82}^{+36065.1}_{-2509.9}$)  & (${16977.40}^{+184385.2}_{-13675.2}$)  & ${28190.57}^{+7133.6}_{-7615.3}$  & ${26628.15}^{+9958.7}_{-7175.0}$ \\ 
    C2157.636m           & ${76.00}\pm{20.00}$ & (${1.28}^{+2.86}_{-1.13}$)  & (${1.95}^{+1.88}_{-1.51}$)  & ${27.32}^{+19.42}_{-23.75}$  & ${29.30}^{+30.18}_{-24.53}$ \\ 
    O163.1679m           & ${60.00}\pm{20.00}$ & (${0.25}^{+0.26}_{-0.25}$)  & (${1.36}^{+26.86}_{-1.05}$)  & ${77.21}^{+44.25}_{-41.66}$  & ${85.29}^{+46.83}_{-41.94}$ \\ 
    Si234.8046m          & ${65.00}\pm{12.00}$ & (${7.69}^{+6.00}_{-4.40}$)  & (${4.79}^{+4.95}_{-1.64}$)  & ${52.70}^{+27.57}_{-17.67}$  & ${54.34}^{+27.82}_{-20.43}$ \\ 
    Fe217.9314m          & $<3.00$ & (${0.38}^{+0.39}_{-0.36}$)  & (${0.05}^{+0.51}_{-0.04}$)  & ${0.79}^{+1.10}_{-0.55}$  & ${0.94}^{+0.74}_{-0.78}$ \\ 
    Fe225.9811m          & ${13.00}\pm{8.00}$ & (${0.80}^{+0.92}_{-0.77}$)  & (${0.11}^{+0.94}_{-0.08}$)  & ${6.35}^{+5.62}_{-2.57}$  & ${6.61}^{+3.63}_{-3.71}$ \\ 
    CO2600.05m          & ${1.50}\pm{0.50}\times10^{-04}$ & (${7.79}^{+3.2\times10^{+11}}_{-7.78}\times10^{-18}$)  & (${1.19}^{+4.\times10^{+03}}_{-1.12}\times10^{-20}$)  & ${5.06}^{+1.4\times10^{+03}}_{-5.05}\times10^{-09}$  & ${7.05}^{+\times10^{+04}}_{-7.05}\times10^{-08}$ \\ 
    H22.12099m           & $<2.00$ & (${2.02}^{+1.2\times10^{+14}}_{-1.89}\times10^{-19}$)  & (${1.16}^{+18.77}_{-0.65}\times10^{-21}$)  & ${4.52}^{+10.76}_{-4.52}$  & ${0.10}^{+0.09}_{-0.09}$ \\ 
    H228.2130m           & $<7.00$ & (${2.02}^{+1.2\times10^{+14}}_{-1.89}\times10^{-19}$)  & (${1.16}^{+18.77}_{-0.65}\times10^{-21}$)  & ${1.29}^{+128.18}_{-1.29}$  & ${0.28}^{+5.37}_{-0.28}$ \\ 
    H212.2752m           & $<7.50$ & (${2.02}^{+1.2\times10^{+14}}_{-1.89}\times10^{-19}$)  & (${1.16}^{+21.41}_{-0.60}\times10^{-21}$)  & ${0.14}^{+0.87}_{-0.14}$  & ${0.59}^{+0.91}_{-0.59}$ \\ 
    H217.0300m           & $<7.00$ & (${2.02}^{+1.2\times10^{+14}}_{-1.89}\times10^{-19}$)  & (${1.16}^{+18.90}_{-0.65}\times10^{-21}$)  & ${0.17}^{+1.56}_{-0.17}$  & ${0.71}^{+2.61}_{-0.71}$ \\ 
    O16300+63A           & ${8.50}\pm{2.00}$ & (${1.62}^{+1.99}_{-1.57}$)  & (${0.19}^{+1.91}_{-0.16}$)  & (${3.54}^{+3.39}_{-2.14}$)  & (${4.37}^{+2.26}_{-3.33}$) \\ 
    O27320+30A           & ${6.30}\pm{2.00}$ & (${4.91}^{+1.63}_{-1.39}$)  & (${4.98}^{+2.29}_{-0.90}$)  & (${4.76}^{+1.51}_{-1.62}$)  & (${4.86}^{+1.27}_{-1.02}$) \\ 
    O34363.21A           & ${65.90}\pm{6.59}$ & (${58.86}^{+15.18}_{-15.15}$)  & (${50.96}^{+8.26}_{-8.92}$)  & (${53.71}^{+9.86}_{-8.24}$)  & (${55.66}^{+7.90}_{-11.46}$) \\ 
    S24068.60A           & ${3.70}\pm{1.00}$ & (${1.97}^{+0.61}_{-0.49}$)  & (${1.91}^{+0.71}_{-0.34}$)  & (${2.02}^{+0.80}_{-0.37}$)  & (${2.13}^{+0.45}_{-0.43}$) \\ 
    S36312.06A           & ${6.70}\pm{2.00}$ & (${9.96}^{+1.78}_{-1.46}$)  & (${9.96}^{+1.16}_{-1.06}$)  & (${9.83}^{+0.89}_{-1.42}$)  & (${9.56}^{+1.16}_{-1.33}$) \\ 
    Ar44740.12A          & ${4.50}\pm{3.00}$ & (${4.66}^{+2.70}_{-2.37}$)  & (${3.00}^{+1.16}_{-0.91}$)  & (${3.49}^{+1.45}_{-1.01}$)  & (${3.74}^{+1.36}_{-1.12}$) \\ 
    Fe34987.33A          & ${7.40}\pm{3.00}$ & (${0.26}^{+0.09}_{-0.07}$)  & (${0.25}^{+0.09}_{-0.06}$)  & (${0.25}^{+0.06}_{-0.06}$)  & (${0.23}^{+0.05}_{-0.06}$) \\ 
    Fe54227.19A          & ${1.80}\pm{0.70}$ & (${5.41}^{+4.26}_{-3.12}$)  & (${4.04}^{+1.69}_{-1.70}$)  & (${4.47}^{+1.85}_{-1.46}$)  & (${4.19}^{+1.02}_{-1.44}$) \\ 
      \hline
\end{tabular} \\
 }
\label{tab:lines}
    \tablefoot{Fluxes normalized to H$\beta=1000$ need to be scaled by $4.56\times10^{36}$ to convert to luminosity in erg\,s$^ {-1}$. Values in parentheses correspond to predicted observations. Model $\mathcal{M_{\rm I,opt}}$ considers only optical ionized gas observations, $\mathcal{M_{\rm I,opt+IR}}$ considers infrared lines as additional observations, $\mathcal{M_{\rm NA}}$ considers all lines tracing the ionized and neutral atomic gas (CO is only used as an upper limit), and $\mathcal{M_{\rm NM}}$ considers the neutral molecular gas. }
  \end{table*}

\begin{figure}
   \centering
\includegraphics[width=0.5\textwidth]{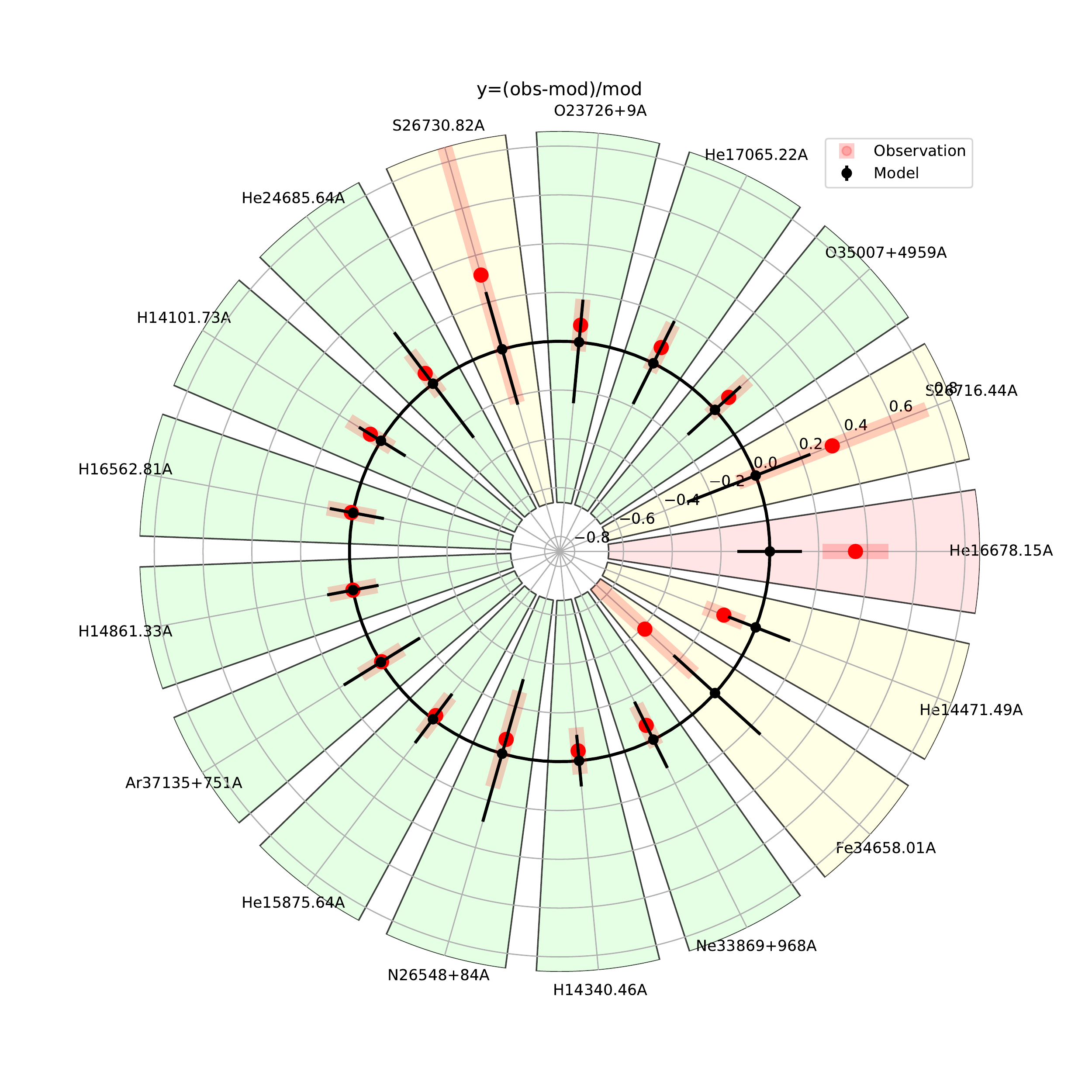}
\caption{Comparison between observed and modeled fluxes for the ionized gas model $\mathcal{M_{\rm I,opt}}$ of \izw. Plotted values in red are the difference between observations and models normalized by the model value. MULTIGRIS produces this plot in cartesian and polar coordinates, we show here the latter for illustration and because it is better adapted to a large number of observations. The black circle, black dots, and orthogonal black lines indicate the model and the red points and red error bars indicate the observations. The color of the sector is green if observations and models agree within errors and yellow if the model agrees within the observed error bar. }
         \label{fig:matchplotp08}
       \end{figure}

\begin{figure*}
   \centering
\includegraphics[width=0.35\textwidth]{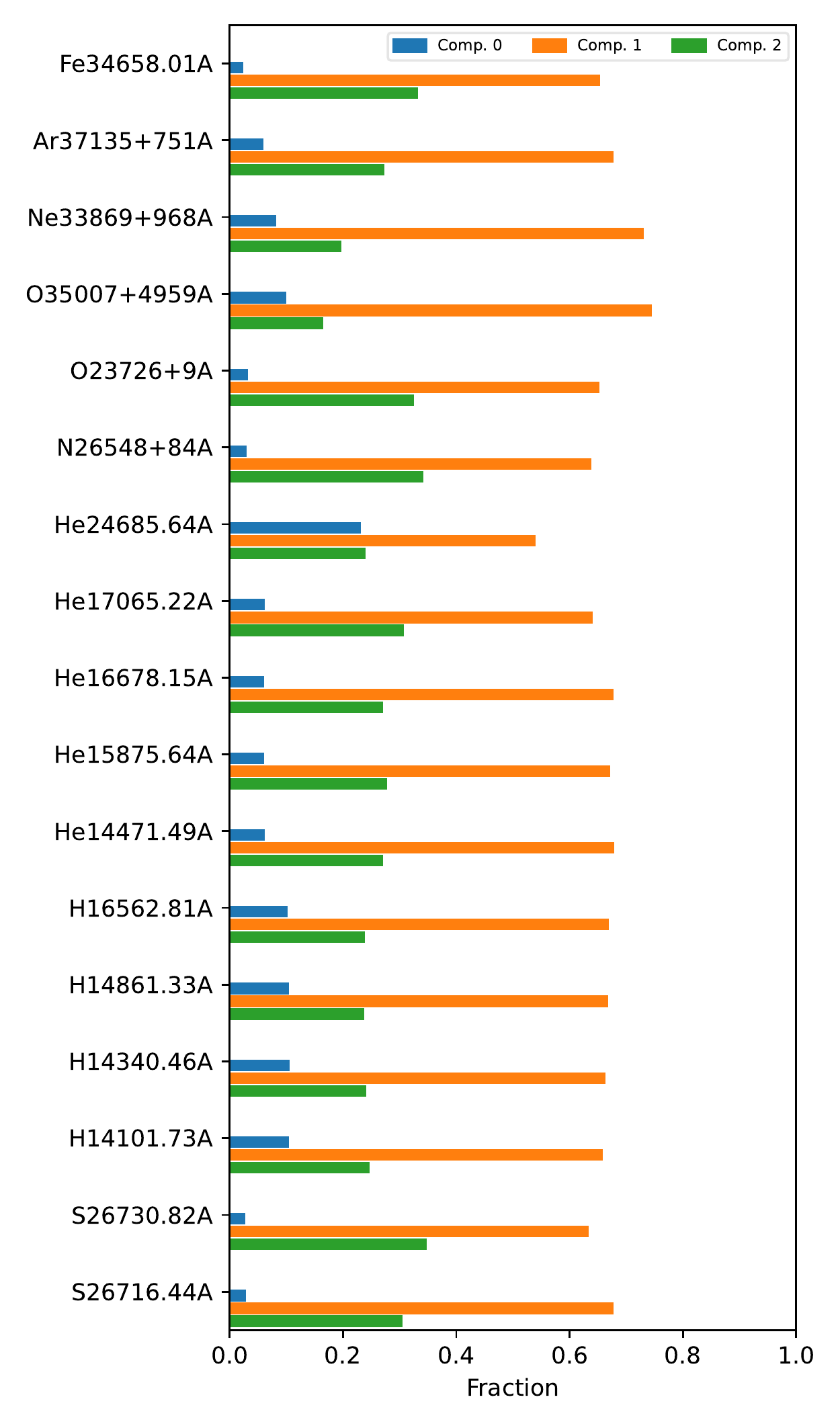}
\includegraphics[width=0.3\textwidth]{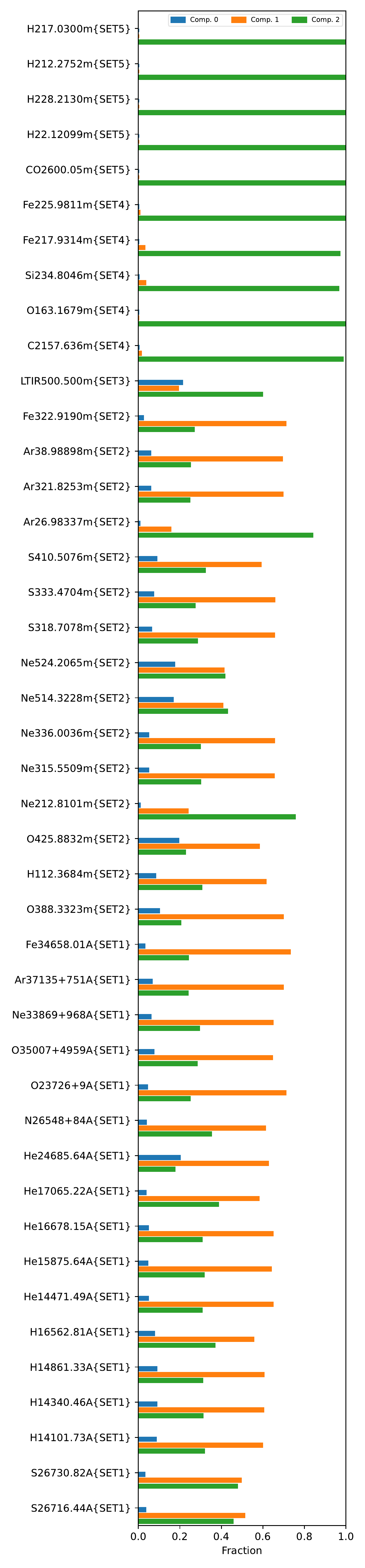}
\includegraphics[width=0.3\textwidth]{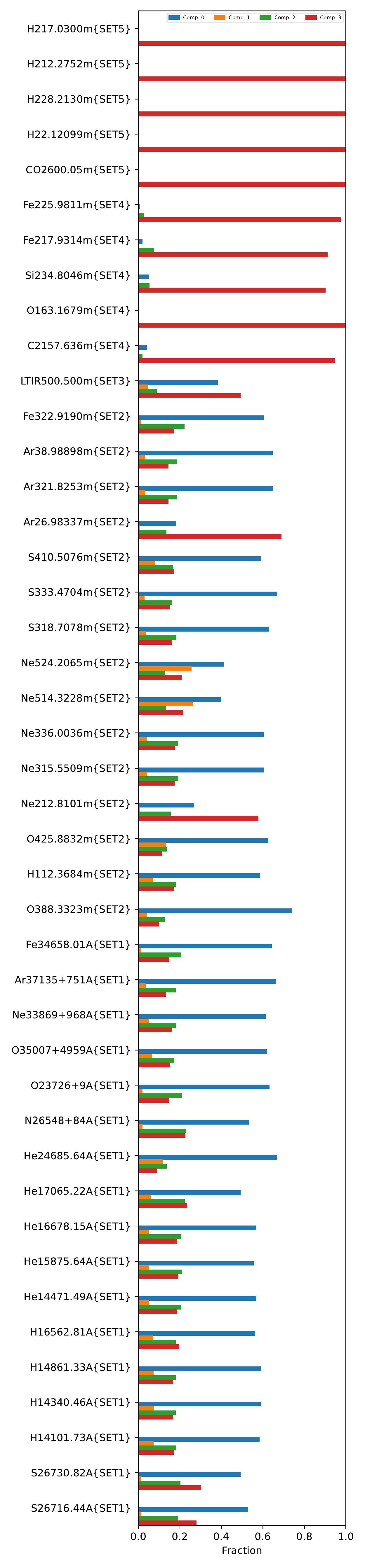}
\caption{Origin of each line emission in the MULTIGRIS model for the ionized gas model $\mathcal{M}_{\rm I,opt}$ (left), the neutral atomic gas model $\mathcal{M}_{\rm NA}$ (middle), and the neutral molecular gas model $\mathcal{M}_{\rm NM}$ (right). }
         \label{fig:fracplotp08}
       \end{figure*}

       We show the comparison between the observations and the predictions of model $\mathcal{M}_{\rm I,opt}$ in Table\,\ref{tab:lines} and Figure\,\ref{fig:matchplotp08}. Results for parameters are shown in Table\,\ref{tab:i18_results}. \citetalias{Pequignot2008a} derived one sector (a) as a matter-bounded ``veil'' with covering factor (CF) $44$\% to account for the fact that absorbing gas is expected for any line of sight from the cluster, best corresponding to our sector \#1, one sector as a matter-bounded sector with CF=$30$\%, best corresponding to our sector \#2, and one sector as a radiation-bounded sector with CF=$26$\% best corresponding to our sector \#3. The inner radius in model $\mathcal{M}_{\rm I,opt}$ is $\approx6^{+25}_{-5}\times10^{20}$\,cm as compared to a fixed value of $5.6\times10^{20}$\,cm in \citetalias{Pequignot2008a} ($18.2$\,Mpc). 
Sector \#1 contributes little to the total line emission but contributes significant fraction of He\2\ $4686$\AA, which is the brightest line in this sector (Fig.\,\ref{fig:fracplotp08}), as in \citetalias{Pequignot2008a}.

\begin{table*}
  \caption{\izw\ model parameters.}\label{tab:i18_results}
  \begin{tabular}{l|cccccc}
      \hline
    \hline
      Parameter                                & $\mathcal{M_{\rm I,opt}}$  & $\mathcal{M_{\rm I,opt+IR}}$ &  $\mathcal{M_{\rm NA}}$  & $\mathcal{M_{\rm NM}}$ & \citetalias{Pequignot2008a}/\citetalias{Lebouteiller2017a}                                      \\
    \hline
    \hline
 $w$  & $0.32^{+0.49}_{-0.30}$ & $0.52^{+0.36}_{-0.49}$ & $0.51^{+0.40}_{-0.41}$ & $0.64^{+0.24}_{-0.36}$\\ 
      & $0.24^{+0.48}_{-0.24}$ & $0.20^{+0.53}_{-0.19}$ & $0.39^{+0.39}_{-0.37}$ & $0.16^{+0.28}_{-0.14}$\\ 
      & $0.28^{+0.50}_{-0.27}$ & $0.16^{+0.68}_{-0.14}$ & $0.07^{+0.11}_{-0.06}$ & $0.10^{+0.26}_{-0.10}$\\ 
      &  &  &  & $0.07^{+0.11}_{-0.05}$\\ 
\hline
 $U_0$  & $-2.08^{+1.63}_{-1.51}$ & $-3.06^{+0.54}_{-0.44}$ & $-2.82^{+1.21}_{-0.68}$ & $-3.00^{+0.44}_{-0.43}$\\ 
        &$-3.03^{+1.16}_{-0.97}$ & $-2.46^{+1.70}_{-1.22}$ & $-2.98^{+1.35}_{-0.54}$ & $-2.25^{+1.39}_{-1.13}$\\ 
        & $-1.72^{+1.72}_{-1.44}$  & $-2.89^{+1.49}_{-1.02}$ & $-1.78^{+1.23}_{-1.55}$ & $-2.82^{+1.10}_{-0.65}$\\ 
        &  &  &  & $-2.27^{+1.41}_{-1.21}$\\ 
\hline
 $n_0$ [$\log$ cm$^{-3}$]  & $1.14^{+1.13}_{-1.09}$ & $1.01^{+0.91}_{-0.91}$ & $1.21^{+1.12}_{-0.69}$ & $1.44^{+0.95}_{-0.82}$\\ 
                           &$2.20^{+1.29}_{-1.15}$  & $1.53^{+1.96}_{-1.35}$ & $1.49^{+1.41}_{-0.92}$ & $1.85^{+1.26}_{-1.24}$\\ 
                           & $0.72^{+1.22}_{-0.71}$ & $2.34^{+1.18}_{-1.70}$ & $2.17^{+1.01}_{-0.64}$ & $1.95^{+1.38}_{-1.13}$\\ 
                           &  &  &  & $2.02^{+1.04}_{-0.75}$\\ 
\hline
 cut\tablefootmark{a}  & $0.42^{+0.45}_{-0.35}$ & $0.65^{+0.33}_{-0.39}$ & $0.36^{+0.38}_{-0.35}$ & $0.59^{+0.28}_{-0.24}$\\ 
                       &$0.59^{+0.52}_{-0.45}$ & $0.25^{+0.44}_{-0.23}$ & $0.72^{+0.38}_{-0.31}$ & $0.52^{+0.58}_{-0.40}$\\ 
                       & $0.93^{+0.20}_{-0.51}$  & $0.76^{+0.35}_{-0.47}$ & $1.32^{+0.51}_{-0.19}$ & $1.01^{+0.66}_{-0.65}$\\ 
                       &  &  &  & $1.56^{+0.83}_{-0.52}$\\ 

\hline\hline
 $Z$ [$\log$ Z$_\odot$]  & $-1.45^{+0.10}_{-0.08}$ & $-1.46^{+0.10}_{-0.09}$ & $-1.44^{+0.07}_{-0.07}$ & $-1.41^{+0.07}_{-0.07}$ & $-1.46$ (\citetalias{Pequignot2008a}; O/H) \\ 
 $Z_{\rm dust}$\tablefootmark{b}  & $-0.17^{+0.66}_{-0.46}$ & $-0.04^{+0.57}_{-0.50}$ & $0.10^{+0.45}_{-0.41}$ & $0.34^{+0.27}_{-0.52}$\\ 
 $L_{\rm X}$ [$\log$ erg\,s$^{-1}$]  & $39.40^{+1.14}_{-0.33}$ & $39.99^{+0.59}_{-0.74}$ & $39.87^{+0.35}_{-0.67}$ & $39.96^{+0.36}_{-0.60}$\\ 
 $T_{\rm X}$ [$\log$ K]  & $6.08^{+0.66}_{-0.58}$ & $6.16^{+0.47}_{-0.64}$ & $5.92^{+0.47}_{-0.37}$ & $6.04^{+0.38}_{-0.50}$\\ 
 $Q({\rm H})$ [$\log$ s$^{-1}$]  & $52.24^{+0.42}_{-0.24}$ & $52.28^{+0.39}_{-0.25}$ & $52.21^{+0.21}_{-0.21}$ & $52.13^{+0.26}_{-0.16}$ & $52.1$ (\citetalias{Pequignot2008a}; $18.2$\,Mpc) \\ 
 $f_{\rm esc}(1-\infty)$ [\%]  & $56.07^{+37.20}_{-34.01}$ & $59.20^{+35.85}_{-19.19}$ & $54.17^{+16.90}_{-21.74}$ & $47.56^{+29.14}_{-14.80}$& $\approx63$, $[60,70]$ (\citetalias{Pequignot2008a}) \\ 
 - $f_{\rm esc}(1-1.8)$ [\%]  & $48.12^{+35.72}_{-31.57}$ & $51.89^{+30.48}_{-26.93}$ & $47.02^{+20.89}_{-27.18}$ & $41.73^{+30.10}_{-22.32}$\\ 
 - $f_{\rm esc}(1.8-4)$ [\%]  & $68.53^{+31.07}_{-42.82}$ & $78.05^{+20.58}_{-18.82}$ & $71.55^{+23.42}_{-18.14}$ & $71.26^{+21.88}_{-18.39}$\\ 
 - $f_{\rm esc}(4-20)$ [\%]  & $20.38^{+35.63}_{-16.27}$ & $26.31^{+33.29}_{-15.58}$ & $24.14^{+33.70}_{-8.43}$ & $22.07^{+14.30}_{-7.92}$\\ 
 - $f_{\rm esc}(20-\infty)$ [\%]  & $79.99^{+19.09}_{-17.88}$ & $87.37^{+11.78}_{-8.55}$ & $81.34^{+7.60}_{-11.75}$ & $78.07^{+9.72}_{-9.59}$\\ 
 $M_{\rm H+}$ [$\log$ M$_\odot$]  & $6.46^{+0.91}_{-0.92}$ & $6.47^{+1.05}_{-0.94}$ & $6.42^{+0.12}_{-0.91}$ & $6.29^{+0.24}_{-0.79}$ & $6.5$ (\citetalias{Pequignot2008a}; $18.2$\,Mpc) \\ 
 $M_{\rm H0}$ [$\log$ M$_\odot$]  & $4.25^{+1.32}_{-1.41}$ & $3.82^{+1.10}_{-1.19}$ & $7.28^{+0.56}_{-0.78}$ & $7.23^{+0.60}_{-0.95}$& $\approx[6.9-7.3]$ (\citetalias{Lebouteiller2017a}) \\ 
 $M_{\rm H2}$ [$\log$ M$_\odot$]  & $-5.67^{+2.78}_{-4.38}$ & $-7.22^{+3.25}_{-3.03}$ & $3.82^{+2.51}_{-2.50}$ & $5.39^{+1.67}_{-4.17}$\\ 
 - $M_{\rm H2,C+}$ [$\log$ M$_\odot$]  & $-5.81^{+2.79}_{-5.25}$ & $-7.40^{+3.10}_{-4.04}$ & $3.82^{+2.41}_{-2.50}$ & $5.38^{+1.50}_{-4.16}$\\ 
 - $M_{\rm H2,C}$ [$\log$ M$_\odot$]  & $-9.23^{+3.15}_{-5.07}$ & $-10.41^{+4.04}_{-3.68}$ & $0.38^{+3.24}_{-2.98}$ & $2.28^{+3.05}_{-4.90}$\\ 
 - $M_{\rm H2,CO}$ [$\log$ M$_\odot$]  & $-19.04^{+3.31}_{-11.29}$ & $-22.34^{+6.66}_{-7.35}$ & $-3.32^{+5.17}_{-4.95}$ & $-0.19^{+4.37}_{-7.74}$\\ 
 $M_{\rm dust}$ [$\log$ M$_\odot$]  & $1.16^{+1.09}_{-1.08}$ & $1.16^{+1.20}_{-1.09}$ & $2.23^{+0.52}_{-0.63}$ & $2.45^{+0.50}_{-0.91}$ &  $2.2$ (\citetalias{Lebouteiller2017a}),  $2.86\pm0.24$\tablefootmark{c} \\ 
 $f_{\rm CII,H+}$  & $0.95^{+0.05}_{-0.03}$ & $1.00^{+0.00}_{-0.05}$ & $0.06^{+0.16}_{-0.05}$ & $0.04^{+0.15}_{-0.04}$\\ 
 $f_{\rm CII,H}$  & $0.05^{+0.03}_{-0.05}$ & $0.00^{+0.05}_{-0.00}$ & $0.94^{+0.06}_{-0.16}$ & $0.93^{+0.05}_{-0.20}$\\ 
 $f_{\rm CII,H2}$  & $0.00^{+0.00}_{-0.00}$ & $0.00^{+0.00}_{-0.00}$ & $0.00^{+0.03}_{-0.00}$ & $0.01^{+0.13}_{-0.01}$\\ 
       \hline
  \end{tabular} \\
  \tablefoot{Models $\mathcal{M}_{\rm I,opt}$ and $\mathcal{M}_{\rm I,opt+IR}$ considers ionized gas observations assuming \citetalias{Pequignot2008a}'s modeling strategy without and with infrared lines respectively. $\mathcal{M}_{\rm NA}$ adds neutral atomic gas observations ([C\2], [O\1], [Si\2], and [Fe\2]). $\mathcal{M}_{\rm NM}$ adds the CO marginal detection of \cite{Zhou2021a} and upper limits on near-IR and mid-IR H$_2$. The escape fractions are given for several energy ranges in Rydberg. The reported values correspond to the median of the posterior distribution together with the confidence interval. Combined quantities (e.g., $M_{\rm H2}$) are calculated for each draw and the final median value can therefore deviate from the combination of median values.  \\
\tablefoottext{a}{1: ionization front (H$^+$--H$^0$ transition), 2: PDR front (H$^0$--H$_2$ transition), 3: C$^0$--CO transition.}
\tablefoottext{b}{Dust-to-gas mass ratio. The value indicated is the deviation in dex from the median relationship in \cite{Galliano2021a}.}
\tablefoottext{c}{Value from \cite{RemyRuyer2015a} assuming graphite composition.}
}
\end{table*}

We find good agreements for the number of ionizing photons $Q_0$ and the ionized gas mass $M({\rm H}^+)$ compared to \citetalias{Pequignot2008a}. The escape fraction is calculated using the ionizing continuum output from Cloudy and is also in good agreement, although it is by far the parameter with the largest relative uncertainty. In \citetalias{Pequignot2008a}, the fraction of photons absorbed within the H\2\ regions, $Q_{\rm abs}/Q$, was estimated a priori to be $0.37\pm0.03$ and at least $0.30$ based on the observed vs.\ expected H$\alpha$ emission and assuming that all ionizing photons are eventually absorbed in the galaxy. This leads to an expected escape fraction $f_{\rm esc}\approx63$\% and at most $70$\% for the H\2\ regions and the best models in \citetalias{Pequignot2008a} agree with these values. The escaping photons are responsible for the diffuse H$\alpha$ halo emission in the galaxy. 

While the dust-to-gas mass ratio has little impact on the predicted emission lines in the H\2\ region, it is interesting that the model is still able to recover the expected value. The value we quote for $Z_{\rm dust}$ corresponds to the scaling relative to the median relationship of $Z_{\rm dust}$ vs.\ $Z_{\rm gas}$ in \cite{Galliano2021a}. \izw\ lies only slightly above that relationship in that study while the SFGX models are allowed to reach up to five times below or above (Table\,\ref{tab:params}). 

The X-ray luminosity we infer is difficult to compare to \citetalias{Pequignot2008a}. In \citetalias{Pequignot2008a} a stellar radiation field including an ad-hoc EUV component was used and an X-ray blackbody was added to explain the neutral gas emission lines. The stellar model BPASS \citep{Eldridge2017a} does not have enough energy in the EUV range to explain the He\2\ emission so there is a need to include the X-ray multicolor blackbody model from the SFGX database. Relatively, He\2\ is the main line emission in sector \#1 (Fig.\,\ref{fig:fracplotp08}). The X-ray luminosity we derive, $39.40^{+1.14}_{-0.33}$ ($\log$\,erg\,s$^{-1}$), agrees with the value measured by \cite{Thuan2004b}, $\approx39.48$, with \textit{Chandra} in the $0.5-10$\,keV range. It is somewhat lower than the XMM-\textit{Newton} value in \cite{Kaaret2013a} of $40.15$ ($\log$\,erg\,s$^{-1}$) but still within uncertainties. 

The fact that we obtain globally similar results to \citetalias{Pequignot2008a} is remarkable considering that we use different pressure profiles and a general prescription for the abundance patterns vs.\ metallicity. However, the automatic approach obviously cannot pretend to reach the same level of fine-tuning as specific studies.

Not surprisingly, the predicted [C\2] in model $\mathcal{M}_{\rm I,opt}$, $1.6^{+2.2}_{-1.2}$, is much lower than the observed value ($76^{+20}_{-20}$) because [C\2] is expected to emit deeper while we restricted ourselves to models reaching the ionization front. This is an indication that the fraction of observed [C\2] in the ionized gas is small, which is confirmed by models accounting for the neutral gas (Sect.\,\ref{sec:newmodels}).

\subsection{New topological models}\label{sec:newmodels}

We now attempt to update the results in \citetalias{Lebouteiller2017a} who converted the \citetalias{Pequignot2008a} modeling approach to Cloudy. The use of the latter was motivated by the need to account for dust in order to compare the relative importance of the photoelectric effect heating and X-ray photoionization in the neutral gas. To this effect \citetalias{Lebouteiller2017a} introduced several neutral atomic and molecular lines in the model but parameters were adjusted manually and there was no convergence test. Apart from an exploratory model with four sectors including a deep sector with constant pressure to investigate the molecular gas, the sector configuration in \citetalias{Lebouteiller2017a} was precisely the same as in \citetalias{Pequignot2008a}.

For all the models described below, in addition to new IR constraints probing the ionized or neutral gas, we also relax some of the constraints on the topology, notably by letting the inner radius free for each sector. Furthermore, while we fixed the number of components in model $\mathcal{M_{\rm I,opt}}$ to reproduce the results of \citetalias{Pequignot2008a} in Section\,\ref{sec:IGP08}, we now test different number of components. All model predictions are provided for convenience in the appendix (Figs\,\ref{fig:boxplot_IG}, \ref{fig:boxplot_NG}, \ref{fig:boxplot_NGH2}; Table\,\ref{tab:i18_results_plawnormal}).

\subsubsection{Ionized gas: Model $\mathcal{M}_{\rm I,opt+IR}$}\label{sec:newmodelsIG}

\begin{figure*}
   \centering
\includegraphics[width=0.75\textwidth]{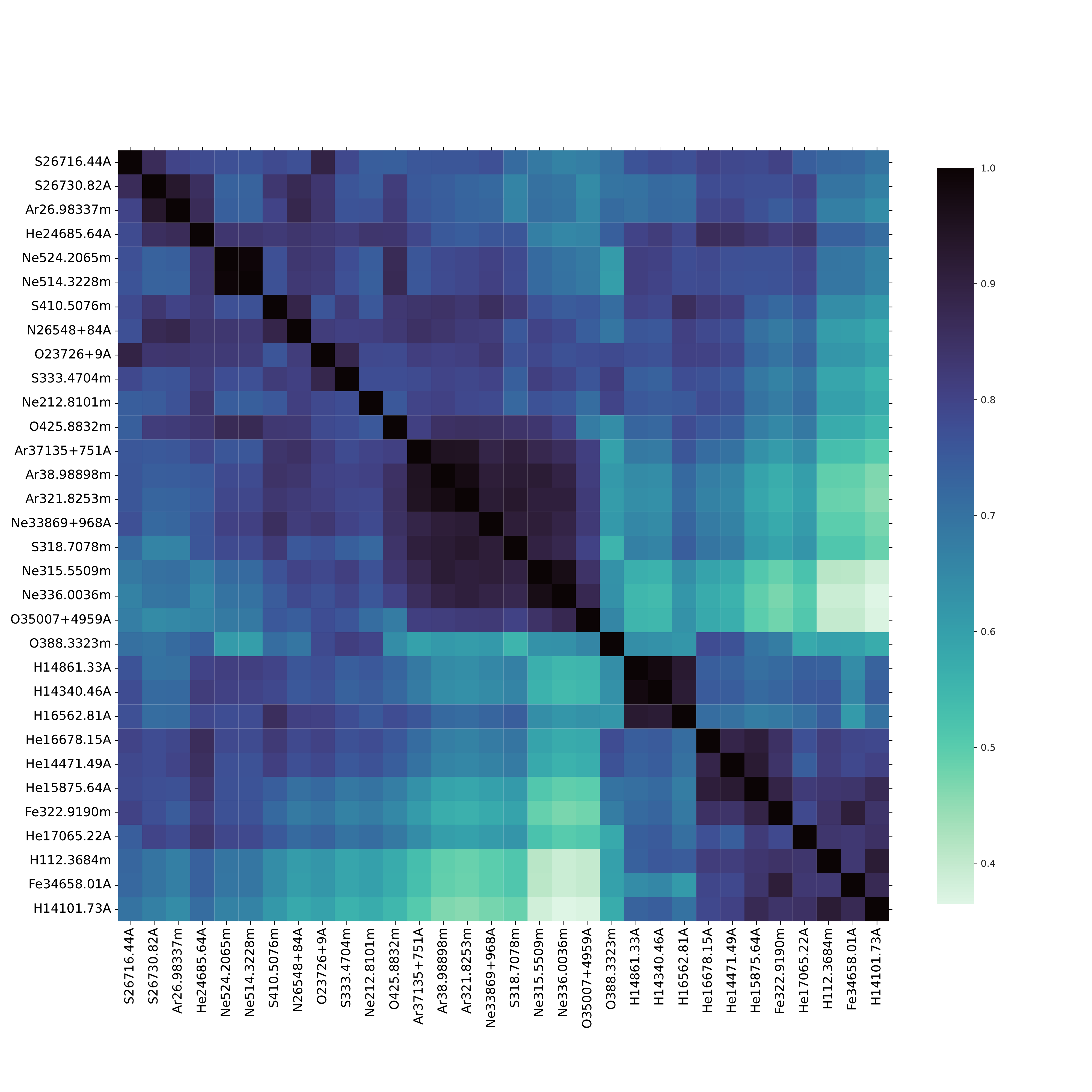}
\caption{Map of the correlation coefficients between predicted values for the ionized gas model $\mathcal{M}_{\rm I,opt}$ of \izw\ (see Figure\,\ref{fig:corr_compa} for the plot description). We can see for instance that [O\3] $88$\mic\ does not correlate with the same parameters compared to other lines. }
         \label{fig:corr_compa_i18IG}
       \end{figure*}

For model $\mathcal{M_{\rm I,opt+IR}}$ we consider the ionized gas again but this time considering IR lines in addition to optical ones. Results are shown in Table\,\ref{tab:lines} and Figure\,\ref{fig:matchplotp08}. Some newly added observations arise from the same physical conditions as the optical lines and thus provide further constraints to the model, but some other observations provide new constraints, such as [O\3] $88$\mic\ for the rather diffuse highly-ionized gas Figure\,\ref{fig:corr_compa_i18IG} confirms that the optical and IR [O\3] lines do not trace the same physical conditions.

The model comparison metrics indicate that the most likely number of sectors is one but this does not account for the fact that the prior odds for a model with a single sector is expected to be low (Sect.\,\ref{sec:model_comparison}). It turns out, however, that the predictions are in good agreement for models with more than one sector (Fig.\,\ref{fig:boxplot_IG}). 
We also tested various combinations with the ionization parameter $U$ and gas density $n$ as power-law distributions and the cut parameter as either a power-law or normal distribution (model $\mathcal{M}_{\rm I,power-law}$). Results are shown in Figure\,\ref{fig:boxplot_IG} and Table\,\ref{tab:i18_results_plawnormal}. Globally, even though those latter models have a somewhat lower probability, the predictions for all secondary parameters are in agreement with the values inferred with a small, discrete, number of components. 

Using the model $\mathcal{M_{\rm I,opt+IR}}$ as an illustration (i.e., with three sectors), the X-ray luminosity is now slightly larger compared to $\mathcal{M_{\rm I,opt}}$.
The predicted value, $L_{\rm X}=39.99^{+0.59}_{-0.74}$ ($\log$\,erg\,s$^{-1}$), is now in better agreement with the XMM-\textit{Newton} value in \cite{Kaaret2013a} of $40.15$\,erg\,s$^{-1}$.

Other notable differences with model $\mathcal{M_{\rm I,opt}}$ include the mass of neutral atomic hydrogen $M$(H$^0$) and the dust mass $M_{\rm dust}$ which become somewhat lower. However, the values have large uncertainties reflecting the fact that model $\mathcal{M_{\rm I,opt+IR}}$ holds little information on the neutral gas.

\subsubsection{Neutral atomic gas: Model $\mathcal{M}_{\rm NA}$}\label{sec:newmodelsNG}

For the neutral atomic gas model $\mathcal{M_{\rm NA}}$, we include [C\2], [O\1], [Si\2], and [Fe\2] as additional constraints. We also include the total IR emission (which, according to \citetalias{Lebouteiller2017a}'s models originates from dust emission in the H\2\ region, a moderate contribution from dust in the H\1\ region, and a small but significant contribution from free-free emission). We allow the cloud depth to reach slightly beyond the H/H$_2$ transition in order to fully consider the neutral atomic layer.

The model comparison metrics (Sect.\,\ref{sec:model_comparison}) indicate that model with a single sector is highly unlikely (Fig.\,\ref{fig:boxplot_NG}). Results with more than one sector are in good agreement. As for the ionized gas model ($\mathcal{M}_{\rm I,opt+IR}$; Sect.\,\ref{sec:newmodelsIG}), we also tested various combinations with $U$ and $n$ as power-law distributions and the cut parameter as a broken power-law (model $\mathcal{M}_{\rm NA,power-law}$). The use of a broken power-law instead of a single power-law is motivated by the fact that the extrapolation of the distribution of depth values for matter-bounded regions does not have to hold for the radiation-bounded regions. Not surprisingly, we find a pivot cut around one, that is at the ionization front, with significantly different slopes below and above (Table\,\ref{tab:i18_results_plawnormal}). The corresponding model $\mathcal{M}_{\rm NA,power-law}$ has a relatively lower probability as compared to the models with a small, discrete, number of components, indicating that more complex distributions may be needed. Nevertheless, the predictions are globally in good agreement with model $\mathcal{M_{\rm NA}}$ (Figure\,\ref{fig:boxplot_IG}).

Parameter results for model $\mathcal{M_{\rm NA}}$ are shown in Tables\,\ref{tab:lines} and\ \ref{tab:i18_results}. Most global parameters (e.g., metallicity and mass) are unchanged when compared to model $\mathcal{M_{\rm I,opt+IR}}$ with only ionized gas observations. However, the mixing weights and depth of the sectors are significantly different compared to the ionized gas model, which explains the change in the escape fraction values, which strongly depends on the topology (Sect.\,\ref{sec:ncompsinfluence}). The escape fraction somewhat decreased compared to $\mathcal{M_{\rm I,opt+IR}}$, and, while the difference is relatively small, this could imply that some of the neutral gas emission arises on somewhat larger spatial scales than the H\2\ region.

The mass of atomic hydrogen is now much larger compared to $\mathcal{M_{\rm I,opt+IR}}$ and in agreement with expectations (Table\,\ref{tab:i18_results}). We find a mass of molecular gas of about $\approx10^{3.8}$\,M$_\odot$. \citetalias{Lebouteiller2017a} had evaluated the molecular gas mass assuming the same topology as $\mathcal{M_{\rm I,opt}}$, that is with the same covering factors as in \citetalias{Pequignot2008a}. This effectively corresponded to a rather diffuse molecular medium as opposed to clumps that were considered in a specific four-sector model. The H$_2$ mass distributed in the large-scale radiation-bounded sector in \citetalias{Lebouteiller2017a} was extremely low, on the order of $1$\,M$_\odot$. We find a much larger value in our model $\mathcal{M_{\rm NA}}$ because the topology is different (sector mixing weight and different inner radii) and the H$_2$ mass is furthermore controlled by the somewhat arbitrary maximum depth criterion (which we fix to slightly beyond the H/H$_2$ transition for $\mathcal{M_{\rm NA}}$). This rather small H$_2$ mass of $\approx10^{3.8}$\,M$_\odot$ exists in the C$^+ $ zone and is therefore CO-dark with a CO-dark H$_2$ fraction of $\approx100$\%\ (but the overall C$^+$ emission is still dominated by the neutral atomic gas). The dust mass is also close to the observed value and the dust-to-gas mass ratio is now slightly larger compared to $\mathcal{M_{\rm I,opt+IR}}$ and in better agreement with \cite{Galliano2021a}, which is not surprising since the dust mass mostly comes from the neutral phase. However, by comparing models $\mathcal{M}_{\rm I,opt}$ (or $\mathcal{M}_{\rm I,opt+IR}$) and $\mathcal{M}_{\rm NA}$ we find that $\approx2/3$ of the TIR emission comes from the ionized phase, which is in agreement with \citetalias{Lebouteiller2017a} who found that TIR can be decomposed as $\approx20\%$ free-free emission, $\approx15\%$ dust in the H\1\ region, and $\approx65\%$ dust in the H\2\ region.

The [C\2] emission is much larger compared to $\mathcal{M_{\rm I,opt+IR}}$ (Table\,\ref{tab:lines}) confirming that most of the predicted [C\2] emission arises in the neutral atomic gas. We find a fraction of [C\2] arising in the neutral atomic gas, $f_{\rm CII,H}$ of about $94\%$. The deep narrow sector \#3 is responsible for most of the [C\2], [O\1], [Si\2], [Fe\2], and H$_2$ line emission and also dominates the [Ar\2] and [Ne\2] emission (Fig.\,\ref{fig:fracplotp08}; Table\,\ref{tab:i18_results}). We note that the relative fluxes predicted for [C\2], [O\1], and [Si\2] differ somewhat compared to observations, which is likely due to the density profile assumed in the models. The model predicts an essentially null CO emission.

\subsubsection{Neutral molecular gas: Model $\mathcal{M}_{\rm NM}$}

We now add molecular gas observations as constraints for model $\mathcal{M_{\rm NM}}$. For CO, we use the recently claimed CO(2-1) detection of \cite{Zhou2021a}, corresponding to a CO(1-0) luminosity of $0.18$\,L$_\odot$ ($\approx3.5\sigma$ detection). Even though the signal is quite uncertain, the new observation is much deeper than previous studies and provides at the very least a useful upper limit. We also include several upper limits for near-IR and mid-IR H$_2$ lines (see \citetalias{Lebouteiller2017a}).

The model comparison metrics (Sect.\,\ref{sec:model_comparison}) indicate that the model with a single sector is highly unlikely (Fig.\,\ref{fig:boxplot_NGH2}). As for model $\mathcal{M_{\rm NA}}$, we also tested various combinations with $U$ and $n$ as power-law distributions and the cut parameter as a broken power-law. The corresponding model $\mathcal{M}_{\rm NM,power-law}$ reproduces better the observed CO but has a much lower probability as compared to the models with a small, discrete, number of components, indicating that more complex distributions of the cut parameter throughout the ionized, neutral atomic, and neutral molecular gas may be needed (Figure\,\ref{fig:boxplot_IG}). 
In the following we discuss the model $\mathcal{M}_{\rm NM}$ composed of 4 sectors. 

In Table\,\ref{tab:i18_results} we can see that the narrowest sector now reaches somewhat deeper, but with a most likely value not quite reaching the H$^0$-H$_2$ transition. Otherwise the other primary parameters are mostly unchanged compared to $\mathcal{M_{\rm NA}}$.

\begin{figure}
   \centering
\includegraphics[width=0.49\textwidth,clip]{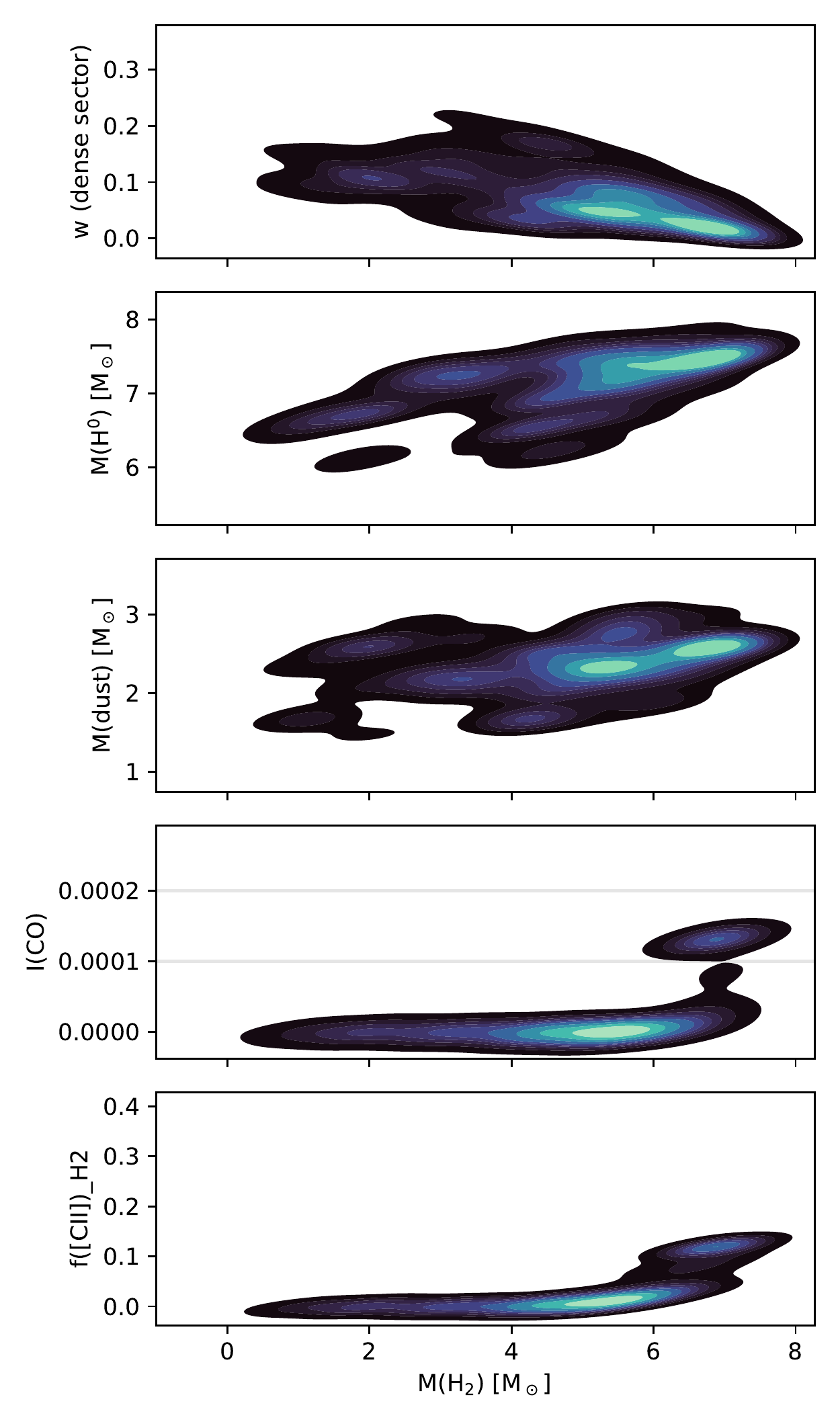}
\caption{Properties of the molecular sector in model $\mathcal{M}_{\rm NM}$. From top to bottom we show as a function of the total predicted H$_2$ mass, the mixing weight of the molecular sector, the mass of H$^0$, of dust, the CO(1-0) intensity (scaled to $4.56\times10^{36}$\,erg\,s$^{-1}$; with the horizontal lines showing the observed range in \citealt{Zhou2021a}), and the fraction of [C\2] originating from the molecular sector. }
         \label{fig:kdeh2a}
       \end{figure}

In Figure\,\ref{fig:kdeh2a} we see that the most likely values for $I({\rm CO})$ are much lower than the observed value even within uncertainties. CO being the only observation constraining such depths and the detection level being so low, the model is mostly constrained by other lines, in particular those arising mostly in the neutral atomic gas as well as TIR. In other words, the CO detection is in fact unlikely within the model topological assumptions in model $\mathcal{M}_{\rm NM}$. There is a solution agreeing with the observed CO intensity, corresponding to an H$_2$ mass $\approx10^7$\,M$_\odot$, but with a relatively low probability of $\lesssim20$\%. This large value coincides with the upper limit of \citetalias{Lebouteiller2017a} based on the observed CO(1-0) upper limit of $<8\times10^{-3}$ (normalized to H$\beta=1000$) from \cite{Leroy2007a}.

\begin{figure}
   \centering
\includegraphics[width=0.49\textwidth,clip]{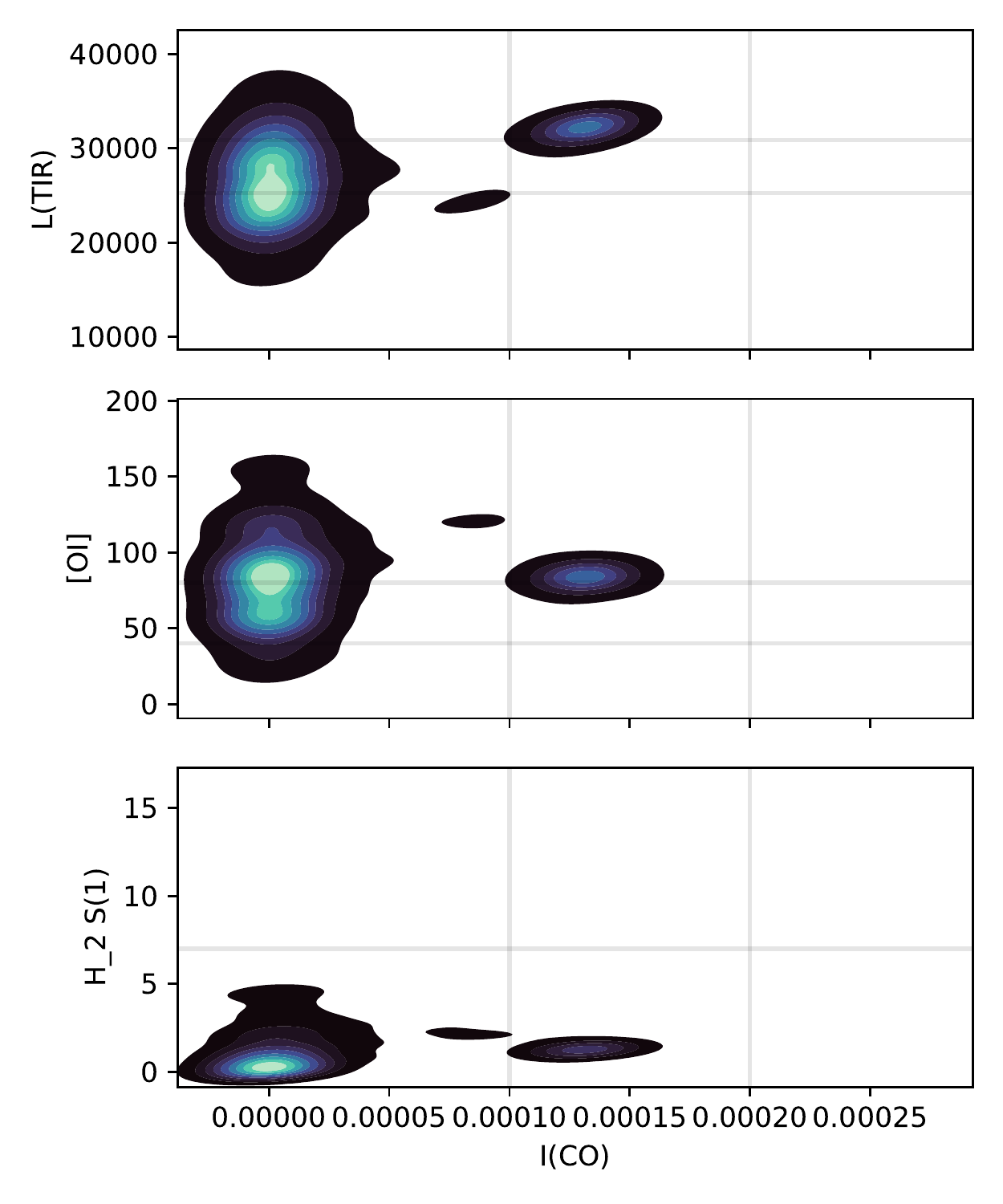}
\caption{Variation of the observed fluxes as a function of $I$({\rm CO}). From top to bottom we show the IR flux, the [O\1] flux, and the H$_2$ S(1) line flux. Fluxes are scaled to $4.56\times10^{36}$\,erg\,s$^{-1}$. Horizontal and vertical lines show the observed uncertainties or the upper limit (for H$_2$).  }
         \label{fig:kdeh2b}
       \end{figure}

When the observed CO flux is well reproduced, the H$^0$ mass becomes somewhat too large, being significantly above $10^7$\,M$_\odot$ (Figure\,\ref{fig:kdeh2a}) and the total IR emission starts deviating from the observed value (Figure\,\ref{fig:kdeh2b}). Adding CO tends to make a deeper narrower sector as the CO emission is better and better reproduced (though these models are unlikely). The mixing weight of the deep sector ought to be lower otherwise it would produce too much [O\1] and H$_2$ in particular. Even with four sectors (model $\mathcal{M}_{\rm NM}$), the CO emission always come from the same sector as [C\2] and [O\1]. In other words, if we need to account for CO then we need a deep narrow sector and this sector produces a significant/dominant emission of [C\2] and [O\1]. This implies that the [C\2] and [O\1] emission should be probing a mostly neutral atomic medium associated to the CO clumps if they do exist, since the maximum fraction of [C\2] associated with H$_2$ is $\approx15\%$ (Fig.\,\ref{fig:kdeh2a}).

       A more stringent observational constraint would be useful either to confirm that the CO emission is in fact much lower or to constrain a better topological model. Nevertheless, the predicted H$_2$ mass is about $10^{5.4}$\,M$_\odot$ and anticorrelates with the covering factor of the dense sector (top panel of Fig.\,\ref{fig:kdeh2a}). As for model $\mathcal{M}_{\rm NA}$, the small H$_2$ mass is better traced by C$^+$ than C$^0$ or CO (Table\,\ref{tab:i18_results}), and the CO-dark H$_2$ gas fraction is therefore $\approx100$\%. However, most of the [C\2] emission is not associated with H$_2$ but with H$^0$, in agreement with \citetalias{Lebouteiller2017a}.

\begin{figure}
   \centering
\includegraphics[width=0.49\textwidth,clip]{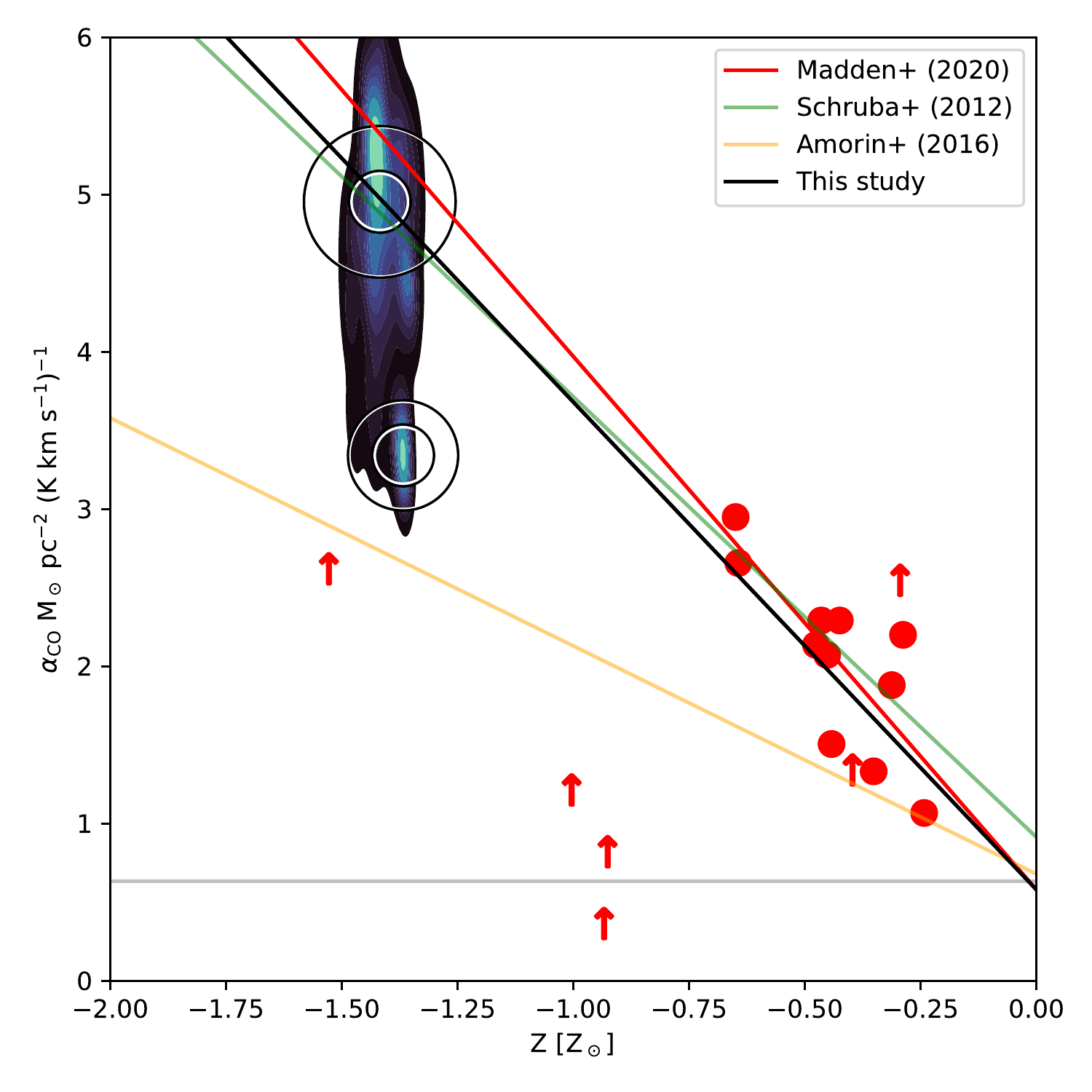}
\caption{Variation of $\alpha_{\rm CO}$ as a function of metallicity compared to \cite{Schruba2012a,Sandstrom2013a,Amorin2016a,Madden2020a}. The red data points and upper limits are from \cite{Madden2020a}. We also show on top the most likely solution (empty black circle with the gray circle indicating the standard deviation) with $M({\rm H}_2)=10^{5.4}$\,M$_\odot$ and CO $3500$ times fainter than in \cite{Zhou2021a} and on bottom the secondary, less likely, solution with $10^{7}$\,M$_\odot$ and CO in agreement with \cite{Zhou2021a}. }
         \label{fig:alphaCO}
       \end{figure}

The CO luminosity we predict is $\approx3500$ times lower than \cite{Zhou2021a}, implying $\alpha_{\rm CO}=M({\rm H}_2)/L({\rm CO})=9\times10^4$\,M$_\odot$\,(K\,km\,s$^{-1}$)$^{-1}$. For the metallicity of \izw, together with the reference solar value for $\alpha_{\rm CO}=4.3$\,M$_\odot$\,pc$^{-2}$\,(K\,km\,s$^{-1}$)$^{-1}$ (e.g., \citealt{Bolatto2013a}), this translates into a slope of $y=-3.1$ with $\alpha_{\rm CO}\propto (Z/Z_\odot)^y$. This coefficient is on the high end of metallicity-dependent $\alpha_{\rm CO}$ studies but quite close to the value derived by \cite{Madden2020a} of $-3.39$ and the extrapolation of \cite{Schruba2012a} (Fig.\,\ref{fig:alphaCO}). If we consider instead the less likely solution at $M({\rm H}_2)\sim10^7$\,M$_\odot$ corresponding to the CO observation in \cite{Zhou2021a}, we obtain a $\alpha_{\rm CO}\approx2200$\,M$_\odot$\,pc$^{-2}$\,(K\,km\,s$^{-1}$)$^{-1}$, implying $y\approx-2$. The calculation and discussion of $\alpha_{\rm CO}$ for the full DGS sample will be discussed in Ramambason et al. (in prep.).

\subsubsection{Escape fraction of ionizing photons}

\begin{figure}
   \centering
\includegraphics[width=0.35\textwidth,clip,trim=0 0 0 0]{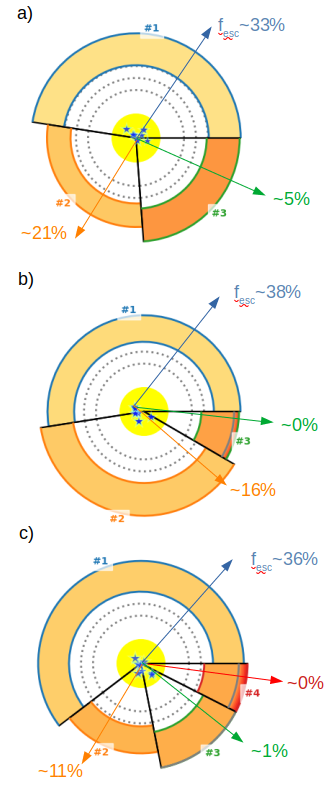}
\caption{Sector configurations for models $\mathcal{M}_{\rm I,opt+IR}$ (a), $\mathcal{M}_{\rm NA}$ (b), and $\mathcal{M}_{\rm NM}$ (c) for \izw. See Figure\,\ref{fig:example_sectors} for the plot description. The arrows indicate the escape fraction ($1-\infty$\,Ryd) for each sector.}
         \label{fig:sectors_i18}
       \end{figure}

In all models, more than half of the global escape fraction is due to sector \#1 which is always matter-bounded, and with the largest mixing weight (Figure\,\ref{fig:sectors_i18}). The observed fluxes correspond to the NW H\2\ region of \izw\ and the global escape fraction therefore also corresponds to that region, while the galaxy escape fraction is probably significantly lower \citepalias{Pequignot2008a}. We refer to \cite{Ramambason2022a} for a more general discussion of the escape fractions in the Dwarf Galaxy Survey. 

\begin{figure}
   \centering
\includegraphics[width=0.5\textwidth,clip]{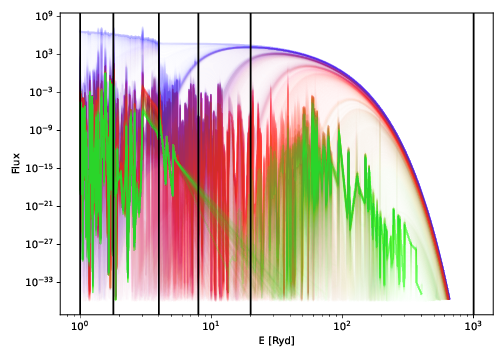}
\includegraphics[width=0.5\textwidth,clip]{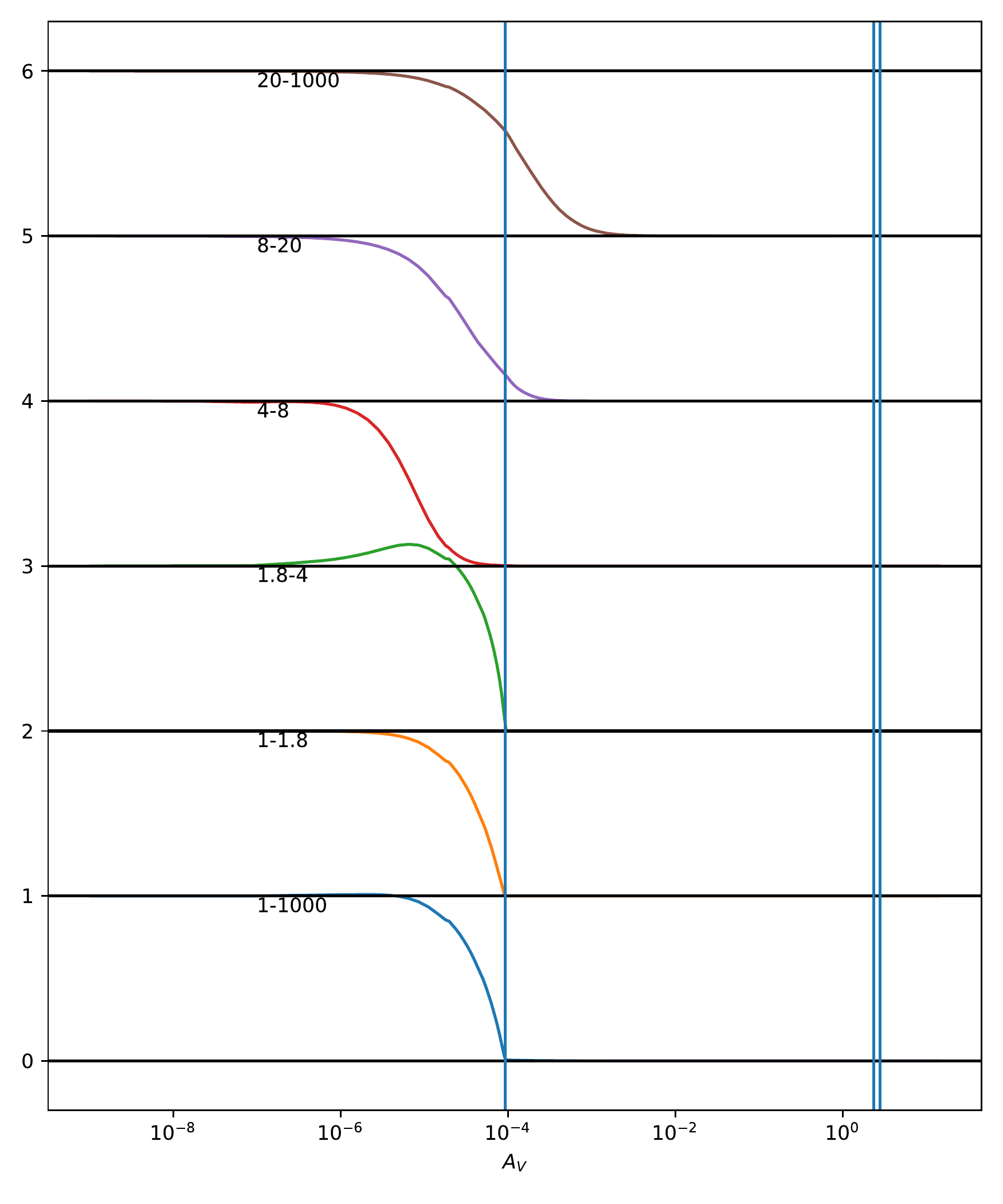}
\caption{Absorption of the ionizing radiation throughout the gas slab. \textit{Top} $-$ Shape of the radiation field spectrum for each zone calculated in Cloudy for the radiation-bounded sector of model $\mathcal{M}_{\rm NM}$, from the illuminated edge of the cloud (blue) to the last zone (green). The vertical lines show the $1$, $1.8$, $4$, $8$, $20$, and $2000$\,Ryd energy values. The spectrum is heavily and rapidly absorbed right above $4$\,Ryd due to He$^{2+}$ absorption. \textit{Bottom} $-$ Evolution of the integrated flux across the cloud depth (normalized to the value at the illuminated edge) for different energy ranges in Rydberg. The vertical lines show the ionization front, PDR, and C/CO transitions from left to right. The radiation field $>20$\,Ryd is able to penetrate in the neutral gas (top panel) but this contributes little to the total escape fraction (bottom panel). The kink for the $1.8-4$\,Ryd range is due to secondary ionization following He recombination.}
         \label{fig:spec_abs}
       \end{figure}

Our models include the presence of an X-ray source associated with the H\2\ region \citepalias{Lebouteiller2017a}. For the first time we are now able to calculate the escape fraction for different photon energy ranges and evaluate the escape fraction of He-ionizing photons, which have received much less attention than H-ionizing photons \citep{Benson2013a}. 
The escape fractions of photons in our models is saved in the energy ranges $1-\infty$, $1-1.8$, $1.8-4$ (He$^0$-ionizing photons), $4-20$ (He$^+$-ionizing photons), and $20-\infty$\,Ryd (Table\,\ref{tab:lines}). In model $\mathcal{M}_{\rm I,opt+IR}$ (Sect.\,\ref{sec:newmodelsIG}) the escape fraction $f_{\rm esc}(1-\infty)\approx59\%$ lies between $f_{\rm esc}(1-1.8)\approx52\%$ and $f_{\rm esc}(1.8-4)\approx78\%$, that is in the range where most of the input radiation field energy lies. The escape fraction $f_{\rm esc}(4-20)$, on the other hand, is remarkably lower ($\approx26\%$), which is due to the absorption by He$^{2+}$ close to the illuminated edge of the cloud (Fig.\,\ref{fig:spec_abs}). Nevertheless, the $f_{\rm esc}(4-20)$ values remains significant due to the fact that the medium gradually becomes more transparent to photons significantly above $4$\,Ryd. As a matter of fact, $f_{\rm esc}(20-\infty)$, which traces the escape of soft to hard X-ray photons is $\approx87\%$ in the model. Even though soft X-ray photons deposit their energy in the neutral gas \citepalias{Lebouteiller2017a}, $f_{\rm esc}(20-\infty)$ remains large even in model $\mathcal{M}_{\rm NA}$ ($\approx80$\%), which is due to the fact that the mixing weight of the radiation-bounded sector is relatively low. Most X-ray photons therefore escape the H\2\ region, but the global X-ray escape fraction reaching the IGM ultimately depends on the distribution of diffuse H\1\ gas. Since low-metallicity star-forming dwarf galaxies are known to harbor bright X-ray sources (e.g., \citealt{BasuZych2016a,Ponnada2019a,Lehmer2020a}), this makes such sources interesting suspects for the IGM heating at $z>10$ \citep{Mirabel2011a}.

\section{Conclusions}

In this study we present a new statistical framework to infer probability density functions (PDFs) of parameters constrained by observations with uncertainties. While the method and associated code, MULTIGRIS, are agnostic to the input grid used, the method was developed in particular for ISM studies with the objective of inferring a topology representing either a discrete number of ISM components around one or several stellar clusters or a continuous distribution of clouds whose physical conditions are parameterized by power- or normal-laws.

The inference is performed using a sequential Monte Carlo method that is well adapted to multipeaked posterior distributions. Primary parameters are sampled using either a nearest neighbor or linear interpolation on the grid values. Built-in priors are available to use predefined configurations flexible enough to various combinations of components.

We illustrate the method with a grid of models calculated with Cloudy. Primary parameters describe the stellar clusters and potential X-ray sources as well as the surrounding ISM distribution. The grid, described in \cite{Ramambason2022a} includes a free parameter defining the cloud depth. While the inference is performed on these primary parameters, the algorithm is able to calculate PDFs of secondary parameters, that is any parameter not considered as an observation and which Cloudy can compute. A large number of random variables and observations can be accommodated. Observations may be gathered in observation sets with associated systematic uncertainties, and upper/lower limits are fully considered. For ISM studies, the extinction may be included as one additional random variable.

We first apply this method to the well-known low-metallicity star-forming dwarf galaxy \izw, whose ISM has been modeled extensively using optical and IR lines in particular. Our objective is to reproduce previous results with a general grid that is not tailored to any specific object and to provide PDFs for the most important parameters. We first consider only ionized gas observations, then add neutral atomic and finally neutral molecular gas observations in order to test the robustness of the diagnostics adding/removing some observations while examine physical parameters related to specific ISM phases. 

We are able to reproduce the metallicity, number of ionizing photons, as well as the mass of H$^+$ and H$^0$, the dust mass and the dust-to-gas mass ratio. We confirm that an X-ray source with luminosity $\approx10^{40}$\,erg\,s$^{-1}$ is necessary to explain the ionized and neutral gas emission, with properties similar to those observed in specific X-ray studies. This implies interesting prospects for constraining the presence, properties, and nature of compact objects from their signatures in the ISM. Such a method is only possible if the topology is well constrained by many observations, with at least some of them themselves influenced by the X-ray source.

We examine the H$_2$ mass, using in particular the mid-IR H$_2$ upper limits as well as the recently claimed detection of CO(1-0) by \cite{Zhou2021a}. With the topology used, we find that the CO(1-0) emission is likely much lower than the observed value. It is possible, however, that CO exists in small dense clumps that we are not able to model. Nevertheless, we derive a most likely mass of H$_2$ of $\approx10^{5.4}$\,M$_\odot$, leading to $\alpha_{\rm CO}=9\times10^{4}$\,M$_\odot$\,pc$^{-2}$\,(K\,km\,s$^{-1}$)$^{-1}$ and to a metallicity coefficient of $y=-3.1$ with $\alpha_{\rm CO}\propto (Z/Z_\odot)^y$, in excellent agreement with \cite{Madden2020a} and \cite{Schruba2012a}. The calculation and discussion of $\alpha_{\rm CO}$ for the full DGS sample will be discussed in Ramambason et al. (in prep.).

The H$_2$ mass is better traced by C$^+$ than C$^0$ or CO and the CO-dark H$_2$ mass is therefore $\approx100$\%. However, the fraction of [C\2] arising from the molecular gas is small and most of the [C\2] emission is associated with H$^0$. Assuming CO clumps exist, the predicted [C\2] and [O\1] emission arise from the same sector as CO, implying that [C\2] and [O\1] lines probe a mostly neutral atomic medium directly associated with the hypothetical CO clumps.

Finally we investigate the escape fraction of ionizing photons and find values around $50-65$\% in agreement with expectations. We decompose the photons by energy range and find that most ($\gtrsim80\%$) soft X-ray photons are able to escape. This has potential implications for the ionization and heating of the IGM as such photons can escape the diffuse H\1\ gas halo.

\begin{acknowledgements}
This work is supported by the FACE Foundation Thomas Jefferson Fund. We would like to thank Frédéric Galliano for insightful and inspiring discussions about Bayesian statistics, and the anonymous referee for constructive feedback and encouraging comments. 
\end{acknowledgements}

%
%

\bibliographystyle{aa} 
\bibliography{/local/home/vleboute/ownCloud/bibtexendum/bibtexendum} 

\begin{appendix}

\section{Other models}\label{sec:othermodels}

In this appendix we show the results for the ionized, neutral-atomic, and neutral-molecular  gas models with a varying number of sectors and with a power- or normal-law distribution for some parameters.

\begin{figure*}
   \centering
\includegraphics[width=0.7\textwidth]{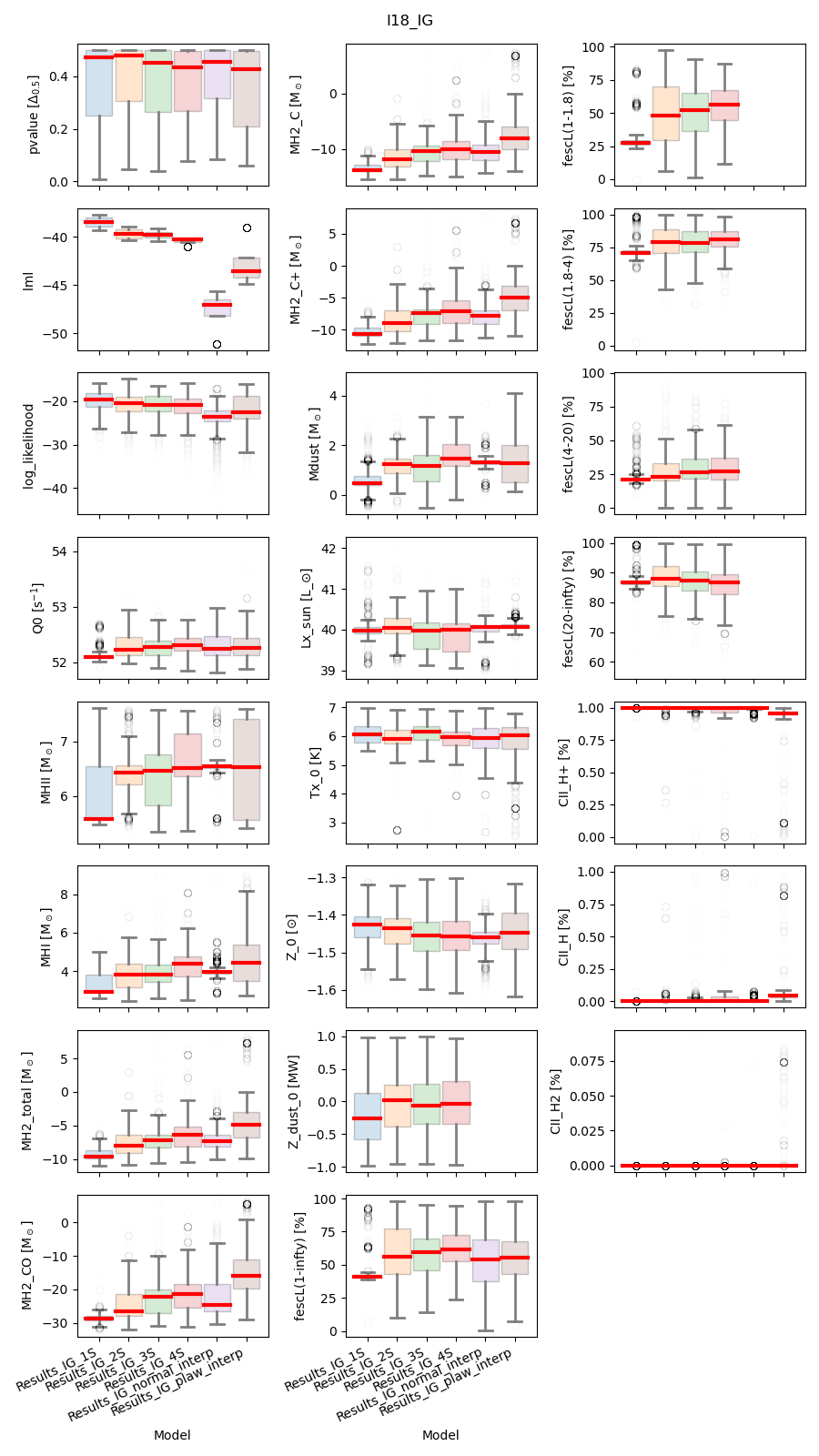}
\caption{Results for model $\mathcal{M}_{\rm I,opt+IR}$. See Figure\,\ref{fig:paramsncomps} for the plot description.} 
         \label{fig:boxplot_IG}
       \end{figure*}
       
\begin{figure*}
   \centering
\includegraphics[width=0.7\textwidth]{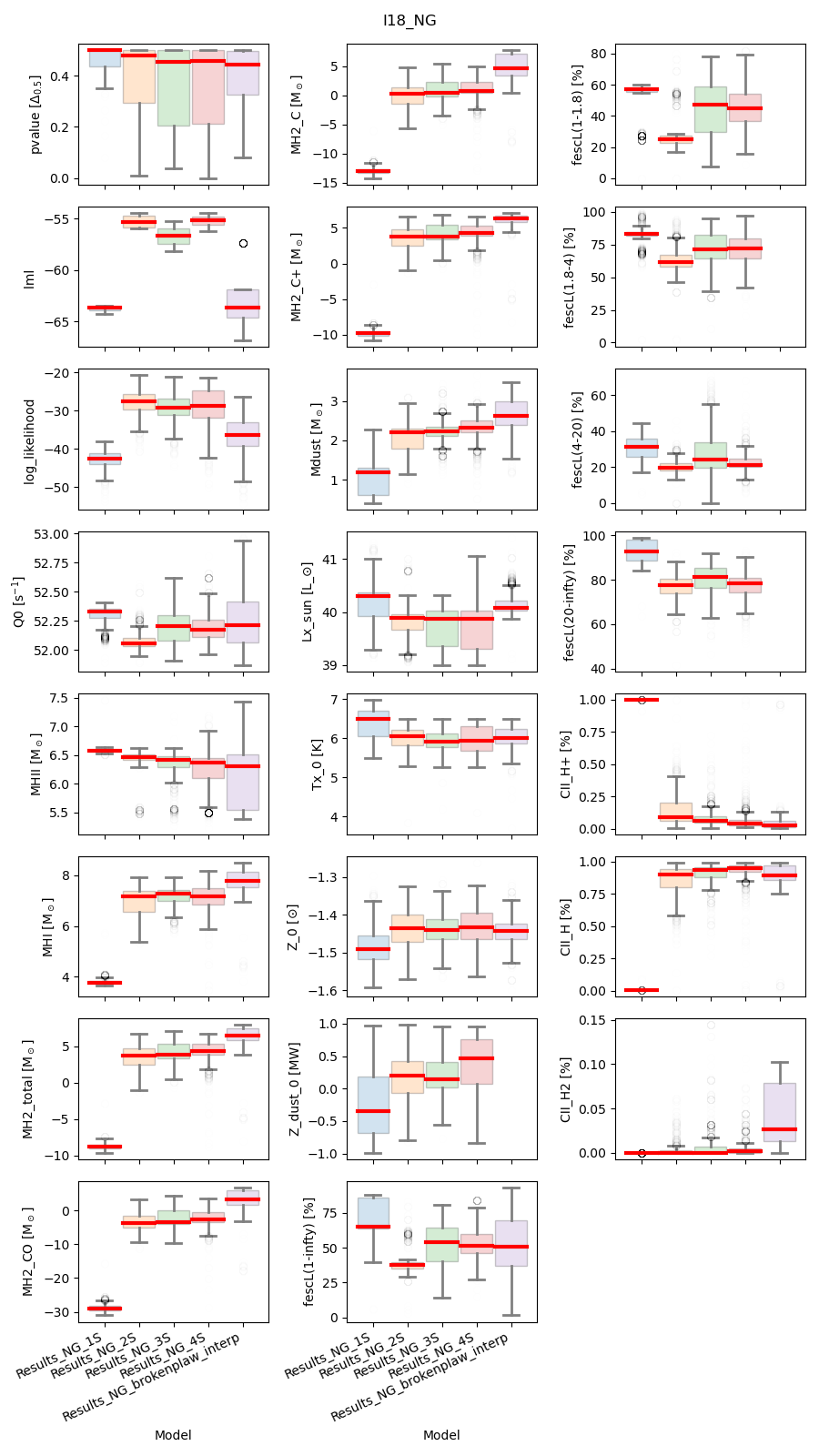}
\caption{Results for model $\mathcal{M}_{\rm NA}$. See Figure\,\ref{fig:paramsncomps} for the plot description. }
         \label{fig:boxplot_NG}
       \end{figure*}

\begin{figure*}
   \centering
\includegraphics[width=0.7\textwidth]{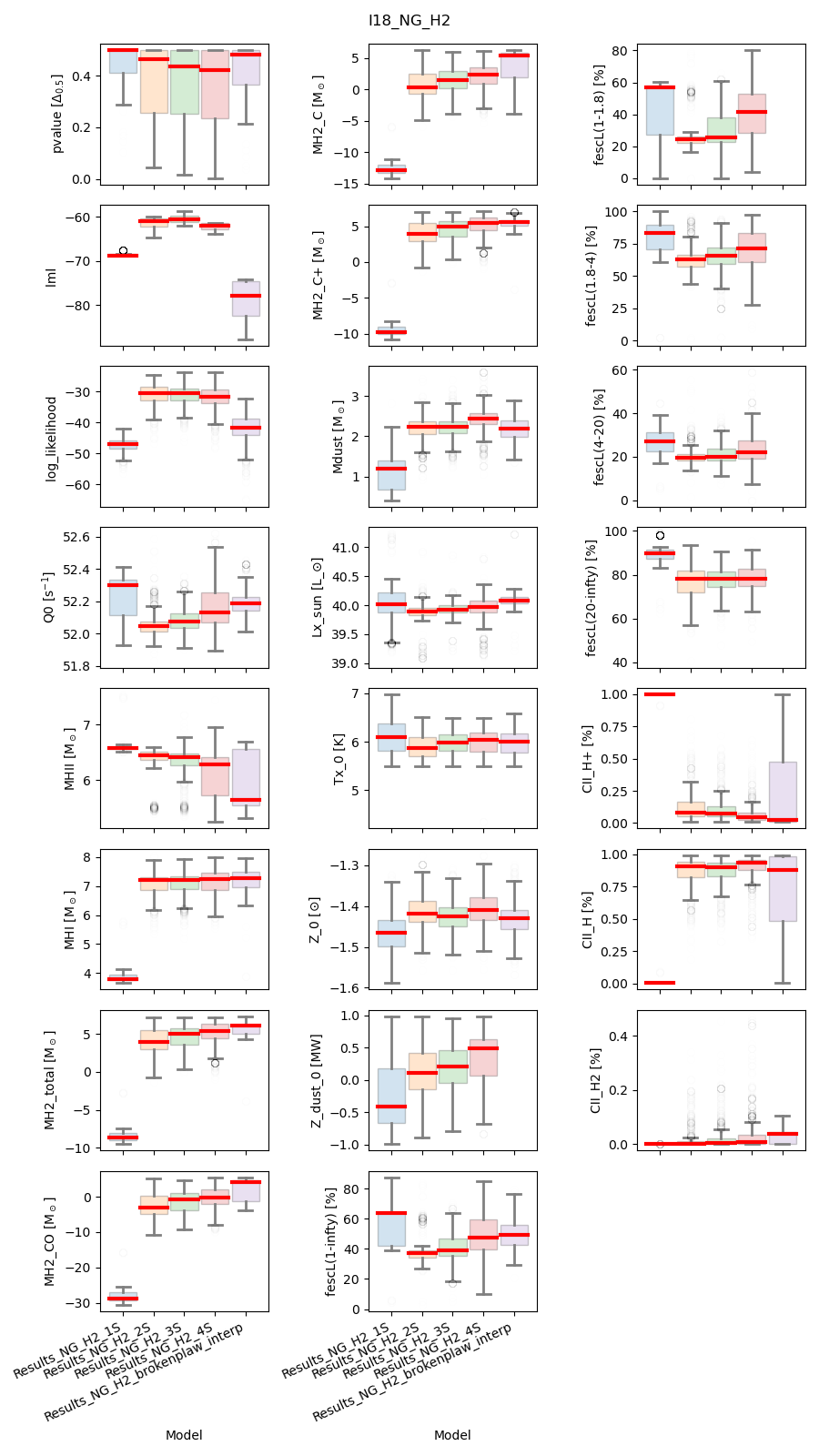}
\caption{Results for model $\mathcal{M}_{\rm NM}$. See Figure\,\ref{fig:paramsncomps} for the plot description. }
         \label{fig:boxplot_NGH2}
       \end{figure*}

\begin{table*}
  \caption{\izw\ model parameters for other considered models.}\label{tab:i18_results_plawnormal}
  \begin{tabular}{l|cccc}
      \hline
    \hline
    Parameter                                & $\mathcal{M_{\rm I,normal}}$ & $\mathcal{M_{\rm I,power-law}}$  &   $\mathcal{M_{\rm NA,power-law}}$  &   $\mathcal{M_{\rm NM,power-law}}$     \\
    \hline\hline
 $U$  & [$-3.34^{+0.58}_{-0.61}$,$-0.62^{+0.51}_{-2.57}$] & [$-3.31^{+0.67}_{-0.43}$,$-1.11^{+0.90}_{-1.52}$] & [$-3.16^{+0.30}_{-0.32}$,$-0.78^{+0.61}_{-0.31}$] & [$-3.08^{+0.57}_{-0.28}$,$-0.60^{+0.47}_{-1.99}$]\\ 
        & $\alpha_U=$$-1.45^{+2.15}_{-2.16}$ & $\alpha_U=$$-1.57^{+1.51}_{-1.34}$ & $\alpha_U=$$-1.63^{+0.34}_{-1.14}$ & $\alpha_U=$$-2.06^{+2.35}_{-0.50}$\\ 
        &  &  &  & \\ 
        &  &  &  & \\ 
\hline
 $n$ [$\log$ cm$^{-3}$]  & [$1.11^{+0.75}_{-1.03}$,$3.30^{+0.61}_{-0.81}$] & [$1.01^{+1.32}_{-0.91}$,$3.13^{+0.79}_{-0.78}$] & [$1.14^{+0.84}_{-0.60}$,$3.69^{+0.26}_{-0.52}$] & [$1.54^{+0.78}_{-0.70}$,$3.24^{+0.49}_{-0.54}$]\\ 
                           & $\alpha_n=$$-3.30^{+2.28}_{-1.67}$ & $\alpha_n=$$-3.48^{+1.74}_{-1.29}$ & $\alpha_n=$$-2.86^{+1.64}_{-1.17}$ & $\alpha_n=$$-2.55^{+1.39}_{-2.04}$\\ 
                           &  &  &  & \\ 
                           &  &  &  & \\ 
\hline
 cut\tablefootmark{a}  & [$0.38^{+0.46}_{-0.28}$,$0.74^{+0.27}_{-0.44}$] & [$0.20^{+0.16}_{-0.19}$,$1.03^{+0.20}_{-0.20}$] & [$0.29^{+0.21}_{-0.29}$,$1.59^{+0.26}_{-0.24}$] & [$0.24^{+0.20}_{-0.17}$,$1.37^{+0.87}_{-0.25}$]\\ 
                       & $\mu_{\rm cut}=$$0.38^{+0.90}_{-0.47}$ & $\alpha_{\rm cut}=$$-0.81^{+1.33}_{-1.29}$ & $\alpha_{\rm cut}=$$-0.31^{+2.44}_{-1.41}$,$-2.11^{+1.35}_{-1.10}$ & $\alpha_{\rm cut}=$$0.76^{+2.50}_{-2.44}$,$-1.92^{+1.91}_{-1.73}$\\ 
                       & $\sigma_{\rm cut}=$$0.77^{+1.75}_{-1.62}$ &  & $x_{\rm cut}\tablefootmark{b}=$$0.94^{+0.31}_{-1.03}$ & $x_{\rm cut}\tablefootmark{b}=$$0.86^{+0.12}_{-0.15}$\\ 
                       &  &  &  & \\ 

\hline\hline
 $Z$ [$\log$ Z$_\odot$]  & $-1.46^{+0.06}_{-0.08}$ & $-1.45^{+0.08}_{-0.12}$ & $-1.44^{+0.06}_{-0.05}$ & $-1.43^{+0.05}_{-0.06}$\\ 
 $L_{\rm X}$ [$\log$ erg\,s$^{-1}$]  & $40.06^{+0.21}_{-0.93}$ & $40.07^{+0.33}_{-0.14}$ & $40.08^{+0.50}_{-0.16}$ & $40.07^{+0.13}_{-0.09}$\\ 
 $T_{\rm X}$ [$\log$ K]  & $5.92^{+0.94}_{-1.25}$ & $6.04^{+0.45}_{-2.54}$ & $6.01^{+0.45}_{-0.32}$ & $5.99^{+0.34}_{-0.48}$\\ 
 $Q({\rm H})$ [$\log$ s$^{-1}$]  & $52.24^{+0.40}_{-0.23}$ & $52.26^{+0.44}_{-0.28}$ & $52.21^{+0.57}_{-0.28}$ & $52.19^{+0.17}_{-0.10}$\\ 
 $f_{\rm esc}(1-\infty)$ [\%]  & $54.04^{+31.41}_{-33.20}$ & $55.33^{+31.06}_{-28.47}$ & $50.83^{+40.23}_{-31.80}$ & $49.16^{+20.31}_{-12.30}$\\ 
 $M_{\rm H+}$ [$\log$ M$_\odot$]  & $6.54^{+0.82}_{-1.01}$ & $6.52^{+1.03}_{-1.04}$ & $6.31^{+0.29}_{-0.82}$ & $5.64^{+0.97}_{-0.15}$\\ 
 $M_{\rm H0}$ [$\log$ M$_\odot$]  & $3.95^{+1.04}_{-1.11}$ & $4.43^{+3.67}_{-1.54}$ & $7.76^{+0.59}_{-0.45}$ & $7.29^{+0.46}_{-0.52}$\\ 
 $M_{\rm H2}$ [$\log$ M$_\odot$]  & $-7.32^{+4.49}_{-1.20}$ & $-4.83^{+12.08}_{-4.66}$ & $6.53^{+1.22}_{-1.13}$ & $6.13^{+0.98}_{-1.36}$\\ 
 - $M_{\rm H2,C+}$ [$\log$ M$_\odot$]  & $-7.87^{+4.91}_{-1.66}$ & $-4.99^{+11.68}_{-5.45}$ & $6.34^{+0.69}_{-0.96}$ & $5.58^{+1.28}_{-0.81}$\\ 
 - $M_{\rm H2,C}$ [$\log$ M$_\odot$]  & $-10.48^{+4.71}_{-2.21}$ & $-7.98^{+14.80}_{-5.33}$ & $4.61^{+2.94}_{-2.07}$ & $5.36^{+0.64}_{-3.70}$\\ 
 - $M_{\rm H2,CO}$ [$\log$ M$_\odot$]  & $-24.56^{+12.41}_{-4.41}$ & $-15.85^{+21.40}_{-11.95}$ & $3.37^{+3.23}_{-3.18}$ & $4.07^{+0.77}_{-5.71}$\\ 
 $M_{\rm dust}$ [$\log$ M$_\odot$]  & $1.30^{+0.77}_{-0.93}$ & $1.27^{+1.59}_{-0.88}$ & $2.62^{+0.60}_{-0.45}$ & $2.18^{+0.43}_{-0.38}$\\ 
 $f_{\rm CII,H+}$  & $0.99^{+0.00}_{-0.05}$ & $0.96^{+0.04}_{-0.85}$ & $0.03^{+0.07}_{-0.02}$ & $0.02^{+0.54}_{-0.02}$\\ 
 $f_{\rm CII,H}$  & $0.01^{+0.05}_{-0.00}$ & $0.04^{+0.78}_{-0.04}$ & $0.89^{+0.09}_{-0.08}$ & $0.88^{+0.11}_{-0.48}$\\ 
 $f_{\rm CII,H2}$  & $0.00^{+0.00}_{-0.00}$ & $0.00^{+0.07}_{-0.00}$ & $0.03^{+0.06}_{-0.02}$ & $0.04^{+0.04}_{-0.04}$\\ 
    \hline 
  \end{tabular}  \\
  \tablefoot{For each parameter $U$, $n$, and cut, the first line shows the lower and upper boundaries. $Z_{\rm dust}$ was fixed to zero (i.e., corresponding to the median relationship in \citealt{Galliano2021a}).}
  \tablefoottext{a}{1: ionization front (H$^+$--H$^0$ transition), 2: PDR front (H$^0$--H$_2$ transition), 3: C$^0$--CO transition.}
  \tablefoottext{b}{Pivot point for the broken power law. } 
  \end{table*}

\end{appendix}
    
\end{document}